\documentclass[10pt]{iopart}
\usepackage{anysize}
\usepackage{float}
\usepackage{wrapfig}
\usepackage[margin=2.2cm]{geometry}
\usepackage{graphicx,subfigure,wrapfig}
\usepackage{amsmath,empheq}
\usepackage{amssymb}
\usepackage{physics}
\usepackage{mathtools}
\usepackage{cancel} 
\usepackage[cal=boondoxo]{mathalfa}
\usepackage[most]{tcolorbox}
\usepackage{cite}
\usepackage{upgreek}
\usepackage{textgreek} 
\tcbset{colback=white!10!white, colframe=black!50!black, 
        highlight math style= {enhanced, 
            colframe=black,colback=black!10!white,boxsep=0pt}
        }

\usepackage{bm}
\usepackage[dvipsnames]{xcolor} 
\usepackage{hhline}
\usepackage{ntheorem}   
\usepackage{mdframed}   

\theoremstyle{break}
\theoremheaderfont{\bfseries}
\newmdtheoremenv[%
linecolor=black,leftmargin=20,%
rightmargin=20,
backgroundcolor=gray!40,%
innertopmargin=0pt,%
ntheorem]{myprop}{Proposition}[section]

\makeatletter
\newsavebox{\@brx}
\newcommand{\llaangle}[1][]{\savebox{\@brx}{\(\m@th{#1\langle}\)}%
  \mathopen{\copy\@brx\kern-0.5\wd\@brx\usebox{\@brx}}}
\newcommand{\rraangle}[1][]{\savebox{\@brx}{\(\m@th{#1\rangle}\)}%
  \mathclose{\copy\@brx\kern-0.5\wd\@brx\usebox{\@brx}}}
\makeatother

\usepackage{tikz}
\usetikzlibrary{shapes.geometric}

\newcommand{\st}[1]{%
    \begin{tikzpicture}[baseline=(char.base)]
\node[draw, regular polygon, regular polygon sides=8, inner sep=1pt, minimum size=1.25em] (char) {\small #1};
    \end{tikzpicture}%
}

\usepackage{etoolbox}
\makeatletter
\newrobustcmd{\fixappendix}{%
  \patchcmd{\l@section}{1.5em}{7em}{}{}%
  \patchcmd{\l@subsection}{2.3em}{7em}{}{}%
}
\makeatother

\usepackage{graphicx}
\usepackage[breaklinks,colorlinks = true,linkcolor = red,urlcolor=blue,citecolor=WildStrawberry]{hyperref}
\usepackage{ulem}
\usepackage[mathscr]{eucal}
\expandafter\let\csname equation*\endcsname\relax
\expandafter\let\csname endequation*\endcsname\relax

\usepackage{amsfonts,psfrag,latexsym}
\usepackage{epstopdf}
\usepackage{color}
\definecolor{dgreen}{rgb}{0,0.7,0}
\definecolor{myc}{rgb}{0.8,0.2,0.3}
\def\redw#1{{\color{red} #1}}

\newcommand{\mc}[1]{\mathcal{#1}}  
\newcommand{\msc}[1]{\mathscr{#1}}

\newcommand{\mb}[1]{\mathbb{#1}}  
\newcommand{\mf}[1]{\mathfrak{#1}} 
\newcommand{\mt}[1]{\mathtt{#1}} 
 
\newcommand{\ovl}[1]{\overline{#1}}

\makeatletter
\newcommand{\lta}[1][]{\savebox{\@brx}{\(\m@th{#1\langle}\)}%
  \mathopen{\copy\@brx\kern-0.5\wd\@brx\usebox{\@brx}}}
\newcommand{\rta}[1][]{\savebox{\@brx}{\(\m@th{#1\rangle}\)}%
  \mathclose{\copy\@brx\kern-0.5\wd\@brx\usebox{\@brx}}}
\makeatother

\def\lla{\left \langle}
\def\rra{\right \rangle}
\def\lsq{\left [}
\def\rsq{\right ]}
\def\lf{\left (}
\def\rf{\right )}

\def \mtx{{\mathtt{x}}}
\def \mtX{{\mathtt{X}}}
\def\mtt{{\mathtt{t}}}

\def\fr{{\mathfrak{f}}}  
\def\fp{\mc{f}}         

\def\dfp{\delta \mc{f}}
\def\drp{\delta \rho}
\def\dphi{\delta \phi}
\def\dFp{\delta F}
\def\dfr{\delta \fr}
\def\drr{\delta \varrho}
\def\dvphi{\delta \varphi}

\def\bfp{\bar{\fp}}
\def\brp{\bar{\rho}}
\def\bphi{\bar \phi}
\def\bFp{\bar{F}}
\def\bfr{\bar{\fr}}
\def\brr{\bar{\varrho}}
\def\bvphi{\bar \varphi}
\def\bFr{\bar{\mathscr{F}}}
\def\bx{\bar{x}}
\def\bX{\bar{X}}
\def\dx{\delta x}
\def\dX{\delta X}
\def\bt{\bar{t}}
\def\bv{\bar{v}}

\begin{document}
\title{Quasiparticle dynamics and diffusive scale hydrodynamics in an inhomogeneous gas of hard rods}
\author{Anupam Kundu}
\address{International Centre for Theoretical Sciences, TIFR, Bengaluru -- 560089, India}
\ead{anupam.kundu@icts.res.in}
\begin{abstract}
We investigate the stochastic dynamics of quasiparticles within a gas of hard rods in one dimension for two choices of initial states: one with long-range  correlations and  the other without it.  We derive analytical results for the phase space density correlations in the former case to complement the known results for the latter case. These results enable us to obtain explicit expressions for the mean, variance, autocorrelation and cross-correlation of individual  quasiparticles, extending previous results to inhomogeneous cases. We also propose two methods for simulating an ensemble of trajectories of a quasiparticle with fixed initial position and velocity, and verify our analytical results on correlations. The long-range correlations introduce a diffusive-scale correction to the mean Euler generalized hydrodynamic equations, modifying the standard local equilibrium form, and our findings reveal that the form of the correction term depends on the long-range correlations present in the system.
\end{abstract}
\maketitle

\tableofcontents

\section{Introduction} 

%
The idea of quasiparticle is an important concept in modern condensed matter and statistical physics. In many-body systems with a very large number of interacting degrees of freedom, the collective behavior often becomes too complex to describe in terms of the original microscopic variables alone. Remarkably, however, the low-energy excitations of such systems can frequently be understood as effective particle-like entities, known as quasiparticles , which propagate through the medium while carrying physical quantities such as energy, momentum, spin, or charge.  This idea forms the basis of Landau's Fermi liquid theory and underlies much of modern condensed matter physics \cite{landau1959theory, PhysRev.127.1423, PinesNozieres1966QuantumLiquids, Shankar1994Renormalization}. 
The quasiparticle framework is a widely-employed and powerful approach, appearing in many contexts such as Fermi liquid theory \cite{landau1959theory,phillips2012advanced}, Boltzmann theory \cite{ziman2001electrons},  collective modes in spin system\cite{van1958spin},  and quantum field theories \cite{peskin2018introduction}.

A key strength of the quasiparticle picture is that it transforms the complex dynamics of interacting many-body systems into the motion and interactions of effective particles. This picture becomes exact in the context of integrable systems where quasiparticles are infinitely lived and provide a perfect basis for expressing the hydrodynamic equations of extensive number of conserved densities elegantly --  known as generalised hydrodynamics (GHD) \cite{doyonlecturenotes, bertini2016transport, castro2016emergent, doyon2025generalised}. 
Although integrability is a specific, fine-tuned characteristic, many near-integrable systems demonstrate unique large-scale relaxation effects. This has generated substantial interest in GHD, which has successfully explained ballistic-scale motion across various systems \cite{kinoshita2006quantum, caux2019hydrodynamics, malvania2021generalized, alba2021generalized}. 

In quantum integrable systems, quasiparticles are identified with stable (infinitely lived) excitations \cite{doyonlecturenotes,Hubner2025Hydrodynamic}, much like quasiparticles in condensed matter systems. Often they are characterized by the solutions (rapidities) of the Bethe equations \cite{Franchini2017IntegrableTechniques,lu2009quasiparticlesxxzmodel}. A similar quasiparticle description also exists for classical integrable systems within the framework of GHD, where quasiparticle density constitutes the elementary hydrodynamic fields \cite{spohn2018interacting, spohn2021hydrodynamic, doyon2019generalized, spohn2024hydrodynamic}. An important question is then how these quasiparticles behave at the level of individual trajectories. This is question of long-standing interest.

The motion of individual objects embedded in a many-body medium has been a central topic since Einstein's pioneering work on Brownian motion \cite{Topping_1956,Haenggi2005BrownianReview}. Similar questions arise in a wide variety of systems, including tracer diffusion in single-file systems \cite{Harris1965SingleFile,Chou2011NonEquilibrium} and in passive micro-rheology \cite{MasonWeitz1995Microrheology}. Motivated by the development of generalized hydrodynamics, analogous questions have recently been addressed for quasiparticles in classical integrable fluids \cite{bulchandani19,lepri25,aggarwal25, olla24}.

Over the past few years, quasiparticle (QP) dynamics has been investigated in several classical integrable fluids, including the hard-rod gas \cite{SciPostPhys.20.6.166}, the box-ball system \cite{olla24}, and the Toda fluid \cite{aggarwal25,chahal2026quasiparticlediffusiontodafluid}. In all these systems, the mean and mean-squared displacement of a tagged QP have been computed and shown to exhibit Brownian behavior at long times. However, these studies are restricted to homogeneous equilibrium backgrounds. In this work, we ask how the dynamics of a tagged QP is modified when the background is instead in an inhomogeneous, non-equilibrium, and non-stationary state. It has recently been shown \cite{doyon2023emergence,doyonBMFT} that, under Euler hydrodynamic evolution, such states dynamically develop long-range (LR) spatial correlations among the conserved densities on Euler (ballistic) space-time scales. These correlations are absent in homogeneous equilibrium states and are expected to modify the fluctuations of a QP  moving in inhomogeneous background. We address this question for the one-dimensional hard-rod gas. In particular, while the mean, variance, autocorrelation, and covariance of tagged QP trajectories have been obtained analytically for homogeneous backgrounds, here we extend these results to inhomogeneous backgrounds.

Since the LR correlations that develop in inhomogeneous backgrounds modify the stochastic dynamics of QPs, they are also expected to affect the hydrodynamic (HD) evolution of the conserved densities. Indeed, these correlations primarily influence the mean and  fluctuations of QP, suggesting that the corrections to the Euler generalized hydrodynamics (GHD) should also be modified. Such LR-correlation-induced corrections have recently been demonstrated in Ref.~\cite{hubner2024diffusive}. In the final part of the paper we study the correction to the Euler GHD equation arising due to the 
emergent LR correlations. 

For a class of initial states described by an almost factorized joint distribution (see Eq.~\eqref{eq:mbP_r}), where correlations arise solely from the non-overlapping constraint, the authors of Ref.~\cite{hubner2026diffusive} microscopically derived the GHD equation on diffusive scales for the one-dimensional hard-rod gas. They demonstrated explicitly how dynamically generated LR correlations invalidate the LE approximation and modify the diffusive correction to the Euler equation \cite{hubner2025diffusive,hubner2025hydrodynamics}.

In this paper, we present a similar microscopic derivation of the diffusive-scale GHD for the hard-rod gas starting from a different class of initial states, namely those factorized in the point-particle coordinates (see Eq.~\eqref{eq:mbP_p}). Unlike the initial state considered in Ref.~\cite{hubner2026diffusive}, this initial state already possesses LR correlations. We show that, in this case as well, the diffusive correction to the Euler GHD differs from the NS term obtained under the LE approximation in Refs.~\cite{boldrighini1997one,doyon2017dynamics}.

\section{Hard rod gas and initial conditions}
We consider a 1d system of $N$ hard rods, each with unit mass and length $a$. The state of the $i$-th rod is described by its position and momentum $\{\mtX_i, v_i\}$ for $i=1, 2, \dots, N$. Given that the mass is unity, the momentum of each rod is equivalent to its velocity. The dynamics is characterized by ballistic motion interrupted by elastic collisions during which the rods exchange their respective momenta.

The rods are initially confined to a region of length $L$ and allowed to move on the infinite line subsequently. We consider the thermodynamic limit, defined by $N \to \infty$ and $L \to \infty$, such that the initial density profile $\varrho_0(\mtX)$ remains a finite-valued function across all space. This system is integrable, possessing $N$ locally conserved quantities $Q_\nu$, which can be chosen as the moments of the individual velocities:
\begin{align}
Q_\nu = \sum_{i=1}^N v_i^\nu, \quad \nu=1, 2, \dots, N. \label{cons-Q}
\end{align}
In the limit $a \to 0$, the interacting hard-rod system reduces to a non-interacting hard-point gas (HPG). The microscopic dynamics of the hard rods can be formally mapped onto a system of hard-point particles, each of unit-mass, using a specific coordinate transformation \cite{percus1969exact,bernstein1988expansion,lebowitz1968time}. Given a configuration of $N$ hard rods $\{\mtX_i, v_i\}$, the corresponding point-particle configuration $\{\mtx_i, v_i\}$ is constructed as:
\begin{equation}
\label{eq:map-hr-hp}
    \mtx_i = \mtX_i - (i-1)a, \quad \text{for } i = 1, \dots, N.
\end{equation}
This mapping effectively removes the inaccessible volume occupied by the rods. In this point-particle representation, the dynamics are greatly simplified: particles move ballistically and, upon collision, exchange velocities without the spatial jumps characteristic of hard rods. Consequently, the system can be evolved as a collection of non-interacting particles, with a final relabeling based on their spatial ordering. Since the transformation in Eq.~\eqref{eq:map-hr-hp} is 
one-to-one, the hard-rod dynamics are exactly solvable by mapping the system back from the point-particle representation. 
The following two types of initial conditions are typically considered in the literature \cite{doyon2023ballistic, percus1969exact, ferrari2023macroscopic, Mrinal_2024_HR}. 
\paragraph{(a) IC `factorized' in hard rod coordinates $({\rm IC}_{\rm fhr})$:} In this case the initial positions and velocities $\{\mtX_i,v_i\}$ of the rods are chosen directly from the following joint probability distribution \cite{doyonlecturenotes,doyon2023ballistic}: 
\begin{align}
\mathbb{P}_{\rm r}(\{\mtX_i,v_i\})=\frac{1}{Z_r}\prod_{i=1}^N\Omega(\mtX_i,v_i)\prod_{i=1}^{N-1}\Theta(\mtX_{i+1}-\mtX_i-a), \label{eq:mbP_r}
\end{align}
where $\Theta(\mtX)$ is the Heaviside theta function,  $Z_{\rm r}$ is the normalization factor, and $\int d\mtX \int dv~\Omega(\mtX,v)=1$. Note that the  joint distribution is `factorized' except for the constraint $\mtX_{i+1}\ge \mtX_i +a$. We assume that the initial state varies slowly over space, characterized by the following choice:
\begin{align}
\Omega(\mtX,v)=e^{-\omega(\mtX/\ell,v)},  \label{Psi(x,v)}
\end{align}
with $\omega(\mtX,v)$ being some suitable function that goes to infinity for $|\mtX| \to \infty$ and $|v| \to \infty$. The parameter $\ell$ characterizes the length scale of spatial variation. 

\paragraph{(b) IC `factorized' in hard point particle coordinates $({\rm IC}_{\rm fhp})$):}
In this case, we first choose the configuration $\{\mtx_i,v_i\}$ in point particle coordinates from a joint distribution $\mb{P}_{\rm p}(\{\mtx_i,v_i\})$ and then transform to hard rod coordinates $\{\mtX_i = \mtx_i+(i-1)a,v_i\}$. The joint distribution is given by \cite{AnupamMFT, mukherjee2026microscopic}
\begin{align}
  \mathbb{P}_{\rm p}(\{\mtx_{i},v_{i}\},0)=N!\prod_{i=1}^{N} \psi\lf \mtx_{i}, v_i\rf ~\prod_{i=1}^{N-1}\Theta(\mtx_{i+1}-\mtx_i)
 ,\label{eq:mbP_p}
 \end{align}
where the $\Theta$ functions ensure the ordering $\{\mtx_{i+1}\ge \mtx_{i}~;~i=1,...,N\}$ and $\int d\mtx \int dv~ \psi(\mtx,v)=1$. The average single particle phase space density and mass density are given, respectively, by $\bar {\mc f}(\mtx,v)=N\psi\lf \mtx,v \rf$ and $\bar \rho(\mtx)=N\int dv~\psi\lf \mtx,v \rf$. In particular, we choose $\psi(\mtx,v)=\psi_0\lf \frac{\mtx}{\sigma},v \rf$ such that $\brp(\mtx)$ varies over a length scale $\sigma$. For the corresponding hard rod gas (obtained using the transformation in Eq.~\eqref{eq:map-hr-hp}), the initial average mass density profile is expressed in terms of $\bar \rho(\mtx)$ as $\brr\big( \mtX(\mtx)\big)=\frac{\brp(\mtx)}{1+a\brp(\mtx)}$
where  
$\mtX(\mtx)=\mt x+a\int d\mt{y}~\Theta(\mtx-\mt{y})\brp(\mt{y})$.  The transformation $\mt X(\mtx)$ is essentially the same mapping as described in Eq.~\eqref{eq:map-hr-hp}, now expressed using the mass density of the point-particles $\brp(\mtx).$ It is easy to realize that for large $N$ and small $a$ with fixed $\frac{Na}{\sigma}$, the average mass density $\brr(\mtX)$ varies over a length scale $\ell = O(\sigma)$ for large $\sigma$. Note that the densities of the hard-rod gas in the initial state ${\rm IC}_{\rm fhp}$ are already correlated over long-distances to start with. However, we believe that after a small time evolution (small compared to the hydrodynamic scale and large compared to the microscopic scale \cite{hubner2026diffusive}) the two initial states ${\rm IC}_{\rm fhr}$ and ${\rm IC}_{\rm fhp}$ become equivalent albeit the structure of long-range correlations are different. 

On macroscopic (Euler) space-time scale, the evolution of system can be  described by hydrodynamic density fields defined by coarse-graining the 
microscopic empirical phase space density $\hat \fr(\mt Z,v,\mt t)$ as 
\begin{subequations}
\label{eq:fr->scaled-fr}
\begin{align}
\begin{split}
    \hat {\bar\fr}(\mt X,v,\mtt)= \frac{1}{\Delta \mt X}\int_{\mt X-\Delta \mt X/2}^{\mt X+\Delta \mt X/2} d\mt Z ~\hat {\mf f}(\mt Z,v, \mt t),~~
    \text{where}~~~\hat {\mf f}(\mt Z,v,\mt t) = \sum_{i=1}^N\delta(\mt Z-\mt X_i(\mtt))\delta(v-v_i).
    \end{split}
    \label{def:mfb_f}
\end{align}
The coarse graining scale $\Delta \mt X$ can be chosen as $\Delta \mt X \sim \ell^\nu$ with $0<\nu<1/2$ \cite{hubner2025hydrodynamics} such that $a \ll \Delta \mt X \ll \ell$. For initial states varying over large length scales $\ell$, as in Eq.~\ref{eq:mbP_r}, it is expected,  using large-deviation argument \cite{doyon2023ballistic},  that the coarse-grained density $\hat {\bar\fr}(\mt X,v,t)$ varies over space-time scale of $O(\ell)$. In other words, one expects the following form 
\begin{align}
  \hat {\bar\fr}(\mt X,v,\mtt) = \fr\lf \frac{\mt X}{\ell},v,\frac{\mt t}{\ell}\rf, \label{f_r-scaling}   
\end{align}
\end{subequations} 
where $\fr\lf X,v,t\rf$ is the scaling density. Hence, it seems convenient to work entirely in the scaled  space-time coordinates
\begin{align}
X=\mtX/\ell,~~t =\mtt/\ell, \label{scaled-coordinates}
\end{align} 
and density $\fr(X,v,t)$. A similar coarse-grained phase space density of point particles can also be defined as 
\begin{align}
  \begin{split}
    \mc {f}(x,v,t)= \frac{1}{\Delta \mt x}\int_{\ell x-\Delta \mt x/2}^{\ell x+\Delta \mt x/2} d\mt z ~\hat {\mc f}(\mt z,v, \mt t=\ell t), 
    ~~\text{where}~~~\hat {\mc f}(\mt z,v,\mt t) = \sum_{i=1}^N\delta(\mt z-\mt x_i(\mtt))\delta(v-v_i).
    \end{split}
    \label{def:mfb_p}  
\end{align}
Ideally, coarse-grained fields should carry an additional label $\ell$, however, in the limit of large $\ell$, it is expected that they converge to  smooth fields \cite{doyon2023ballistic}, independent of $\ell$. Hence, we drop this label to make the notation simple.
For each microscopic configuration $\{X_i,v_i\}$ randomly chosen from the ensemble, one can construct a coarse-grained density field $\fr(X,v,t)$ which  fluctuates randomly. For generic integrable systems, the GHD provides the evolution of the mean density field $\bfr(X,v,t) = \lla \mf{f}( X,v, t)\rra$ (averaged over initial configurations), starting from a mean profile $\bfr( X,v,0)$ \cite{doyon2025generalised, alba2021generalized}. For large $\ell$,  the average density evolves according to the Euler GHD equation \cite{hubner2025hydrodynamics, doyon2023ballistic}
\begin{align}
\partial_{ t} \bfr( X,v, t)+\partial_{ X}\big(v_{\rm eff}(X,v,t) \bfr(X,v, t)\big)=0, \label{eq:eghd}
\end{align}
on ballistic space-time scale ($ X \sim  t$), where the effective velocity is 
\begin{align}
v_{\rm eff}(X,v,t)&= 
\frac{v-a\int dv~v~\bfr ( X,v, t)}{1-a \brr( X, t)},~~~{\rm with}~~~\brr( X,t) = \int dv ~\bfr( X,v,  t).
\label{def:v_eff}
\end{align}
As will be shown in the next section, the displacement of a QP, initially located at $X_q$ with velocity $v_q$, can be expressed as a functional of the phase-space density $\fr(X,v,t)$. The contribution arising from the Euler evolution of the mean density profile $\bfr(X,v,t)$ follows directly from the solution of the Euler equation. Our goal, however, is to characterize QP dynamics beyond this Euler-coordinate description. Since the displacement depends nonlinearly on the fluctuating density field, its statistical properties receive contributions from the space-time correlations of the density fluctuations. Indeed, for inhomogeneous initial states, the system develops long-range  spatial correlations, which modify the mean displacement, variance, and other correlations compared with the homogeneous case. For the initial state ${\rm IC}_{\rm fhr}$, the contributions of LR correlations to the mean and variance were recently computed in Ref.~\cite{hubner2026diffusive,hubner2025hydrodynamics}. We extend the framework developed for the ${\rm IC}_{\rm fhr}$ initial state to investigate QP dynamics for ${\rm IC}_{\rm fhp}$. We derive explicit analytical expressions for the mean and variance of the QP position and show that the same framework naturally extends to obtain analogous expressions for the covariance and autocorrelation of both the QP position and velocity.

The second main goal of the paper is to find the diffusive scale correction to the Euler equation~\eqref{eq:eghd}.   It has been shown that, for homogeneous equilibrium state the diffusive correction is given by the following Navier-Stokes equation \cite{boldrighini1997one, doyon2017dynamics}6 
\begin{subequations}
\label{eq:hd_diff-Boldrigni}
 \begin{align}
 \partial \bfr(X,v,t)+\partial_X\big(v_{\rm eff}(X,v,t) \bfr(X,v, t)\big) =\frac{1}{2\ell}\partial_X \int du ~\msc{D}(v,u) \partial_X\bfr(X,u,t),
\end{align}
where 
\begin{align}
 \msc{D}(v,u) = \frac{1}{1-a \brr(X,t)}\lf \delta(u-v) \int dw~|v-w|~\bfr(X,w,t) - |u-v|~ \bfr(X,v,t)\rf.  
 \label{def:mscD}
\end{align}   
\end{subequations}
However, in inhomogeneous state,  the LR correlations affect the mean and correlations in the QP dynamics in a nontrivial way. Consequently, such correlations are expected to  get manifested into the hydrodynamic equation as correction at diffusive scale. Recently, it has been shown that for an inhomogeneous local equilibrium state, the above diffusive scale equation is modified \cite{hubner2026diffusive, hubner2025hydrodynamics}. The usual derivation of the diffusive scale correction under the assumption of local equilibrium fails because of the emergence of the LR correlation on the Euler space-time scale during evolution \cite{doyon2023emergence}. The presence of such long range correlations indicates that the local state of the system is not in equilibrium form, and consequently such correlations modify the evolution conserved densities in integrable systems at the diffusive space-time scale \cite{hubner2025diffusive}. Based on a microscopic description, such modifications have been rigorously computed for hard rods for  ${\rm IC}_{\rm fhr}$ [see Eq.~\eqref{eq:mbP_r}] initial state \cite{hubner2026diffusive, hubner2025hydrodynamics}. Explicit forms of the correction on diffusive space-time scale have been obtained. In the second part of the paper we derive a similar diffusive scale correction for the other initial ensemble ${\rm IC}_{\rm fhp}$ described by Eq.~\eqref{eq:mbP_p}.

\paragraph{Organization of the paper:} In Sec.~\ref{sec:lr-C}, we derive the LR correlations generated in case of inhomogeneous initial states. For generic integrable systems, such LR correlations have been predicted first in \cite{doyon2023emergence} using ballistic macroscopic fluctuation theory (BMFT) and later  have been computed  explicitly for hard rod gas in \cite{doyon2023ballistic, AnupamMFT, kethepalli2025ballistic} for ${\rm IC}_{\rm fhr}$ initial state directly in the hard-rod picture ({\it i.e.,} without going to the point-particle picture). In this paper we compute this correlation using correlations in the associated point particle densities (or, in other words, densities of contracted coordinates \cite{hubner2025hydrodynamics}). In sec.~\ref{sec:qp-dyna}, we discuss how to express the position of the tagged QP as a non-linear functional of the phase space density and show how the LR correlations determine the statistical properties of QP dynamics. 
The explicit calculations of the mean, variance, covariance, and two point correlation of the QPs  are presented in Sec.~\ref{sec:computation-statistics}, and are verified numerically in Sec.~\ref{sec:num-veri}. 
In Sec.~\ref{sec:hd_diff}, we derive  the diffusive scale correction to the Euler GHD equation explicitly and show how the LR correlations govern the form of this correction.  Finally, we conclude our paper in Sec.~\ref{conclusion}. Some details of the calculations are deferred to the Appendix to keep the presentation focused on the main findings.

\section{Long-range correlation}
\label{sec:lr-C}
In this section, we study the correlation of the phase space density on the Euler (ballistic) space-time scale. For that we follow \cite{hubner2026diffusive} and define the height fields
\begin{align}
\begin{split}
\varphi(X,v,t) &= \int dY~\Theta(X-Y)\fr(Y,v,t), \\
\phi(x,v,t) &= \int dy~\Theta(x-y)\mc{f}(y,v,t),
\end{split}
\label{def:height-field}
\end{align}
which evolve according to 
\begin{align}
\begin{split}
\partial_t \varphi(X,v,t) &= - v_{\rm eff}[\fr](X,v,t)~ \partial_X \varphi(X,v,t), \\
\partial_t \phi(x,v,t) &= - v ~\partial_x \phi(x,v,t),
\end{split}
\label{evo-eq-phi}
\end{align}
where $v_{\rm eff}[\fr]$ is given in Eq.~\eqref{def:v_eff}. It is easy to see that these two height fields are related  
\begin{align}
\phi(x_t(X),v,t) =\varphi(X,v,t),~~\text{and}~~\varphi(X_t(x),v,t)=\phi(x,v,t),
\end{align}
where
\begin{subequations}
\begin{align}
x_t(X)&=X - a\int dv~\varphi(X,v,t) = X-a\int dv \int dY \Theta(X-Y)\fr(Y,v,t), \label{trans:X->x-hf} \\
X_t(x)&=x+a\int dv~\phi(x,v,t)= x+a\int dv \int dy \Theta(x-y)\fp(y,v,t). \label{trans:x->X-hf}
\end{align} 
\label{trans:X<->x-hf}
\end{subequations}
These coordinates  evolve according to 
\begin{align}
\frac{d x_t(X)}{dt} &= a \int dv~v~\fr(X,v,t), ~~
\frac{d X_t(x)}{dt} = a \int dv~v~\fp(x,v,t). 
\label{eq:dX/dt-&-dx/dt}
\end{align}
Given the fluctuations of the phase space densities
$\fr(X,v,t) =\bfr(X,v,t) +\dfr(X,v,t)$ and $\fp(x,v,t) = \bar \fp (x,v,t) +\dfp (x,v,t)$, the fluctuations of the  height fields, at linear order, are given by
\begin{subequations}
\label{del-phi-relations}
\begin{align}
\dvphi(X,v,t) &= \int du~\left (\delta(v-u) -a \frac{\bar \fp(\bx_t(X),v,t)}{1+a \brp(\bx_t(X),t)} \right)~\dphi(\bx_t(X),u,t), \label{del-phi-relations-1}\\
\dphi(x,v,t) &= \int du~\left (\delta(v-u) +a \frac{\bfr(\bar X_t(x),v,t)}{1-a \brr(\bar X_t(x),t)} \right)~\dvphi(\bar X_t(x),u,t),
\label{del-phi-relations-2}
\end{align}
\end{subequations}
where, $\bfp(x,v,t) = \lla \fp(x,v,t)\rra$ and $\bfr(X,v,t) = \lla \fr(X,v,t)\rra$. Similarly, we denote the mean mass densities by $\brp(x,t)=\langle \rho(x,t)\rangle$ and $\brr(x,t)=\langle \varrho(X,t)\rangle$. Furthermore, the transformations $\bx_t(X)$ and $\bX_t(x)$ used in Eq.~\eqref{del-phi-relations}, are given by Eq.~\eqref{trans:X<->x-hf} with $\fr(X,v,t), \fp(x,v,t)$ are replaced, respectively, by $\bfr(X,v,t), \bfp(x,v,t)$ on the right hand side. In terms of the height field fluctuations the correlations of the phase space densities are defined as 
\begin{align}
\mc{C}_{\rm r}(\ell X,v,\ell t; \ell Y, u,\ell t')= \lla \dfr(X,v,t)\dfr(Y,u,t') \rra  &= \partial_X \partial_Y \lla \dvphi(X,v,t) \dvphi(Y,u,t')\rra, \label{def:C_r} \\
\mc{C}_{\rm p}(\ell x,v,\ell t; \ell y, u,\ell t')= \lla \dfp(x,v,t)\dfp(y,u,t') \rra &= \partial_x \partial_y \lla \dphi(X,v,t) \dphi(Y,u,t')\rra. 
\label{def:C_p}
\end{align}
It was shown that the correlation of  hard rod density has the following scaling form \cite{doyon2023ballistic, AnupamMFT} 
 \begin{subequations}
 \label{corr-scling+structure}
\begin{align}
\mc{C}_{\rm r}(\ell X,v,\ell t; \ell Y, u,\ell t')&= \frac{1}{\ell}\msc{C}_{\rm r}(X,v,t;Y,u,t').
\end{align}
We will below show that for both choices of initial states, the scaled correlation at equal time $t=t'$ has two parts -- one singular and the other non-singular but  long-ranged:
\begin{align}
\msc{C}_{\rm r}(X,v,t;Y,u,t) = \delta(X-Y)\msc{C}^{\rm r}_{\rm gge}(X,u,v) + \msc{C}^{\rm r}_{\rm lr}(X,u;Y,v;t).
\label{C_r-st}
\end{align}
 \end{subequations}
The singular part $\msc C_{\rm gge}^{\rm r}(X,u,v)$  
represents the contribution of local GGE approximation of the microscopic distribution in a fluid cell and the long range part $\msc C_{\rm lr}^{\rm r}(X,u;Y,v;t)$ comes from the fact that these distributions in two far apart fluid cells are weakly correlated [at $O(1/\ell)$] \cite{doyon2023emergence}. For point particle gas, on the other hand,  the LR part exists only for ${\rm IC}_{\rm fhr}$ initial condition and one has $\mc{C}_{\rm p}(\ell x,v,\ell t; \ell y, u,\ell t)= \frac{1}{\ell}\msc{C}_{\rm p}(x,v,t;y,u,t)$ with $\msc{C}_{\rm p}(x,v,t;y,u,t) = \delta(X-Y)\msc{C}^{\rm p}_{\rm gge}(x,u,v) + \msc{C}^{\rm p}_{\rm lr}(x,u;y,v;t)$. 
In the following, we will use relations between fluctuations of hard-rod and point particles densities in Eq.~\eqref{del-phi-relations} to compute these correlations in hard-rod densities from point-particle densities. 

\subsection{Hard rod correlation from point particle picture}
We start by computing $\langle \dvphi(X,v,t)\dvphi(Y,u,t') \rangle$. Using the relation between $\dvphi$ and $\dphi$ in 
Eq.~\eqref{del-phi-relations-1}, we first get  
\begin{align}
&\langle \dvphi(X,v,t)\dvphi(Y,u,t') \rangle = \int dp_1 \int dp_2 \left (\delta(v-p_1) -a \frac{\bar \fp(\bx_t(X),v,t)}{1+a \brp(\bx_t(X),t)} \right) \notag \\ 
&~~~~~~~~~~\times~\left (\delta(u-p_2) -a \frac{\bar \fp(\bx_{t'}(Y),u,t')}{1+a \brp(\bx_{t'}(Y),t')} \right)~\langle \dphi(\bx_t(X),p_1,t)\dphi(\bx_{t'}(Y),p_2,t')\rangle, \notag \\
&~~~~~~~~~~= \int dp_1 \int dp_2 \left (\delta(v-p_1) -a \bfr(X,v,t) \right)\left (\delta(u-p_2) -a \bfr(Y,u,t') \right) \notag \\
&~~~~~~~~~~~~~~~~~~~~~~~~\times ~\langle \dphi(\bx_t(X)-p_1t,p_1,0)\dphi(\bx_{t'}(Y)-p_2t',p_2,0)\rangle, \label{corr-vrphi-1} 
\end{align}
where $\bx_t(X)$ is given in Eq.~\eqref{trans:X<->x-hf} with $\varphi(X,v,t)$ replaced by $\bvphi(X,v,t) = \lla \varphi(X,v,t) \rra$.
This correlation for the ${\rm IC}_{\rm fhr}$ initial state (`factorized' in hard rod coordinates) has been computed previously in detail using various methods \cite{doyon2023emergence, doyon2023ballistic, AnupamMFT}. In this paper, we mainly focus  on computing this correlation for the  initial state ${\rm IC}_{\rm fhp}$ `factorized' in point particle coordinates (see Eq.~\eqref{eq:mbP_p}). For comprehensiveness, we also present the results for  initial  ${\rm IC}_{\rm fhr}$  in 
\ref{app:eq:<dfp-dfp>(0)-hrIC}.

\paragraph{\bf Computation of correlation for ${\rm IC}_{\rm fhp}$:}
For this case, it is easy to show that the two point correlation of the point particle densities is given by
\begin{align}
\langle \dfp (y,v,0)\dfp (z,u,0)\rangle= \frac{1}{\ell} \delta(v-u)~\delta(y-z)~\bar \fp (y,v,0)-\frac{\bfp(z_1,p_1,0)\bfp(z_2,p_2,0)}{N}, \label{def:eq-<ff>-pp}
\end{align}
which implies
\begin{align}
\begin{split}
\langle \dphi(x,v,0)\dphi(z,u,0)\rangle &= \int dx_1 \int dx_2 ~\Theta(x-x_1)\Theta(z-x_2)~\langle \dfp (x_1,v,0)\dfp (x_2,u,0)\rangle, \\
&=\frac{1}{\ell} \delta(v-u)\bphi(\min(y,z),v,0) 
-\frac{ \bar \phi(x,v,0)\bar \phi(z,u,0)}{N}
\end{split}
\label{eq:corr-phi}
\end{align}
with $\bar\phi(y,w,0) = \int dx~\Theta(y-x)~\bar \fp (x,w,0)$. 
The  terms proportional to $1/N$ on the right hand sides of the above two equations are finite size correction to the leading order terms in the thermodynamic limit. These corrections would become important for numerical verification (see sec.~\ref{sec:num-veri}). For simplicity of the presentation,  we neglect these finite size corrections in the following discussion. 

Using Eq.~\eqref{eq:corr-phi} in Eq.~\eqref{corr-vrphi-1}, we find 
\begin{align}
\begin{split}
 \langle \dvphi(Y,v,t)\dvphi(Z,u,t') \rangle 
 &=  \frac{1}{\ell}  
 \int dp_1 \left (\delta(v-p_1) -a \bfr(Y,v,t) \right)\left (\delta(u-p_1) -a \bfr(Z,u,t') \right)  \\ 
 &~~~~~~~~~~~~~~~~~~~\times~~~~\bphi(\min(\bx_t(Y)-p_1t,\bx_{t'}(Z)-p_1t'),p_1,0),  \\
 &=\frac{1}{\ell}  
 \int dp_1 \left (\delta(v-p_1) -a \bfr(Y,v,t) \right)\left (\delta(u-p_1) -a \bfr(Z,u,t') \right)  \\ 
&~~~~~~~~~~~~~~~~~~~\times~~~~\bvphi(\bX_t(\min(\bx_t(Y),\bx_{t'} (Z)-p_1(t'-t))),p_1,t),
\end{split}
 \label{corr-vrphi-3}
\end{align}
where $\bX_t(x)$ is given in Eq.~\eqref{trans:x->X-hf} with $\fp(y,v,t)$ inside the integrand replaced by $\bfp(y,v,t)$. 
Inserting this result in Eq.~\eqref{def:C_r} we find the following expression for the unequal time correlation $\msc C_{\rm r}(Y,v,t;Z,u,t')=\ell \lla \dfr(Y,v,t)\dfr(Z,u,t') \rra $:
\begin{align}
\begin{split}
\msc C_{\rm r}(Y,v,t;Z,u,t')= \int dw \big[ &\delta\lf \bx_t(Y)-wt-\bx_{t'}(Z)+wt'\rf (1-a\brr(Z,t))\bfr(Y,w,t) \\
&~~\times~\left\{ \lf \delta(v-w)-a\bfr(Y,v,t)\rf \lf \delta(u-w)-a\bfr(Z,u,t')\rf\right\} \\
&-a\Theta\lf \bx_t(Y)-wt-\bx_{t'}(Z)+wt'\rf \lf \delta(u-w)-a\bfr(Z,u,t')\rf  \\ 
&~~~~~~~~~~~~~~~~~~~~~\times~~\bfr(Z,w,t')\partial_Y\bfr(Y,v,t) \\
&-a\Theta\lf \bx_t(Z)-wt'-\bx_{t'}(Y)+wt\rf \lf \delta(v-w)-a\bfr(Y,v,t)\rf \\ 
&~~~~~~~~~~~~~~~~~~~~~\times~~
\bfr(Y,w,t)\partial_Z\bfr(Z,u,t') \\
&+a^2 \bphi(\bX_t(\min(\bx_t(Y),\bx_t(Z)-w(t'-t)),w,t) \\
&~~~~~~~~~~~~~~~~~~~~~~~~~~~~~~~~
\times~~\lf \partial_Y\bfr(Y,v,t)\rf \lf\partial_Z\bfr(Z,u,t') \rf\big].
\end{split}
\label{def:<f_r(Yvt)f_r(Zut')>_ICpp}
\end{align}
It is easy to see that for $t=t'$ the expression simplifies and in fact, it has the structure of Eq.~\eqref{C_r-st} with 
\begin{subequations}
    \label{def:<dfrdfr>-pp-ini}
\begin{align}
\msc{C}_{\rm gge}^{\rm r}(X,v,u)
=&\lsq \delta(u-v)\bfr(X,v,t) -a(2-a\rho(X,t))\bfr(X,v,t)\bfr(Y,u,t)\rsq 
\label{mscC_gge-<dfrdfr>-hpIC}
\end{align}
and
\begin{align}
\begin{split}
\msc{C}^{\rm r}_{\rm lr}(X,v;Y,u;t)&=~~~a\Theta(X-Y)\Big[ \left(\partial_X\bfr(X,v,t)\right)\left(\partial_Y\bfr(Y,u,t)\right)\bFr(Y,t) \\ 
&~~~~~~~~~~~~~~~~~~~~~~~~~~~~~ 
-\left(\partial_X\bfr(X,v,t)\right)~(1-a\brr(Y,t))\bfr(Y,u,t)  \Big ]
 \\
&~~~+~a\Theta(Y-X)\Big [
\left(\partial_X\bfr(X,v,t)\right)\left(\partial_Y\bfr(Y,u,t)\right)\bFr(X,t) \\ 
&~~~~~~~~~~~~~~~~~~~~~~~~~~~~~ 
-\left(\partial_Y\bfr(Y,u,t)\right)~(1-a\brr(X,t))\bfr(X,v,t) \Big ] 
\end{split}
\label{mscC_lr-<dfrdfr>-hpIC}
\end{align}
\end{subequations}
where $\bFr(X,t)=\int dZ \int dv ~\Theta(X-Z)\bfr(Z,v,t)$. 
We observe that the LR part has a jump at $X=Y$. We point out that the mass density-density correlation can be obtained by integrating the phase-space density correlation $\msc C_{\rm r}(Y,v,t;Z,u,t')$ in Eq.~\eqref{def:<f_r(Yvt)f_r(Zut')>_ICpp} over the velocities, which agrees with the result obtained using both BMFT and microscopic computations in Ref:~\cite{mukherjee2026microscopic}.

In summary, the equal-time correlation $\msc C_{\rm r}(X,v;Y,u;t)$ of the hard rod gas has a singular and a non-singular part, as stated in Eq.~\eqref{C_r-st}. The singular part represents the GGE correlation. The non-singular part represents LR correlation, which has a discontinuity at $X=Y$. Writing Eq.~\eqref{mscC_lr-<dfrdfr>-hpIC} in the following form 
\begin{align}
 \msc C^{\rm r}_{\rm lr}(X,v;Y,u;t) = \Theta(Y-X)\msc C^{\rm r,+}_{\rm lr}(X,v,Y,u,t) + 
 \Theta(X-Y) \msc C^{\rm r,-}_{\rm lr}(X,v,Y,u,t),
 \label{ex:C^lr-discont}
\end{align}
we identify 
\begin{align}
\msc C^{\rm r,+}_{\rm lr}(X,v,Y,u,t) =    a\Big [
\left(\partial_X\bfr(X,v,t)\right)\left(\partial_Y\bfr(Y,u,t)\right)\bFr(X,t) 
-\left(\partial_Y\bfr(Y,u,t)\right)~(1-a\brr(X,t))\bfr(X,v,t) \Big ],
\end{align}
and $\msc C^{\rm r,-}_{\rm lr}(X,v,Y,u,t)= \msc C^{\rm r,+}_{\rm lr}(Y,u,X,v,t)$.
The presence of such jumps in the equal-time correlation functions was first predicted in \cite{doyon2023emergence} for generic integrable systems and was explicitly demonstrated for hard rod gas in \cite{hubner2025diffusive,hubner2026diffusive} for ${\rm IC}_{\rm fhr}$. Here we have explicitly demonstrated the presence of such a jump at $X=Y$ for initial state ${\rm IC}_{\rm fhp}$. 

Near the jump, the correlation function can be split into  symmetric and  anti-symmetric parts
\begin{subequations}
\label{decompose-C-sym-asymp}
  \begin{align}
\msc C^{\rm r}_{\rm lr}(X,v;Y,u;t) \overset{X \approx Y}{ =} 
\msc C^{\rm r,sym}_{\rm lr}(X,v,u,t) + \text{sgn}(X-Y)\msc C^{\rm r,asym}_{\rm lr}(X,v,u,t),
\end{align}
where 
\begin{align}
  \msc C^{\rm r,sym}_{\rm lr}(X,v,u,t)=& \frac{1}{2}\lf\msc C^{\rm r,+}_{\rm lr}(X,v,X,u,t)+\msc C^{\rm r,-}_{\rm lr}(X,v,X,u,t) \rf, \\
  \msc C^{\rm r,asym}_{\rm lr}(X,v,u,t)=& \frac{1}{2}\lf\msc C^{\rm r,+}_{\rm lr}(X,v,X,u,t)-\msc C^{\rm r,-}_{\rm lr}(X,v,X,u,t) \rf.
\end{align}  
\end{subequations}
Note that the un-equal time correlation does not have such singularities. 
These features of the correlation, both equal and un-equal time, will be useful for deriving diffusive scale hydrodynamics later in sec.~\ref{sec:hd_diff}.

\noindent
\paragraph{\bf Computation of correlation for ${\rm IC}_{\rm fhr}$:} As shown in \ref{app:eq:<dfp-dfp>(0)-hrIC}, a similar calculation can also be done for the ${\rm IC}_{\rm fhr}$ initial condition (Eq.~\eqref{eq:mbP_r}) in which the hard rod coordinates are directly sampled. In this case also the equal time correlation has two parts -- a sigular GGE contribution and a discontinuous LR correlation as in Eq.~\eqref{C_r-st}. Furthermore, the LR part can again be decomposed into a symmetric and an antisymmetric part near the jump at $X=Y$, as was shown previously in \cite{hubner2026diffusive, hubner2025hydrodynamics}.


\section{QP dynamics}
\label{sec:qp-dyna}
In the hard-rod gas, a QP is a rod tagged with a specific initial position $X_q$ and velocity $v_q$. As the physical rods collide and exchange their velocities, the tag also jumps from one rod to another. Consequently, a QP consumes an instantaneous displacement equal to the length of the rod at each collision. Hence, the ballistic motion of a QP is interrupted by  stochastic jumps at random collisions with other rods. In this section, we provide exact analytical expression of the mean, variance, covariance and other two-point correlations of individual tagged QPs moving inside a  gas of hard-rods distributed starting from an arbitrary inhomogeneous initial state of type ${\rm IC}_{\rm fhp}$.

We start by considering the  motion of a QP rod starting at the location $\mt X_q(0)=\mtX_q$ with velocity $v_q$.  For a given sample of the initial configuration of the position and the velocities of other rods, the location $\mtX_q(\mtt)\equiv \mtX_{\mtt}(\mtX_q(0),v_q)$ of the QP  at time $t$ can be written as 
\begin{align}
\mtX_q(\mtt) &= \hat \mtx(\mt X_q(0)) +v_q \mtt +a\sum_{r\ne q} \Theta\left[\hat \mtx(\mtX_q(0))+v_q\mtt-\hat \mtx(X_r(0))-v_r\mtt \right],
\label{def:X_q-dis}\\
\text{where,}~~ 
\hat \mtx(\mt X_q)&=\mtX_q - a\sum_{r\ne q}\Theta\left[\mtX_q-\mtX_r\right]. \label{trans:X->x-dis}
\end{align}
The above way of defining the dynamics of the QP is natural for ${\rm IC}_{\rm fhr}$ as the coordinates in the initial states are chosen directly in hard-rod coordinates only. For the ${\rm IC}_{\rm fhp}$ initial state, the coordinates are first chosen from the point particle picture and then transformed to the hard rod coordinates. In this case,  one would like to start with a definition of $\mtX_q(t)$ in terms of the coordinates of the point particles. Formally, it can be written as 
\begin{align}
\mtX_q(\mtt) =& \mtX_q(0) - a \sum_{r\ne q}\Theta[\mtX_q(0)-\hat \mtX(\mt x_r(0))] +v_q\mtt+a\sum_{r\ne q}\Theta[\mtX_q(\mtt) -\hat \mtX(\mt x_r(\mtt))], \label{def:X_q-dis-pp}\\
\text{where,}~~ 
\hat \mtX(\mtx_q)&=\mtx_q+ a\sum_{r\ne q}\Theta\left[\mtx_q-\mtx_r\right]. \label{trans:x->X-dis}
\end{align}
Note $\mt X_q(\mtt)$ appears on both sides of Eq.~\eqref{def:X_q-dis-pp}. 
It is easy to see that  
 $\sum_{r\ne q}\Theta[\mtX_q-\hat \mtX(\mtx_r)] = \sum_{r\ne q}\Theta[\hat \mtx(\mtX_q)-\mtx_r] = \sum_{r\ne q}\Theta[\hat \mtx(\mtX_q)-\hat \mtx(\mtX_r)] $, for any configuration at any time because $\hat \mtX(\hat \mtx(X_q))=\mtX_q$ and $\hat \mtx(\hat \mtX(\mtx_q))=\mtx_q$. Hence, the last sum on the rhs of Eq.~\eqref{def:X_q-dis-pp} can be written as 
 $a\sum_{r\ne q}\Theta[\mtX_q(\mtt) -\hat \mtX(\mt x_r(\mtt))] = 
 a\sum_{r\ne q}\Theta[\hat \mtx(\mtX_q(\mtt)) -\hat\mtx(\hat \mtX(\mt x_r(\mtt)))]= a\sum_{r\ne q}\Theta[\hat \mtx(\mtX_q(\mtt)) -\mt x_r(\mtt)]$. Using $\hat \mtx(\mtX_q(\mtt))=\hat \mtx(\mtX_q(0))+v_q \mtt$ and similarly, $\mtx_r(\mtt) = \hat \mtx(\mtX_r(0))+v_r \mtt$, we see that
 the expressions of $\mtX_q(\mtt)$ in Eqs.~\eqref{def:X_q-dis-pp} and ~\eqref{def:X_q-dis} are exactly the same.
Hence, one can work with the definition in Eq.~\eqref{def:X_q-dis} for both choices of the initial states. 

In the thermodynamic limit, the Eq.~\eqref{def:X_q-dis} can be written in terms of the empirical single particle phase space density $\fr(X,v,t)$, defined in Eq.~\eqref{f_r-scaling}, as follows [see Appendix~\ref{prf:def:X_q-cont}]
\begin{align}
X_t(X_q,v_q) &= x_0(X_q) +v_q t +a\int dY \int dv~\fr(Y,v,0)\Theta\left(x_0(X_q)+v_qt-x_0(Y)-vt\right), 
\label{def:X_q-cont}
\end{align}
where $x_t(X)$ is given in Eq.~\eqref{trans:X->x-hf}.
Clearly $X_0(X_q,v_q)=X_q$.  Note that  the expression on the right hand side of Eq.~\eqref{def:X_q-cont} involves several fluctuating terms, such as $\fr(Y,v,t)$, $x_t(Y)$ and $x_t(X_q)$.  We separate the fluctuations from the mean of each stochastic contributions as
\begin{align}
    \fr(Y,u,t) &= \bfr(Y,u,t) + \dfr(Y,u,t),~~\text{and}~~x_t(X) = \bx_t(X) + \dx_t(X), \\
    \text{where}~~&~~\bx_t(X)=\langle x_t(X)\rangle= X - a\int dY~\int du~\Theta(X-Y)\bfr(Y,u,t), \\
    \text{and}~~&~~\dx_t(X) = -a \int dY \int du~\Theta(X-Y)~\dfr(Y,u,t). \label{def:dx_0(X)}
\end{align}
with $\bfr(Y,v,t) = \langle \fr(Y,v,t) \rangle$. 
Expanding  to quadratic orders in fluctuations, one can write 
\begin{subequations}
\label{ex:X_t-expanded}
\begin{align}
X_t(Z,v) \approx X_t^{\rm eu}(Z,v) +\Delta X_t(Z,v) + \widehat{\Delta X}_t(Z,v), \label{ex:X_t-expanded-1}
\end{align}
where
\begin{align}
X_{t}^{\rm eu}(Z,v) = \bx_0(Z) +vt + a\int dY \int du~\bfr(Y,u,0)\Theta\left(\mc U^0_{t}(Z,v;Y,u)\right) , \label{X_eu(t)}
 \end{align}
with 
\begin{align}
   \mc U^t_{\tau}(Z,v;Y,u) = \bx_t(Z) +v {\tau} - \bx_t(Y)-u {\tau}, 
   \label{def:mcU^t_tau}
\end{align}
and 
\begin{align}
\begin{split}
   \Delta X_{t}(Z,v)&= \dx_0(Z) +a \int dY\int du~\Theta \left(\mc U^0_{t}(Z,v;Y,u)\right) {\dfr}(Y,u,0),  \\
   &+a \int dY \int du ~  \delta \left( {{\mc U}}^0_{t}(Z,v;Y,u) \right) \big(\dx_0(Z) -\dx_0(Y)\big) \bfr(Y,u,0), \\
   \end{split}
   \label{def:DelX_(X_q,v_q)(t)-hr-pic}
\end{align}
and
\begin{align}
\begin{split}
 \widehat{\Delta X}_t(Z,v) =&~a  \int dY \int du~\delta\big(\mc U^0_{t}(Z,v;Y,u)\big)~
 \big(\dx_0(Z) -\dx_0(Y)\big) ~\dfr(Y,u,0)  \\
 &~ + \frac{a}{2} \int dY \int du~\delta'\big(\mc U^0_{t}(Z,v;Y,u)\big)~\bfr(Y,u,0)~
  \big(\dx_0(Z) -\dx_0(Y)\big )^2. 
  \end{split}
  \label{def:hat_X_t}
\end{align}
\end{subequations}
Here $\int dZ~\delta'(Z)g(Z)=-\int dZ~\delta(Z)\frac{dg(Z)}{dZ}$. From Eqs.~\eqref{ex:X_t-expanded-1} and \eqref{def:hat_X_t}, it is evident that the displacement depends non-linearly on the fluctuations. Hence there will be a contribution to the QP statics from the long-range correlations of the phase-space density.  In the following we will use the expression of $X_t(X_q,v_q)$ in Eq.~\eqref{ex:X_t-expanded} to compute the mean, variance and other two-point correlations of  position and velocity of the $q^{\rm th}$ QP. 

\section{Computation of mean, variance and correlations}
\label{sec:computation-statistics}

\subsection{Mean: $\lla X_t(X_q,v_q)|X_q,v_q\rra$}
To compute the mean position of the $q^{\rm th}$ QP starting from location $X_q$ with velocity $v_q$, we need to perform a conditional average $\lla X_t(X_q,v_q)|X_q,v_q\rra$. This quantity is essentially 
\begin{align}
\lla X_t(X_q,v_q)|X_q,v_q\rra =&    \frac{\lla \int dZ \int dv~ \fr(Z,v,0) X_t(Z,v)\delta(Z-X_q)\delta(v-v_q)\rra}{\bfr(X_q,v_q,0)} \notag \\ 
=& 
 \frac{\lla \fr(X_q,v_q,0) X_t(X_q,v_q)\rra}{\bfr(X_q,v_q,0)}.
 \label{def:cond_<X-t>-1}
\end{align}
Inserting the expression of $X_t(X_q,v_q)$ from Eq.~\eqref{ex:X_t-expanded} and splitting $\fr(Z,v,0)=\bfr(Z,v,0)+\dfr(Z,v,0)$, one finds
\begin{align}
 \begin{split}
 \langle X_t(X_q,v_q)|X_q,v_q \rangle 
 &= X_t^{\rm eu}(X_q,v_q)
  + a  \int dv~
 \frac{\lla \left(\dx_0(X_q) -\dx_0(\bX_0(\bx_t^{q,v}))\right) ~\dfr(\bX_0(\bx_t^{q,v}),v,0) \rra}{1-a\brr(\bX_0(\bx_t^{q,v}),0)}
\\
 &~+ \frac{a}{2}  \int dv~\partial_{\bx_t^{q,v}}\left[\frac{\bfr(\bX_0(\bx_t^{q,v}),v,0)}{1-a\brr(\bX_0(\bx_t^{q,v}),0))}
 \left \langle \big(\dx_0(X_q) -\dx_0(\bX_0(\bx_t^{q,v}))\big )^2\right \rangle\right], 
 \end{split}
 \label{def:<X_t(X_q,v_q)>-hr-pic-pp-ini-bis}
\end{align}
where 
\begin{align}
\bx_t^{q,v}=\bx_0(X_q)+v_qt-vt, \label{def:bx_t^qv}
\end{align}
 and $X_{\rm eu}(t)$ is given in Eq.~\eqref{X_eu(t)}. To evaluate the mean we need two types of correlations $\lla \dfr(Y,v,0)\dx_0(Z)\rra$ and $\lla \dx_0(Y)\dx_0(Z)\rra$. Performing appropriate integrals over the velocities and derivatives with respect to position arguments, one can compute these correlations from the correlations in Eq.~\eqref{corr-vrphi-3}. One finds, 
\begin{align}
&\langle \dfr(Y,v,0)\dx_0(Z) \rangle =- \frac{a}{\ell} (1-a \brr(Z,0))~\partial_Y \left[ \bvphi(\min(Z,Y),v,0) - a~ \bfr(Y,v,0)\bFr(\min(Z,Y),0)\right],
\label{eq:<dfr-dx_0>-pp-ini}
\end{align}
where $\bvphi(Y,v,0)=\int dX\Theta(Y-X)\bfr(X,v,0)$ and $\bFr(Y,0)=\int dv ~\bvphi(Y,v,0)=\int dX\Theta(Y-X)\brr(X,0)$. Similarly, one can show that 
\begin{align}
\langle \dx_0(X)\dx_0(Y) \rangle 
&= \frac{a^2}{\ell}(1-a \brr(X,0))(1-a\brr(Y,0))~\bFr(\min(X,Y),0).
\label{eq:<dx_0(X)dx_0(Y)>-pp-ini}
\end{align}
Inserting the expressions from Eqs.~\eqref{eq:<dfr-dx_0>-pp-ini} and \eqref{eq:<dx_0(X)dx_0(Y)>-pp-ini} in Eq.~\eqref{def:<X_t(X_q,v_q)>-hr-pic-pp-ini-bis} one can evaluate $\langle X_t(X_q,v_q)|X_q,v_q \rangle$ to order $O(1/\ell)$. Taking derivative with respect to time, one finds the velocity of the QP as 
\begin{align}
\begin{split}
\bar{\mc v}_t(X_q,v_q) = \frac{d \langle X_t(X_q,v_q)|X_q,v_q \rangle}{dt} =& \frac{d X_t^{\rm eu}(X_q,v_q)}{dt} +O\lf \frac{1}{\ell}\rf
= v_{\rm eff}(X_q,v_q,t) + O\lf \frac{1}{\ell}\rf
\end{split}
\label{exp:v_q}
\end{align}
where $v_{\rm eff}(X,v,t)$ is given in Eq.~\eqref{def:v_eff}.

\subsection{Variance: $\msc C_{XX}^{qq}(t,t)$}
In this section we compute the conditional variance where $\Delta X_t(X_q,v_q)$ is given in Eq.~\eqref{def:DelX_(X_q,v_q)(t)-hr-pic}. Before proceeding we note that this expression can be simplified if the fluctuations are expressed in terms of point particle phase space density $\dfp(y,v,0)$. It is easy to show that the fluctuation in hard rod density $\dfr(Y,v,0)$ can be expressed in terms of $\dfp(y,v,0)$ at $y=\bx_0(Y)$ as follows
\begin{align}
\dfr(Y,v,0)
=& \Big[\frac{1}{(1+a\brp(y,0))}\partial_{y} \Big\{\dphi(y,v,0)-a\frac{\bfp(y,v,0) \dFp(y,0)}{1+a\brp(y,0))} \Big\}\Big]_{y=\bx_0(Y)},
\label{eq:dmscf(t)-1-a} 
\end{align}
where $\delta F(y,0)=\int dz \int dv \Theta(y-z)\dfp(z,v,0) = \int dz \Theta(y-z)\drp(z,0)$. Inserting the above expression in Eq.~\eqref{def:DelX_(X_q,v_q)(t)-hr-pic} and simplifying we get 
  \begin{align}
  \begin{split}
\Delta X_t(X_q,v_q)   
=& ~  \dx_0(X_q) +a \int dy\int dv~\delta \left(\bx_t^{q,v}
- y\right)\Big[\dphi(y,v,0)+\dx_0(X_q)\bfp(y,v,0)\Big],\\
=&~\frac{\dx_0(X_q)-\dx_t(\bX_t(\bx_t^{q,0}))}{1-a\brr(\bX_t(\bx_t^{q,0}),t)}. 
\end{split}
   \label{def:delX_(X_q,v_q)(t)-pp-pic-1}
\end{align}
where recall $\bx_t^{q,v}$ from Eq.~\eqref{def:bx_t^qv}. Now the conditional variance can be computed easily 
\begin{align}
\lla \Delta X_t(X_q,v_q)^2|X_q,v_q\rra=&  \frac{\lla \fr(X_q,v_q,0) \Delta X_t(X_q,v_q)^2\rra}{\bfr(X_q,v_q,0)} \notag \\
=& \lla\Delta X_t(X_q,v_q)^2\rra + \frac{\lla \dfr(X_q,v_q,0) \Delta X_t(X_q,v_q)^2\rra}{\bfr(X_q,v_q,0)}.
\end{align}
Note the second term in the above equation involves three-point correlation which according to the large-deviation theory 
\cite{doyon2023ballistic} is of order smaller than $O(1/\ell)$ and  can be neglected. Hence we have 
\begin{align}
\text{Var}[X_t(X_q,v_q)]=\lla \Delta X_t(X_q,v_q)^2|X_q,v_q\rra
 =&\frac{\lla \big[\dx_0(X_q)-\dx_t(\bX_t(\bx_t^{q,0}))\big]^2\rra}{(1-a\brr(\bX_t(\bx_t^{q,0}),t))^2} + o\lf \frac{1}{\ell}\rf.
 \label{ex:var(X_t)-1}
\end{align}
Note the expression in Eq.~\eqref{ex:var(X_t)-1} is valid for both types of initial states, ${\rm IC}_{\rm fhr}$ and  ${\rm IC}_{\rm fhp}$, 
given in Eqs.~\eqref{eq:mbP_r} and \eqref{eq:mbP_p}.  On just requires to evaluate the correlation $\langle \dx_t(Y)\dx_{t'}(Z) \rangle$ which can be obtained  by integrating the correlation $ \langle \dvphi(Y,v,t)\dvphi(Z,u,t') \rangle $.  For the ${\rm IC}_{\rm fhp}$ state explicit computation can be performed by integrating the correlation
in Eq.~\eqref{corr-vrphi-3} over the velocities.  One gets 
\begin{align}
 \langle \dx_t(Y)\dx_{t'}(Z) \rangle 
 &=  \frac{a^2}{\ell}  (1-a\brr(Y,t)) (1-a\brr(Z,t'))~ \left[\int du~\bphi(\min(\bx_t(Y)-ut,\bx_{t'}(Z)-ut'),u,0) \right].
 \label{ex:<dx_tdx_tp>}
\end{align}
For initial state ${\rm IC}_{\rm fhr}$, an equivalant expression is provided in Eq.~\eqref{ex:<dx_tdx_tp>-fhr} of Appendix~\ref{app:eq:<dfp-dfp>(0)-hrIC}. 

Recall, we here focus on finding explicit expressions only for the ${\rm IC}_{\rm fhp}$ initial state. 
Inserting the correlation from Eq.~\eqref{ex:<dx_tdx_tp>}  in Eq.~\eqref{ex:var(X_t)-1} and simplifying, we get 
\begin{align}
\begin{split}
\msc C_{XX}^{qq}(t,t)=&\lla \Delta X_t(X_q,v_q)^2|X_q,v_q\rra
 =~\frac{a^2}{\ell}\Bigg[\frac{(1-a \brr(X_q,0))^2~\bFp(\bx_0(X_q),0) }{(1-a\brr(\bX_0(\bx_t^{q,0}),t))^2} +\bFp(\bx_t^{q,0},t)\\
 &~~~~~~~~~~~~~~~~~~~~~~~~~
 -2\frac{(1-a \brr(X_q,0))~\int du~\bphi(\min(\bx_t^{q,u},\bx_{0}^{q,u}),u,0)}{(1-a\brr(\bX_0(\bx_t^{q,0}),t))} \Bigg],
 \end{split}
 +{\rm o}\lf\frac{1}{\ell}\rf, 
 \label{ex:var(X_t)-fn}
\end{align}
where $\bx_t^{q,u}=\bx_0(X_q)+(v_q-u)t$, $\bphi(x,w,t)=\int dz ~\Theta(x-z)~\bfp(z,w,t)$ and $\bFp(x,t)=\int dw~\bphi(x,w,t)$. 


\subsection{Position cross-correlation: $\msc C_{XX}^{q_1q_2}(t_1,t_2)$}
Following a similar method one can compute the cross-correlation of the position of two different QPs (labelled by $q_1$ and $q_2$). This correlation is defined as 
\begin{align}
\begin{split}
\msc C_{XX}^{q_1q_2}(t_1,t_2)=&\lla \Delta X_{t_1}(X_{q_1},v_{q_1})\Delta X_{t_2}(X_{q_2},v_{q_2})|X_{q_1},v_{q_1};X_{q_2},v_{q_2}\rra \\
=&\frac{\lla \fr(X_{q_1},v_{q_1},0)\fr(X_{q_2},v_{q_2},0)\Delta X_{t_1}(X_{q_1},v_{q_1})\Delta X_{t_2}(X_{q_2},v_{q_2})\rra}{\lla \fr(X_{q_1},v_{q_1},0)\fr(X_{q_2},v_{q_2},0)\rra}.
 \end{split}
 \label{def:acorr(X_t1X_t2)-0}
\end{align}
Neglecting higher order correlations, one has
\begin{align}
\begin{split}
\msc C_{XX}^{q_1q_2}(t_1,t_2)=&\lla \Delta X_{t_1}(X_{q_1},v_{q_1})\Delta X_{t_2}(X_{q_2},v_{q_2})\rra + o\lf \frac{1}{\ell}\rf\\
=&\frac{\lla \big[\dx_0(X_{q_1})-\dx_{t_1}(\bX_{t_1}(\bx_{t_1}^{q_1,0}))\big]\big[\dx_0(X_{q_2})-\dx_{t_2}(\bX_{t_2}(\bx_{t_2}^{q_2,0}))\big]\rra}{(1-a\brr(\bX_{t_1}(\bx_{t_1}^{q_1,0}),t_1))(1-a\brr(\bX_{t_2}(\bx_{t_2}^{q_2,0}),t_2))}+ o\lf \frac{1}{\ell}\rf.
 \end{split}
 \label{def:acorr(X_t1X_t2)}
\end{align}
Once again using the result from Eq.~\eqref{ex:<dx_tdx_tp>} and performing some simplifications one finds 
\begin{align}
\msc C_{XX}^{q_1q_2}(t_1,t_2)&= \frac{a^2}{\ell (1-a\brr(\bX_{t_1}(\bx_{t_1}^{q_1,0}),t_1))(1-a\brr(\bX_{t_2}(\bx_{t_2}^{q_2,0}),t_2))} \displaybreak[3] \notag \\
&~ \times~\Big[ (1-a \brr(X_{q_1},0))(1-a \brr(X_{q_2},0))
\int du~\bphi(\min(\bx_{0}^{q_1,u},\bx_{0}^{q_2,u}),u,0)  \displaybreak[3] \notag  \\ 
&~~~~~~  -(1-a \brr(X_{q_2},0))(1-a\brr(\bX_{t_1}(\bx_{t_1}^{q_1,0}),t_1))
\int du~\bphi(\min(\bx_{t_1}^{q_1,u},\bx_{0}^{q_2,u}),u,0) \label{ex:ccorr(X_t1X_t2)-fn}\\ 
&~~~~~~  -(1-a \brr(X_{q_1},0))(1-a\brr(\bX_{t_2}(\bx_{t_2}^{q_2,0}),t_2))
\int du~\bphi(\min(\bx_{t_2}^{q_2,u},\bx_{0}^{q_1,u}),u,0) \notag \\
&~~~~~~  +(1-a\brr(\bX_{t_1}(\bx_{t_1}^{q_1,0}),t_1))(1-a\brr(\bX_{t_2}(\bx_{t_2}^{q_2,0}),t_2))
\int du~\bphi(\min(\bx_{t_1}^{q_1,u},\bx_{t_2}^{q_2,u}),u,0)\Big],\notag \\ 
&~~~~~ +o\lf\frac{1}{\ell}\rf. \notag
\end{align}
The corresponding expression for ${\rm IC}_{\rm fhr}$ can be obtained in the same way using the correlation $\langle \dx_t(Y)\dx_{t'}(Z) \rangle $ Eq.~\eqref{ex:<dx_tdx_tp>-fhr} of Appendix~\ref{app:eq:<dfp-dfp>(0)-hrIC}.

For homogeneous initial state $\bfp(x,v)=\brp_0h(u)$  or equivalently for $\bfp(x,v)=\brr_0h(u)$ with $\brr_0=\frac{\brp_0}{1+a\brp_0}$ where $h(u)=h(-u)$, one can get more explicit expressions for the co-variance $\msc C_{XX}^{q_1q_2}(t_1,t_2)$. In particular for $t_1=t_2=t>0$ we get 
\begin{align}
\msc C_{XX}^{q_1q_2}(t,t)&|_{\rm hom}= \frac{a^2\brr_0}{\ell}\Bigg\{\min\lsq X_{q_1},X_{q_2}\rsq+\min\lsq X_{q_1}+\frac{v_{q_1}t}{1-a\brr_0},X_{q_2}+\frac{v_{q_2}t}{1-a\brr_0}\rsq  -\lsq X_{q_1}+X_{q_2}\rsq, \notag \allowdisplaybreaks[3]\\ 
&~~~~~~ \quad \quad  -\int_{-\infty}^\infty du ~\Theta\lsq{ut-(X_{q_1}-X_{q_2})-\frac{v_{q_1}t}{1-a\brr_0}}\rsq~h(u)\lsq X_{q_1}-X_{q_2}+\frac{v_{q_1}t}{1-a\brr_0}\rsq \\
&~~~~~~ \quad \quad  -\int_{-\infty}^\infty du~ \Theta\lsq{ut-(X_{q_2}-X_{q_1})-\frac{v_{q_2}t}{1-a\brr_0}}\rsq~h(u)\lsq X_{q_2}-X_{q_1}+\frac{v_{q_2}t}{1-a\brr_0}\rsq\Bigg\}, \notag \\ 
&~~~~~~+ o\lf \frac{1}{\ell}\rf. \notag 
\end{align}

\subsection{Position auto-correlation: $\msc C_{XX}^{qq}(t_1,t_2)$}
The position autocorrelation, defined as 
\begin{align}
\begin{split}
\msc C_{XX}^{qq}(t_1,t_2)=&\lla \Delta X_{t_1}(X_q,v_q)\Delta X_{t_2}(X_q,v_q)|X_q,v_q\rra 
 \end{split}
 \label{def:acorr(X_t1X_t2)}
\end{align}
is essentially obtained by considering $q_1=q_2$ in Eq.~\eqref{ex:ccorr(X_t1X_t2)-fn}. We get 
\begin{align}
\msc C_{XX}^{qq}(t_1,t_2)=& \frac{a^2}{\ell (1-a\brr(\bX_{t_1}(\bx_{t_1}^{q,0}),t_1))(1-a\brr(\bX_{t_2}(\bx_{t_2}^{q,0}),t_2))} ~\Big[ (1-a \brr(X_q,0))^2~\bFp(\bx_0(X_q),0) \notag \\ 
&~~~\quad \quad -(1-a \brr(X_q,0))(1-a\brr(\bX_{t_1}(\bx_{t_1}^{q,0}),t_1))
\int du~\bphi(\min(\bx_{t_1}^{q,u},\bx_{0}^{q,u}),u,0) \notag \\ 
&~~~\quad \quad -(1-a \brr(X_q,0))(1-a\brr(\bX_{t_2}(\bx_{t_2}^{q,0}),t_2))
\int du~\bphi(\min(\bx_{t_2}^{q,u},\bx_{0}^{q,u}),u,0) \label{ex:acorr(X_t1X_t2)-fn} \\
&~~~\quad \quad +(1-a\brr(\bX_{t_1}(\bx_{t_1}^{q,0}),t_1))(1-a\brr(\bX_{t_2}(\bx_{t_2}^{q,0}),t_2))
\int du~\bphi(\min(\bx_{t_1}^{q,u},\bx_{t_2}^{q,u}),u,0)\Big].
\notag
\end{align}
As before, one gets a simpler and more explicit expression for the homogeneous state 
\begin{align}
\msc C_{XX}^{qq}(t_1,t_2)|_{\rm hom} = \min(t_1,t_2)~a^2\brp_0~\int_{-\infty}^\infty du~|v_q-u|h(u), 
\end{align}
as was obtained previously in \cite{SciPostPhys.20.6.166, ferrari2023macroscopic}. 


\subsection{Velocity correlation: $\msc C_{vv}^{q_1q_2}(t_1,t_2)$}
The velocity $v_t(X_q,v_q)=\frac{dX_t(X_q,v_q)}{dt}$ of a QP at time $t$ 
can be formally obtained  by taking time derivative of the right hand side of Eq.~\eqref{def:X_q-cont}. Consequently the velocity correlation can be obtained by taking time derivatives of the position correlation function as
\begin{align}
\msc C_{vv}^{q_1q_2}(t_1,t_2)=&\lla \Delta v_{t_1}(X_{q_1},v_{q_2})\Delta v_{t_2}(X_{q_2},v_{_2}q)|X_{q_1},v_{q_1};X_{q_2},v_{q_2}\rra 
= \partial_{t_1}\partial_{t_2} \msc C_{XX}^{q_1q_2}(t_1,t_2). \label{ex:acorr(v_t1v_t2)-fn}
\end{align}
For homogeneous initial state the velocity auto-correlation function becomes delta correlated in time
\begin{align}
\msc C_{vv}^{q_1q_1}(t_1,t_2)|_{\rm hom}=&\delta (t_1-t_2)~a^2 \brp_0 ~ \int_{-\infty}^\infty du~|v_{q_1}-u|h(u).   
\end{align}

\section{Numerical verification of QP dynamics}
\label{sec:num-veri}
In this section we provide numerical verification of the analytical results for the variance and correlation in case of ${\rm IC}_{\rm fhp}$ presented in the previous section. Recall that for ${\rm IC}_{\rm fhp}$ in Eq.~\eqref{eq:mbP_p}, the initial configurations of the point particles are first chosen and then they are transformed to hard-rod coordinates using $\{\mtX_i = \mtx_i+(i-1)a,v_i;~i=1,2,...,N\}$. We want to investigate the  trajectory of a QP rod starting from $\mtX_q=\ell X_q$ with velocity $v_q$. For that one requires to construct an ensemble of initial configurations in which each sample has a rod at $(\mtX_q,v_q)$, which seems to be a difficult task. However one can construct the trajectory approximately in two ways: 
\begin{itemize}
\item Method 1: Treat the location $X_t(X_q,v_q)$ of the $q^{\rm th}$ QP in Eq.~\eqref{def:X_q-cont} as a function(al) of $(X_q,v_q)$ and $\fr(Y,v,t)$, and discretise it. In the discretised form, one can write the following function $\hat X_t(X_q,v_q;\{\mtX_i,v_i\}) = \frac{\hat \mtX_t(\mtX_q=\ell X_q,v_q;\{\mtX_i,v_i\})}{\ell}$ where
    \begin{flalign}
        \hat \mtX_t(\mtX_q,v_q;\{\mtX_i,v_i\}) =&\hat \mtx(\mtX_q;\{\mtX_i\}) +v_qt +\sum_{j=1}^N\Theta \lf \hat \mtx_0(\mtX_q;\{\mtX_i\}) +v_qt - \mtX_j -v_jt\rf  \\
        \text{with,}~~~~~&~\hat \mtx(\mtX_q;\{\mtX_i\}) = \mtX_q-a \sum_{i=1}^N\Theta(\mtX_q-\mtX_i).
    \end{flalign}
In the large $N$ limit, one can approximate the location $X_t(X_q,v_q)$ of a physical QP starting at $(\mtX_q,v_q)$ by the function $\hat X_t(X_q,v_q;\{\mtX_i,v_i\})$ which can be easily evaluated for any given  configuration $\{\mtX_i,v_i\}$ of $N$ rods at any time $t$.

\item Method 2: In this method we  introduce an actual rod at $\mtX_q$ with velocity $v_q$ in the background of other rods at $t=0$. We first choose the configuration $\{\mtx_i,v_i\}$ for the point particles from Eq.~\eqref{eq:mbP_p}. Note the positions in this configuration are ordered $\mtx_{i+1}\ge \mt X_i$,  which can then be easily transformed to hard-rod coordinates using $\{\mtX'_i=\mtx_i+a(i-1)a,v_i\}$. Next we find the label $i^*$ such that $\mtX'_{i^*} < X_q < \mtX'_{i^*+1}$. We then modify  the configuration of the $(i^*+1)^{\rm{th}}$ point particle to $\mtx_{i^*+1}=\mtX_q-i^*a$ and $v_{i^*+1}=v_q$. Hence we now have a modified configuration $(\mtx_1,v_1;\mtx_2,v_2;...;\mtx_{i^*+1}=\mtX_q-i^*a,v_q;...;\mtx_N,v_N)$ which now transforms to the following 
hard-rod configuration $\lf\mtX_1=\mtx_1,v_1;\mtX_2=\mtx_2+a,v_2;...;\mtX_q,v_q;....;\mtX_N=\mtx_N+a(N-1),v_N\rf$. Thus we can create an ensemble of initial states such that  every initial configuration contains a rod exactly at $(X_q,v_q)$. Though the modified ensemble becomes slightly different from ${\rm IC}_{\rm fhp}$, we expect it to produce correct hydrodynamic behaviour for the QP starting from $(\mtX_q,v_q)$ in the theromydnamic limit (as numerically verified below). 
\end{itemize}
\begin{figure}[t]
    \centering
    \includegraphics[width=12.0cm, height=6.0cm]{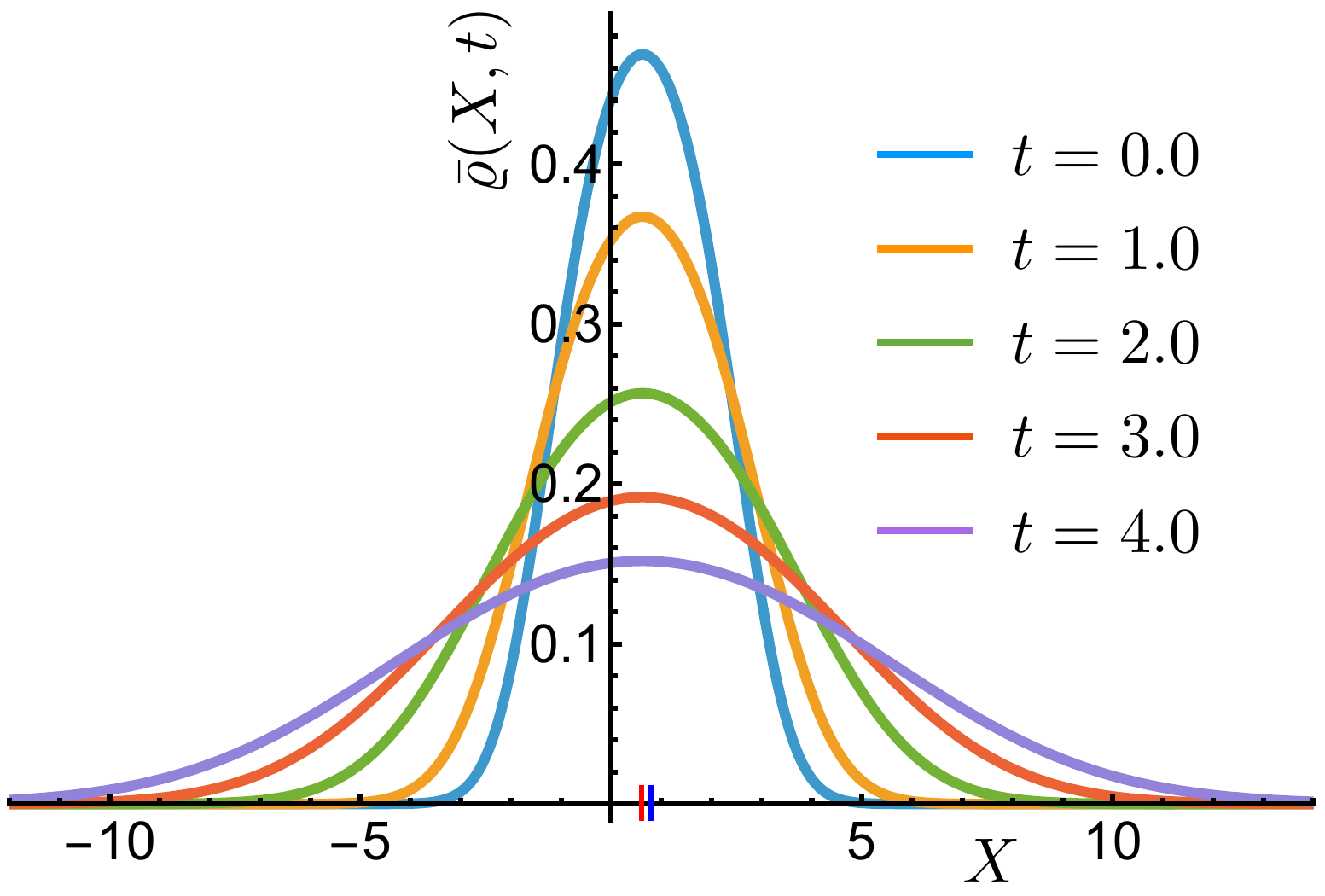}
    \caption{The evolution of the average mass density of hard rod gas starting from initial state ${\rm IC}_{\rm fhp}$ with $\psi(\mtx,v)$ given in Eq.~\eqref{psi(x,v)-sp}. The density profile has a peak at $X=\alpha*a$ and decays with Gaussian tails on both sides. For all our simulations we consider  $\alpha=0.8$, $T=1.0$ and $a=0.71$. }
    \label{fig:density}
\end{figure}

\begin{figure}[h]
    \centering
    \includegraphics[width=14.0cm, height=4.0cm]{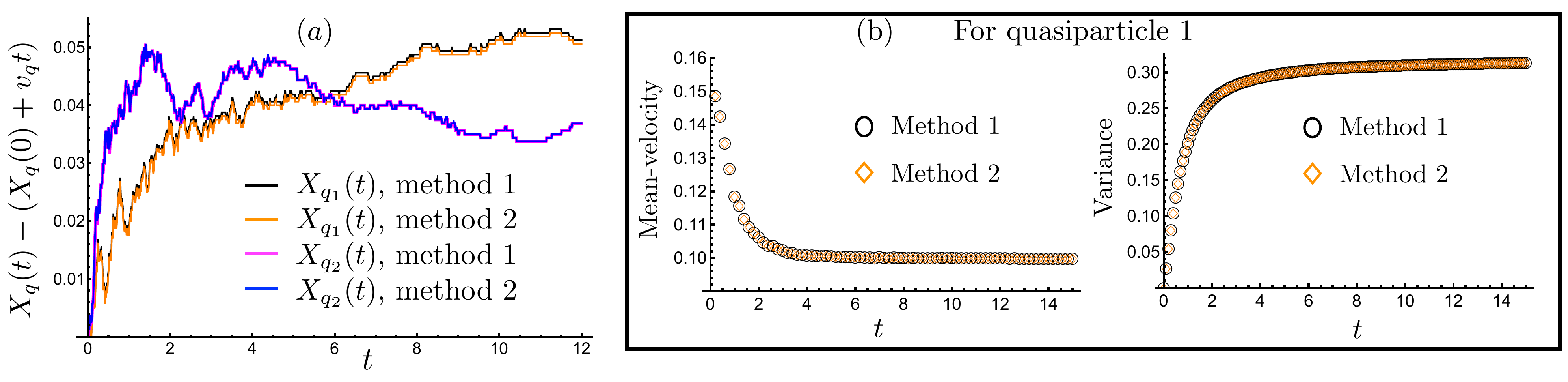}
    \caption{(a) In this we compare the trajectories of two QPs generated by the two methods.. The QPs start, respectively, from positions $X_{q_1}=0.6, X_{q_2}=0.8$ with velocities $v_{q_1}=0.1$ and $v_{q_2}=0.2$. Their initial locations are also indicated by the red and blue vertical lines in density plot of fig.~\ref{fig:density}.  (b) Comparison of the mean velocity and variance of the QP-$1$ computed using method-$1$ and $2$. The QP starts from $\mtX_q=0.6 \sigma$ ({\it i.e.,} $X_q=0.6$) with velocity $v_q=0.1$.}
    \label{fig:comparison}
\end{figure}

\noindent
For numerical evaluation we choose the initial point particle configurations from the ${\rm IC}_{\rm fhp}$  state with
\begin{align}
\begin{split}
\psi(\mtx,v) &= \uppsi(\mtx)\mc h(v),~~\text{where}, \\
 \uppsi(\mtx) &=  \frac{1}{2 \pi  \sigma^2}e^{-\frac{\mtx^2}{2 \sigma^2}},~~  
 \mc h(v)= \frac{1}{\sqrt{2 \pi T}}e^{-\frac{v^2}{2T}},~~\text{and}~~\sigma=\alpha N a.
 \end{split}
\label{psi(x,v)-sp}
\end{align}
Evolution of the corresponding mean density of hard-rod gas is shown in fig.~\ref{fig:density}. Starting from the initial configurations chosen from this state, the QPs moves stochastically. Sample of such stochastic trajectories of two QPs are presented in fig.~\ref{fig:comparison}a obtained using both methods. A comparison of the two methods is provided in fig.~\ref{fig:comparison}. The figure ~\ref{fig:comparison}a compares the trajectories while the plots in fig.~\ref{fig:comparison}b provide  comparison of the mean velocity and variance of a  QP obtained using the two methods. We find that both methods agree both for individual trajectories and at average level. In the following, to numerically verify the theoretical predictions made in the previous section, we only  use method 2.

\noindent
Since numerical simulations are done for large but finite $N$, one needs to take the finite size correction into account for numerical verification. Incorporating this correction, the correlation in Eq.~\eqref{ex:<dx_tdx_tp>-wth-fsc} gets modified to
\begin{align}
 \langle \dx_t(Y)\dx_{t'}(Z) \rangle 
 &=  \frac{a^2}{\ell}  (1-a\brr(Y,t)) (1-a\brr(Z,t'))~\notag \\ 
 &~~\times~\left[\int du~\bphi(\min(\bx_t(Y)-ut,\bx_{t'}(Z)-ut'),u,0) \redw{-\frac{\bFp(\bx_t(Y),t)\bFp(\bx_{t'}(Z),t')}{N}}\right], 
 \label{ex:<dx_tdx_tp>-wth-fsc}
\end{align}
using which in Eq.~\eqref{def:acorr(X_t1X_t2)}, the result in Eq.~\ref{ex:ccorr(X_t1X_t2)-fn} becomes 
\begin{align}
\msc C_{XX}^{q_1q_2}(t_1,t_2&)= \frac{a^2}{\ell }~\Bigg[ 
\frac{(1-a \brr(X_{q_1},0))(1-a \brr(X_{q_2},0))}{(1-a\brr(\bX_{t_1}(\bx_{t_1}^{q_1,0}),t_1))(1-a\brr(\bX_{t_2}(\bx_{t_2}^{q_2,0}),t_2))} \notag \\
& ~~~~~~~\times~\left\{
\int du~\bphi(\min(\bx_{0}^{q_1,u},\bx_{0}^{q_2,u}),u,0) -
\frac{\ell \bFp(\bx_0(X_{q_1}),0)\bFp(\bx_0(X_{q_2}),0) }{N}
\right\}  \label{ex:ccorr(X_t1X_t2)-fn-wth-fsc} \\ 
&~~~~~~~  -\frac{(1-a \brr(X_{q_2},0))}{(1-a\brr(\bX_{t_2}(\bx_{t_2}^{q_2,0}),t_2))}
\left\{\int du~\bphi(\min(\bx_{t_1}^{q_1,u},\bx_{0}^{q_2,u}),u,0) \right. \notag \\
&~~~~~~~~~~\left.-\frac{\ell \bFp(\bx_{t_1}(X_{q_1}),t_1)\bFp(\bx_0(X_{q_2}),0) }{N}
\right\} -\frac{(1-a \brr(X_{q_1},0))}{(1-a\brr(\bX_{t_1}(\bx_{t_1}^{q_2,0}),t_1))}\notag \\
&~\times~\left\{
\int du~\bphi(\min(\bx_{t_2}^{q_2,u},\bx_{0}^{q_1,u}),u,0)  -\frac{\ell \bFp(\bx_{0}(X_{q_1}),0)\bFp(\bx_{t_2}(X_{q_2}),t_1) }{N}
\right\}\notag \\ 
&~+\left\{ \int du~\bphi(\min(\bx_{t_1}^{q_1,u},\bx_{t_2}^{q_2,u}),u,0) 
-\frac{\ell \bFp(\bx_{t_1}(X_{q_1}),t_1)\bFp(\bx_{t_2}(X_{q_2}),t_1) }{N}
\right\} \Bigg]+o\lf\frac{1}{\ell}\rf. \notag
\end{align}

\begin{figure}[t]
    \centering
    \includegraphics[width=15.0cm, height=6.5cm]{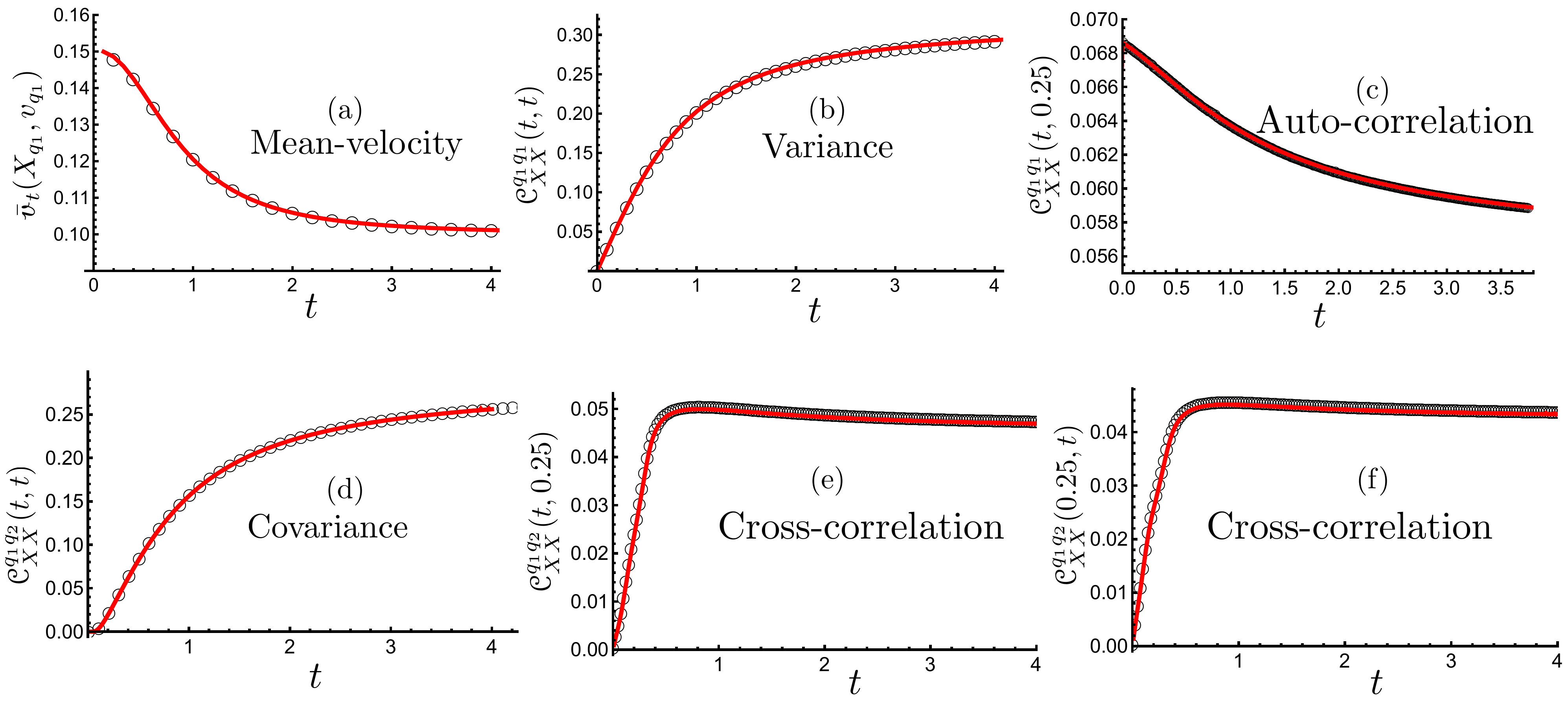}
    \caption{Numerical verification of the analytical predictions of mean velocity, and two-point correlations given in Eq.~\eqref{exp:v_q} and Eq.~\eqref{ex:ccorr(X_t1X_t2)-fn-wth-fsc}, respectively. The black circles represent simulation data and the solid red lines represent theoretical results.   The plots (a), (b) and (c) on the top panel represent quantities involving individual QP, namely,  effective velocity $v_{\rm eff}(X_{q_1},v_{q_1},t)$, variance: $\msc C_{XX}^{q_1q_1}(t,t) = \lla \Delta X_t(X_q,v_q)^2\rra$ and auto-correlation: $\msc C_{XX}^{q_1q_1}(t,t') = \lla \Delta X_{t}(X_{q_1},v_{q_1})\Delta X_{t'}(X_{q_1},v_{q_1})\rra$. The plots in the bottom panel represent quantities related to correlations between two QPs -- covariance: $\msc C_{XX}^{q_1q_2}(t,t) = \lla \Delta X_{t}(X_{q_1},v_{q_1})\Delta X_{t}(X_{q_2},v_{q_2})\rra$ and  cross-correlation $\msc C_{XX}^{q_1q_2}(t,t') = \lla \Delta X_{t}(X_{q_1},v_{q_1})\Delta X_{t'}(X_{q_2},v_{q_2})\rra$. The parameters used in these plots are $N=2000,~X_{q_1}(0)=0.6,~X_{q_2}(0)=0.8,~v_{q_1}(0)=0.1$ and $v_{q_2}(0)=0.2$ in addition to those mentioned in the caption of fig.~\ref{fig:density}. The simulation data are obtained averaging over $R=10^7$ initial configurations.}
    \label{fig:mean-corr-var}
\end{figure}


\noindent
In fig.~\ref{fig:mean-corr-var} we numerically verify our analytical prediction for the time dependence of mean, variance and other two-point correlations of two QPs $q_1$ and $q_2$ which start from positions $x_{q_1}=0.6$ and $X_{q_2}=0.8$ with velocities $v_{q_1}=0.1$ and $v_{q_2}=0.2$. In order to a sense of the initial surrounding of these QPs, we indicate the initial locations of these two QPs in the density plot fig.~\ref{fig:density} with red and blue vertical lines on the x-axis.  In figure~\ref{fig:mean-corr-var}, the black circles represent numerical simulation data and the solid red lines represent analytical results from Eq.~\eqref{exp:v_q} and Eq.~\eqref{ex:ccorr(X_t1X_t2)-fn-wth-fsc}. The excellent agreement between theory and simulation results validates our analytical results. The first plot in (a) represents the mean velocity of  QP $q_1$. At time $t \to 0^+$,  the QP acquires an effective velocity due to collisions with other rods in the surrounding. Note the time $t=0^+$ represents very very small time duration in the hydrodynamic time scale but much larger in microscopic time units (characterised by the mean free time between collisions). Since from fig.~\ref{fig:density} that the density profile has a peak at $X=\alpha a=0.568$ and decays on both sides, the QP starts moving with a higher effective velocity. With time the speed gets reduced because the Qp  encounters less and less collisions as it evolves. This happens because, as time progresses, the rods, irrespective of their initial positions, get ordered according to increasing initial velocities and consequently after some time none of the rods experience any further collisions.  As a result each QP finally (at late times) moves with their respective initial velocities, as can be seen from fig.~\ref{fig:mean-corr-var}a. For the same reason the variance (b), auto-correlation (c), covariance (d) and corss-correlation in plots (e) and (f), show late time saturation after an initial increase with time. The initial increase and late time saturation saturation of the variance can be easily understood from the  fluctuations in the QP trajectories in fig.~\ref{fig:comparison}(a).  One can observe that the QP receives a large number of collisions at small times making the trajectory fluctuate a lot. As time increases, the rate of collisions decreases and consequently the fluctuations in the trajectory gets reduced. Same line of explanation holds for the time dependence of other two-point correlations of the positions of the QPs.

\begin{figure}[h]
    \centering
    \includegraphics[width=14.0cm, height=7.5cm]{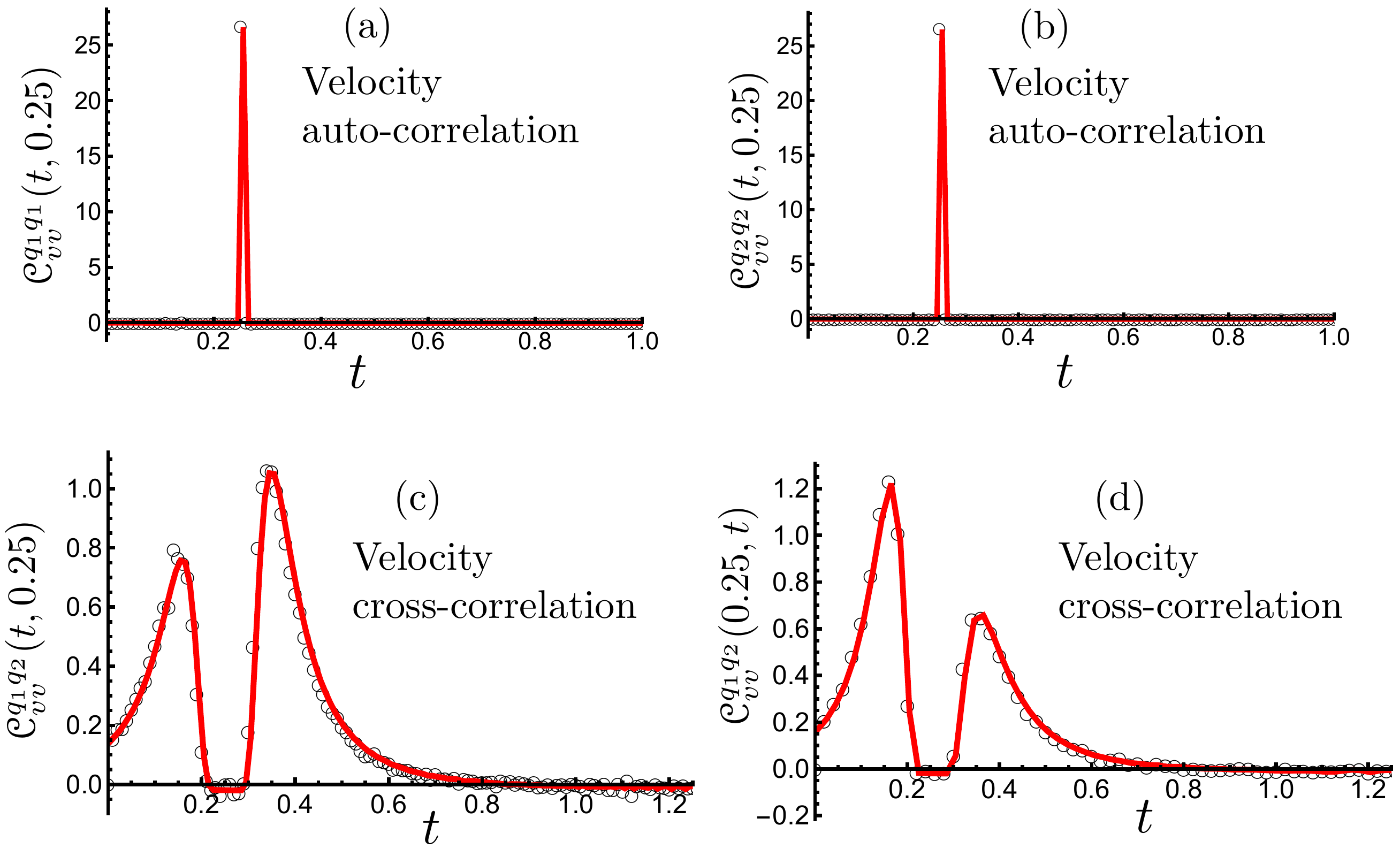}
    \caption{Numerical verification of the analytical predictions of two-point velocity correlations $\msc C_{vv}^{q_1q_2}(t_1,t_2) 
= \partial_{t_1}\partial_{t_2} \msc C_{XX}^{q_1q_2}(t_1,t_2)$ (see  Eq.~\eqref{ex:acorr(v_t1v_t2)-fn}) of the two QPs with $\msc C_{XX}^{q_1q_2}(t_1,t_2)$ given in Eq.~\eqref{ex:ccorr(X_t1X_t2)-fn-wth-fsc}.  The time derivatives are evaluated numerically. The parameters used in these plots are same as fig.~\ref{fig:mean-corr-var}.}
    \label{fig:vell-corr}
\end{figure}

In fig.~\ref{fig:vell-corr}, we numerically verify the two-point velocity correlation $\msc C_{vv}^{q_1q_2}(t_1,t_2)$ defined in Eq.~\eqref{ex:acorr(v_t1v_t2)-fn}. Once again we observe excellent agreement between theory and simulation. We observe from figs.~\ref{fig:vell-corr}(a) and  \ref{fig:vell-corr}(b) that, the velocity of individual QPs at different times are $\delta(t-t')$ correlated as in case of QPs moving in a homogeneous background \cite{ferrari2023macroscopic}. The $\delta$-correlation result for inhomogeneous case was  proved earlier in \cite{hubner2026diffusive,hubner2025hydrodynamics} for ${\rm IC}_{\rm fhr}$ initial state. However, unlike the homogeneous case,  the cross-correlation of the velocities of two different QPs are not delta correlated in time anymore. In fact, for the particular choice of the parameters,  we observe the cross-correlation to be small at equal time and has peaks on both sides of this point. The physical reasoning of this structure is not clear at the moment.

\section{Hydrodynamic equation on diffusion scale}
\label{sec:hd_diff}
Hydrodynamic theories of many-particle systems are often derived phenomenologically by assuming local equilibrium (LE). Under this assumption, the coarse-grained currents can be expanded in gradients of the local conserved densities, leading to a closed set of hydrodynamic equations \cite{doyonlecturenotes,denardis2023hydrodynamic}. For integrable systems, however, a microscopic derivation of hydrodynamics has been achieved for the one-dimensional hard-rod gas \cite{percus1969exact,boldrighini1983one,boldrighini1997one,hubner2026diffusive}. The Euler GHD equation for this system was first derived in Ref.~\cite{percus1969exact} and rigorously established in Ref.~\cite{boldrighini1983one}. The first diffusive correction to the Euler equation, commonly referred to as the Navier--Stokes (NS) term, was derived in Ref.~\cite{boldrighini1997one} under the LE assumption. In generic non-integrable systems, the NS equation accurately describes the evolution of local observables on diffusive scales for a broad class of initial states. This picture, however, breaks down in integrable systems, where the diffusive correction differs from the NS term because of the emergence of LR correlations on the Euler space-time scale \cite{hubner2025diffusive}. The existence of such LR correlations in many-particle integrable systems was first established in Ref.~\cite{doyon2023emergence}. These correlations arise dynamically from initial fluctuations that are coherently transported over macroscopic distances by the Euler evolution \cite{doyon2023emergence}. Assuming that these are the only relevant fluctuations on Euler scales and that local observables fluctuate only through appropriate functions of the conserved densities, a large-deviation framework, known as ballistic macroscopic fluctuation theory (BMFT), was developed in Ref.~\cite{doyon2023ballistic}. BMFT has subsequently been used to compute the probabilities of large fluctuations of local observables and currents on macroscopic space-time scales \cite{doyon2023ballistic,AnupamMFT,kethepalli2025ballistic}. The role of LR correlations in thermalization has also been investigated in generalized fluids \cite{urilyon2026simulating}.

Here we are interested in computing the diffusive scale correction to the Euler equation \eqref{eq:eghd} for ${\rm IC}_{\rm fhp}$ initial state. 
We start by defining $\Psi(t)$ for a smooth function $\Psi(X,v)$ as 
\begin{align}
\Psi(t) 
=  \lla \sum_{q=1}^N \Psi(X_t(X_q,v_q),v_q) \rra
= \int dZ \int dv~\Psi(Z,v) ~\lla\sum_{q=1}^N\delta(Z-X_t(X_q,v_q))(v-v_q) \rra. \label{def:psi(t)} 
\end{align}
Evolving $\Psi(t)$ infinitesimally from $t$ to $t+dt$ we get
\begin{align}
\Psi(t+dt) =& \lla \sum_{q=1}^N \Psi(X_{t+dt}^q,v_q) \rra =  \lla \sum_{q=1}^N \Psi(X_{t}^q +dX_t(X_t^q,v_q)) \rra \cr
=&  \int dZ \int dv~ \lla \fr(Z,v,t)\Psi(Z+dX_{t}(Z,v),v) \rra  \\
=& \int dZ \int dv~ \lla \lf\bfr(Z,v,t)+\dfr(Z,v,t)\rf\Psi(Z+\lla dX_{t}(Z,v)\rra +\delta X_t(Z,v),v) \rra, 
\label{Uln(t+dt)}
\end{align}
where recall $\bfr(Z,v,t) = \lla \fr(Z,v,t) \rra$.
Expanding the right hand side to quadratic order in $dX_t$ we get 
\begin{align}
\Psi(t+dt) 
&= \int dZ \int dv~\bfr(Z,v,t) \Psi(Z,v,t) \displaybreak[3] \notag \\
 &~+ \int dZ \int dv \bigg \langle \lf \bfr(Z,v,t) +\dfr(Z,v,t) \rf \big[\langle dX_{t}(Z,v) \rangle  +  \delta X_{t}(Z,v)  \big ] \bigg \rangle \partial_{Z} \Psi(Z,v) \displaybreak[3]  \label{Uln(t+dt)-1} \\
 &~+ \int dZ\int dv~\bigg \langle \lf \bfr(Z,v,t) +\dfr(Z,v,t) \rf \left[\lla dX_{t}(Z,v) \rra + \delta X_{t}(Z,v)\right]^2 \bigg \rangle \frac{1}{2}\partial^2_{Z} \Psi(Z,v) \notag \\ 
 &~~~~+ O(dX_t^3).
 \displaybreak[3]
 \notag
\end{align}
On the other hand, from the definition \eqref{def:psi(t)} of $\Psi(t+dt)$, we get 
\begin{align}
  \Psi(t+dt) =&  \int dZ \int dv~\bfr(Z,v,t+dt) \Psi(Z,v) \notag \\
  =& \int dZ \int dv~\Psi(Z,v) \Big(\bfr(Z,v,t) + dt \partial_t \bfr(Z,v,t) + O(dt^2) \Big) 
   \label{Uln(t+dt)-2}
\end{align}
Equating the right hand sides of Eqs.~\eqref{Uln(t+dt)-1} and \eqref{Uln(t+dt)-2}, and taking the limit of small $dt$, we get
\begin{align}
\begin{split}
\partial_t \bfr(Z,v,t) =& -\partial_Z\Big[ \bv_t(Z,v)\bfr(Z,v,t) 
   - \frac{a^2}{2\ell}\partial_Z \lf D_t(Z,v) \bfr(Z,v,t)\rf  \Big],
\end{split}
\label{fhd-0}
\end{align}
where 
\begin{align}
\bv_t(Z,v) =& \bigg[ \frac{\bfr(Z,v,t)\lla dX_{t}(Z,v) \rra+\lla \delta X_t(Z,v)\dfr(Z,v,t) \rra}{\bfr(Z,v,t)~dt}\bigg ]_{dt \to 0} =\bigg[ \frac{\lla dX_{t}|Z,v \rra}{dt}\bigg ]_{dt \to 0}, \label{def:barv_t}\\
 D_t(Z,v) =& \bigg[\frac{ \ell \lla \dX_{t}^2(Z,v) \rra}{a^2{dt}}\bigg ]_{dt \to 0},\label{def:D_t}
\end{align}
In \ref{derv:barv_t-and-D_t} we show that 
\begin{align}
\bv_t(Z,v) =& \Big(v_{\rm eff}(Z,v,t) +\frac{1}{\ell} \ovl{\mc{j}}_{\rm d}(Z,v,t) \Big)~+O(dt), \label{dv_t-cond-av-2-dt} \\
 D_t(Z,v) =& \frac{1}{1-a\brr(Z,t)}\int du |v-u|~\bfr(Z,u,t),
   \label{ex:mcj_f^gge}
\end{align}
where the function $v_{\rm eff}(Z,v,t)$ is given in \eqref{def:v_eff} and 
\begin{align}
    \ovl{\mc{j}}_{\rm d}(X_t^q,v_q,t)=& \lf \ovl{\mc{j}}_{\rm d}^{\rm gge}(X_t^q,v_q,t) +\ovl{\mc{j}}_{\rm d}^{\rm lr}(X_t^q,v_q,t) \rf. 
 \label{def:ovl(mcj)_d-1} 
\end{align}
with 
\begin{subequations}
\label{exp:mcj}
    \begin{align}
   \ovl{\mc{j}}_{\rm d}^{\rm lr}(Z,v,t) =&  ~\frac{a}{\bfr(Z,v,t)(1-\brr(Z,t))} ~\int du~\int du'\int dw~(v-u) \left[\delta(u-u')+\frac{a\bfr(X_t^q,u,t)}{1-\brr(Z,t)}\right] \\ 
    &~~~~~~~~~~~~\times~~\left[\delta(v-w)+\frac{a\bfr(Z,v,t)}{1-\brr(Z,t)}\right]
    ~ \msc{C}_{\rm lr}^{{\rm{r,sgn}}(v_q-u)}(Z,w;Z,u';t), 
    \label{ex:mcj_d^lr} \\
      \ovl{\mc{j}}_{\rm d}^{\rm gge}(Z,v,t) =&
   \frac{a^2}{2} \Bigg[\frac{\partial_{Z}\int du~|v_q-u|\bfr(Z,u,t)}{1-a\brr(Z,t)} 
 + \partial_Z\lf \frac{\int du~|v_q-u|\bfr(Z,u,t)}{1-a\brr(Z,t)}\rf\Bigg], \label{ex:mcj_d^gge}
\end{align}
\end{subequations}
For homogeneous initial state $\ovl{\mc j}_{\rm d}^{\rm lr} =0$ and the rest of the term simplifies to reproduce the Navier-Stokes equation 
\eqref{eq:hd_diff-Boldrigni} derived previously in \cite{boldrighini1997one,doyon2017dynamics} making a local equilibrium approximation for the statistical state in a fluid cell. However, for the inhomogeneous initial state, this GHD equation gets modified due to the presence of LR correlation \cite{doyon2023emergence, doyon2023ballistic} that gets generated through coherent transport of initial fluctuations in distant parts of the system  through Euler evolution. This fact was first proved in \cite{hubner2025diffusive} for generic integrable systems and later demonstrated for the hard-rod gas through explicit microscopic calculations \cite{hubner2026diffusive, hubner2025hydrodynamics} for ${\rm IC}_{\rm fhr}$. For this initial state they showed that the  precise cancelation below happens
\begin{align}
\partial_Z\Big[\ovl{\mc j}_{\rm d}^{\rm gge}(Z,v,t)\bfr(Z,v,t) - \frac{a^2}{2}\partial_Z \big(D_t(Z,v)\bfr(Z,v,t) \big) + \ovl{\mc j}_{\rm d}^{\rm lr,asym}(Z,v,t)\bfr(Z,v,t) \Big]=0,
\label{cancellation}
\end{align}
where a further  decomposition $\ovl{\mc j}_{\rm d}^{\rm lr}=\ovl{\mc j}_{\rm d}^{\rm lr,sym}+\ovl{\mc j}_{\rm d}^{\rm lr,asym}$ has been used. This decomposition  is possible because the correlation $\msc C^{\rm r}_{\rm lr}(X,v;Y,u;t)$ can be decomposed in this way as shown in Eq.~\eqref{decompose-C-sym-asymp}. For the initial state ${\rm IC}_{\rm fhp}$,  using the explicit expressions of the correlation $\msc C^{\rm r}_{\rm lr}(X,v;Y,u;t)$ in Eq.~\eqref{def:<dfrdfr>-pp-ini},  it is also possible to 
easily show that the same cancelation occurs as in Eq.~\eqref{cancellation}. (see \ref{app:cancellation}). Hence, for the inhomogeneous case of ${\rm IC}_{\rm fhp}$ initial state, the GHD equation \eqref{fhd-0} for $\bfr(Z,v,t)$ becomes
\begin{subequations}
\label{ex:HD-hr-IC_fhp-final}
\begin{tcolorbox}[ams gather]
  \partial_t\bfr(Z,v,\bt) +\partial_Z\left(v_{\rm eff}(Z,v,\bt)\bfr(Z,v,\bt)\right) = -\frac{1}{\ell} \partial_Z \Big[ \ovl{\mc j}_{\rm d}^{\rm lr, sym}(Z,v,t)\bfr(Z,v,t) \Big],
  \label{eq:bfr-2}
\end{tcolorbox}
\noindent
where $v_{\rm eff}(Z,v,t)$ is given in Eq.~\eqref{def:v_eff} and 
\begin{align}
\ovl{\mc j}_{\rm d}^{\rm lr, sym}(Z,v,t) =& ~\frac{a}{\bfr(Z,v,t)(1-\brr(Z,t))} ~\int du~\int du'\int dw~(v-u) 
\notag  \\ 
    &~~\times~\left[\delta(u-u')+\frac{a\bfr(Z,u,t)}{1-\brr(Z,t)}\right]~\times~\left[\delta(v_q-w)+\frac{a\bfr(Z,v,t)}{1-\brr(Z,t)}\right] \label{def:mcj_d^lr} \\
    & ~~\times~\Big[a^2 \lf \partial_Z \bfr(Z,u,t)\rf \lf \partial_Z \bfr(Z,w,t)\rf \bFr(Z,t) - \frac{a}{2}(1-a\brr(Z,t))\partial_Z\big( \bfr(Z,u,t) \bfr(Z,w,t)\big)\Big],\notag
\end{align}
with $\bFr(X,t)=\int dZ' \int dw'~\Theta(Z-Z')\bfr(Z,w',t)$. The expression of $\ovl{\mc j}_{\rm d}^{\rm lr, sym}(Z,v,t)$ can be further simplfied to 
\begin{align}
\ovl{\mc j}_{\rm d}^{\rm lr, sym}&(Z,v,t) = ~\frac{a}{\bfr(Z,v,t)(1-\brr(Z,t))} ~\int du~(v-u)  
   \label{def:mcj_d^lr-smpl} \\ 
    &~\times~\Bigg\{\Big[a^2 \lf \partial_Z \bfr(Z,u,t)\rf \lf \partial_Z \bfr(X,v,t)\rf \bFr(Z,t) - \frac{a}{2}(1-a\brr(Z,t))\partial_Z\big( \bfr(Z,u,t) \bfr(Z,v,t)\big)\Big],  \notag\\
    & ~+~\frac{a \bfr(Z,v,t)}{1-\brr(Z,t)}\Big[a^2 \lf \partial_Z \brr(Z,t)\rf \lf \partial_Z \bfr(Z,u,t)\rf \bFr(Z,t) - \frac{a}{2}(1-a\brr(Z,t))\partial_Z\big( \brr(Z,t) \bfr(Z,u,t)\big)\Big], \notag\\
     & ~+~\frac{a \bfr(Z,u,t)}{1-\brr(Z,t)}\Big[a^2 \lf \partial_Z \brr(Z,t)\rf \lf \partial_Z \bfr(Z,v,t)\rf \bFr(Z,t) - \frac{a}{2}(1-a\brr(Z,t))\partial_Z\big( \brr(Z,t) \bfr(Z,v,t)\big)\Big], \notag \\
      & ~+~\frac{a^2 \bfr(Z,v,t)\bfr(Z,u,t)}{(1-\brr(Z,t))^2}\Big[a^2 \lf \partial_Z \brr(Z,t)\rf \lf \partial_Z \brr(Z,t)\rf \bFr(Z,t) - \frac{a}{2}(1-a\brr(Z,t))\partial_Z\big( \brr(Z,t)^2 \big)\Big] \Bigg\}.\notag
\end{align}
\end{subequations}
The Eq.~\eqref{ex:HD-hr-IC_fhp-final} describes HD  of hard-rod gas on diffusion scale for ${\rm IC}_{\rm fhp}$ initial state and this is our third main result. For ${\rm IC}_{\rm fhr}$ [see Eq.~\eqref{eq:mbP_r}], the corresponding HD equation for $\bfr(Z,v,t)$ was obtained  in \cite{hubner2026diffusive,hubner2025hydrodynamics}  however the structure of the diffusion scale term is different from that of ${\rm IC}_{\rm fhp}$ as the long range parts of the correlation are different in these two cases. 

\section{Conclusion}
\label{conclusion}
We study the stochastic motion of a QP in a gas of hard rods. We studied the fluctuations and correlations of  QP dynamics at individual level for two choices of initial states, one with LR correlation and the other without it. The calculation required us to derive the LR correlations explicitly -- while the LR correlations were known for the initial state ${\rm IC}_{\rm fhr}$ \cite{hubner2026diffusive, doyon2023ballistic, hubner2025hydrodynamics}, we provide new results for the initial state ${\rm IC}_{\rm fhp}$. These results for LR correlations are further used to derive a general expression  the mean, variance,  autocorrelation and cross-correlation of  QPs at any time $t$, valid for both initial states. While, for the ${\rm IC}_{fhp}$ initial state, we provide  explicit expressions of these quantities, for the ${\rm IC}_{\rm fhr}$ initial state we provide the necessary ingredients to evaluate  general expression easily. As reported previously \cite{hubner2026diffusive}, we also observe that the mean location of the QP at time $t$ receives a diffusive scale correction to the prediction from the mean Euler GHD equation through LR correlations. This correction essentially contributes to the diffusion scale term of the GHD equation satisfied by the mean phase space density $\bfr(Z,v,t)$ and, along with the fluctuations, modifies it from the LE form. However, as argued in \cite{hubner2026diffusive} the diffusive scale equation for $\bfr(Z,v,t)$ does not produce any entropy as combined with the evolution of the two-point correlation, the hydro-scale dynamics is still time reversal symmetric. 

An interesting and important direction to explore would be to achieve a fluctuating hydrodynamic description of hard-rod system valid at mesoscopic scale. According to recent studies \cite{hubner2025hydrodynamics, doyon2025hydrodynamic} the Euler equation \eqref{eq:eghd} with $\bfr(Z,v,t)$ replaced by a fluctuating density field $\fr(Z,v,t)$ provides the correct fluctuating hydrodynamics, including diffusive scale. The same idea has also been used to develop the BMFT  \cite{doyon2023ballistic}. This essentially means that, hard-rod systems being integrable, there are no bare diffusion and emergent noise at the mesoscopic scale. It would be interesting to demonstrate this more explicitly. Second,  the microcopic dynamics of hard rods starting from ${\rm IC}_{\rm fhp}$ initial state being analytically tractable, it would be interesting to compare the microscopic computation of the current fluctuation with hydrodynamic predictions. Finally, analytical study of QP dynamics in other classical integrable models like Ablowitz-Ladik model would also be an interesting direction to explore.

\section{Acknowledgements}
The author thanks Riddhipratim Basu, Subhro Bhattacharjee, Sthitadhi Roy and Abhishek Dhar for  helpful discussions. He acknowledges the financial support of the ANRF, DST, Government of India, under project ANRF/ARGM/2025/001207/MTR,   and  the support from the DAE, Government of India, under Project No. RTI4001.

\appendix
\addtocontents{toc}{\fixappendix}
\section{Hydrodynamic correlations for ${\rm IC}_{\rm fhr}$ initial state in Eq.~\eqref{eq:mbP_r}}
\label{app:eq:<dfp-dfp>(0)-hrIC}
In this appendix, we present the computation of phase space density correlation 
$\lla \dfr(X,v,t)\dfr(Y,u,t')\rra$ of the hard rod gas for ${\rm IC}_{\rm fhr}$. To do this, we once again follow the `height field' method introduced in \cite{hubner2026diffusive}. 
For this initial condition (Eq.~\eqref{eq:mbP_r}), it has been shown that the initial correlation of the phase space density is given by \cite{doyon2017dynamics,AnupamMFT}
\begin{align}
\lla \dfr(X,v,0)\dfr(Y,u,0) \rra = \frac{\delta(X-Y)}{\ell}\left[\delta(v-u)\bfr(X,v,0) -a(2-a\brr(X,0))\bfr(X,v,0)\bfr(Y,u,0) \right], \label{eq:<dfr-dfr>_t=0-hrIC}
\end{align}
which implies
\begin{align}
&\lla \dvphi(X,v,0)\dvphi(Y,u,0)\rra =\frac{1}{\ell}\int dX_1\int dX_2~\Theta(X-X_1)\Theta(Y-X_2) \lla \dfr(X_1,v,0)\dfr(X_2,u,0)\rra, \notag \\
&~~=\frac{1}{\ell}\int dX_1~\Theta\left(\min(X,Y)-X_1\right)\left[ \delta(v-u)\bfr(X_1,v,0)-a(2-a\brr(X_1,0))\bfr(X_1,v,0)\bfr(X_1,u,0)\right]. \label{dvphi-dvphi-corr-hrIC}
\end{align}
We observe that initially the hard-rod densities in this case are not correlated over space. 
However, 
the phase space densities of the point particle gas, even at $t=0$ are highly correlated. In order to compute this correlation, we start with Eq.~\eqref{del-phi-relations-2} and write
\begin{align}
\lla \dphi(x,v,0)\dphi(y,u,0)\rra &= \int dq_1 \int dq_2 \left[ \delta(v-q_1)+a \bfp(x,v,0)\right]\left[ \delta(u-q_2)+a \bfp(y,u,0)\right] \notag \\ 
&~~~~~~~~~~~~~~~~~~~~~~\times~~~\lla \dvphi(\bX_0(x),q_1,0) \dvphi(\bX_0(y),q_2,0) \rra.
\end{align}
A straightforward calculation yields  
\begin{align}
\begin{split}
\ell\langle \dphi(x,v,0)& \dphi(y,u,0) \rangle \\ 
&~=~~\int dz~\Theta\big(\min(x,y)-z\big)\bfp(z,v,0)\Big[ \delta(v-u)-\frac{a(2+a\brp(z,0))}{(1+a\brp(z,0))^2}\bfp(z,u,0) \Big], \\
&~~~~~+
a\int dz~\Theta\big(\min(x,y)-z\big) \\
&~~~~~~~~~~~~\times~~\Big[ 
\frac{\bfp(x,v,0)\bfp(z,u,0)+\bfp(z,v,0)\bfp(y,u,0)+a\bfp(x,v,0)\bfp(y,u,0)\brp(z,0)}{(1+a\brp(z,0))^2}
\Big].
\end{split}
\label{<dphidphi>(0)-IChr}
\end{align}
Taking derivative with respect to $x$ and $y$, respectively, on the right hand side of Eq.~\eqref{<dphidphi>(0)-IChr}, we get 
\begin{subequations}
\label{eq:<dfp-dfp>(0)-hrIC} 
\begin{align}
\lla \dfp(x,v,0)\dfp(y,u,0)\rra &= \partial_x\partial_y \lla \dphi(x,v,0)\dphi(y,u,0)\rra = 
\frac{1}{\ell} \msc C^{\rm p}(x,v,0;y,u,0)
\end{align}
where 
\begin{align}
\msc C^{\rm p}(x,v,0;y,u,0) = \delta(x-y)\msc{C}^{\rm p}_{\rm gge}(x,v,u) + \msc C_{\rm lr}^{\rm p}(x,v;y,u;0)    
\end{align} 
with 
\begin{align}
\begin{split}
\msc{C}^{\rm p}_{\rm gge}(x,v,u)&= \delta(x-y)\delta(v-u)\bfp(x,v,0),  \\
\msc C_{\rm lr}^{\rm p}(x,v;y,u;0)&=a\Theta(y-x) \left[ \bfr(\bX_0(x),v,0) \partial_y \bfp(y,u,0)+a \partial_x \bfp(x,v,0) \partial_y \bfp(y,u,0)~\Gamma(x,0)\right]  \\ 
&~+a\Theta(x-y) \left[ \bfr(\bX_0(y),u,0) \partial_x \bfp(x,v,0)+a \partial_x \bfp(x,v,0) \partial_y \bfp(y,u,0)~\Gamma(y,0)\right].
\end{split}
\label{mscC-hrIC}
\end{align}
Here, 
\begin{align}
\Gamma(x,t) = \int dz~\Theta(x-z) \frac{\brp(z,t)}{(1+a\brp(z,t))^2}
=\int dZ~\Theta(\bX_t(x)-Z) (1-a\brr(Z,t))^2\brr(Z,t). \label{def:Gamma}
\end{align}
with $\bX_t(x)$  given in Eq.~\eqref{trans:X<->x-hf} with $\phi(x,v,t)$ is replaced by $\bphi(x,v,t) =\lla \phi(x,v,t) \rra$.
\end{subequations}
At a later time $t$, the $\lla \dfp(t)\dfp(t') \rra$ correlation is simply given by 
\begin{align}
\lla \dfp(x,v,t)\dfp(y,u,t') \rra = \lla \dfp(x-vt,v,0) \dfp(y-ut',u,0) \rra. 
\end{align}

\noindent
Recall that our aim is to compute the hard-rod gas correlation 
$\langle \dfr (X,v,t)\dfr(Y,u,t') \rangle$. This can be obtained by 
taking derivatives with respect to $X$ and $Y$ on the right hand side of Eq.~\eqref{corr-vrphi-1}. Note that the $\lla \dphi\dphi\rra$ correlation is given in Eq.~\eqref{<dphidphi>(0)-IChr}. One finds, 
\begin{align}
\msc C_{\rm r}(X&,v,t;Y,u,t')=\ell\langle \dfr(X,v,t)\dfr(Y,u,t') \notag \\
&= (1-a\brr(X,t))(1-a\brr(Y,t')) \int dp_1\Big[ \big( \delta(v-p_1)-a\bfr(X,v,t)\big)\big( \delta(u-p_1)-a\bfr(Y,u,t')\big) \notag \\ 
&\quad \quad \quad \quad ~~~~~~~~~~~~~~~~~~~~~\times~\Big\{\delta\big(\bx_t(X) -\bx_{t'}(Y)-p_1(t-t')\big)\msc C^{\rm p}_{\rm gge}(\bx_t(X)-p_1t,p_1,p_2) \notag \\ 
&~~~~~~~~~~~~~~~~~~~~~~~~~~~~~~~~~~~~~~~~
+\msc C^{\rm p}_{\rm lr}(\bx_t(X)-p_1t,p_1;\bx_{t'}(Y)-p_2t',p_2;0))\Big\} \Big], \notag \\
&~~~~~ -a  \partial_X\bfr(X,v,t)(1-a\brr(Y,t'))\int dp_1\int dp_2 \Big[ \big( \delta(u-p_2)-a\bfr(Y,u,t')\big) \notag \\
&~~~~~~~~~~~~~~~~~~~~~~~~~~~~~~~~~
\times~\ell\lla \dphi(\bx_t(X)-p_1t,p_1,0)\dfp(\bx_{t'}(Y)-p_2t',p_2,0)\rra \Big],             \label{eq:<dfr(t)dfr(t')>_hrIC} \\
&~~~~~ -a \partial_Y\bfr(Y,u,t')(1-a\brr(X,t))\int dp_1\int dp_2 \Big[ \big( \delta(v-p_1)-a\bfr(X,v,t)\big) \notag \\
&~~~~~~~~~~~~~~~~~~~~~~~~~~~~~~~~~
\times~\ell \lla \dfp(\bx_{t}(X)-p_1t,p_1,0)\dphi(\bx_{t'}(Y)-p_2t',p_2,0)\rra \Big], \notag \\
&+a^2~\partial_X\bfr(X,v,t)\partial_Y\bfr(Y,u,t')\int dp_1\int dp_2~\ell\lla \dphi(\bx_t(X)-p_1t,p_1,0)\dphi(\bx_{t'}(Y)-p_2t',p_2,0)\rra, \notag
\end{align}
where $\msc C^{\rm p}_{\rm gge}$ and $\msc C^{\rm p}_{\rm lr}$ are given in Eq.~\eqref{mscC-hrIC}. 
A close inspection shows that for $t=t'$ the correlation in Eq.~\eqref{eq:<dfr(t)dfr(t')>_hrIC} also is a sum of a singular and non-singular part as in Eq.~\eqref{C_r-st}
where the singular part $\msc C^{\rm r}_{\rm gge}$ represents the GGE correlation given explicitly in Eq.~\eqref{mscC_gge-<dfrdfr>-hpIC} and the second term contains the LR part of the correlation. It has been shown that the long range part has a jump at $X=Y$ \cite{hubner2026diffusive,hubner2025hydrodynamics}
\begin{align}
\begin{split}
\msc{C}^{\rm r}_{\rm lr}(X,v;Y,u;t) \overset{X \approx Y}{ =}& {\rm sgn}(X-Y)~\frac{a}{2} (1-a\brr(X,t)) \Big[ \bfr(X,u,t) \partial_X\bfr(X,v,t)-\bfr(X,v,t) \partial_X\bfr(X,u,t) \Big] \\ 
&~~~~~~~~~~~~~~~+ \text{terms continuous at}~X=Y,
\end{split}
\label{eq:mscC(X=Y)-hrIC}
\end{align}
where $\text{sgn}(X)=1, 0,-1$ for $X>0,~X=0$ and $X<0$ respectively. 
Note that the antisymmetric part of $\msc C^{\rm r}_{\rm lr}$ for this initial condition is same as that for  ${\rm IC}_{\rm fhp}$ given in Eq.~\eqref{mscC_lr-<dfrdfr>-hpIC}. 

Integrating the correlation $\msc C_{\rm r}(X,v,t;Y,u,t')$ in Eq.~\eqref{eq:<dfr(t)dfr(t')>_hrIC} over velocities $u$ and $v$, one gets the same expression for the mass density-density correlation $\msc C_{\rm r}^{(00)}(X,t;Y,t')$ as  was obtained in Eq.~(106) of Ref:~\cite{AnupamMFT}. Integrating the correlation further  $\msc C_{\rm r}^{(00)}(X,t;Y,t')$ over $X$ and $Y$, one gets the following correlation 
\begin{subequations}
\label{ex:<dx_tdx_tp>-fhr}
\begin{align}
\lla \delta x_t(Z)\delta x_{t'}(Z')\rra =& a^2 \int_{-\infty}^Z dX \int_{-\infty}^{Z'}dY~ \msc C_{\rm r}^{(00)}(X,t;Y,t') \notag \\
=& a^2(1-a\brr(Z,t))(1-a\brr(Z',t')) ~\mc{H}(Z,t;Z',t')
\end{align}
in case of ${\rm IC}_{\rm fhr}$, where,
\begin{align}
\mc{H}(Z,t;Z'&,t')
=\int dy \int du ~\Theta(x_{t}(Z)-ut-y)\Theta(x_{t'}(Z')-ut'-y) \bfp(y,u,0) \displaybreak[3]\notag \\
&
-a\int dy\int du\int dw ~(1-a\brr(y,0)) \Theta(\bx_{t}(Z)-wt-y)\Theta(\bx_{t'}(Z')-ut'-y) \displaybreak[3] \label{eq:mcA} \\ 
&~\times~\Big{\{}\frac{2-a\brr(\bX_0(y),0)}{(1-a\brr(\bX_0(y),0))^2}\bfp(\bX_0(y),u,0)\bfr(\bX_0(y),w,0)   -\bfr(\bX_0(y),u,0)\bfp(\bx_{t}(Z),w,t)   \displaybreak[3]\notag \\ 
&~~~
 -\bfr(\bX_0(y),w,0)\bfp(\bx_{t'}(Z'),u,t') -a \brr(\bX_0(y),0)\bfp(\bx_{t}(Z),w,t) \bfp(\bx_{t'}(Z'),u,t') \Big{\}}, \notag
\end{align}
\end{subequations}
with,
\begin{align}
\begin{split}
\bx_t(X)&= X-a\int dv \int dY ~\Theta(X-Y)\bfr(Y,v,t),  \\
\bX_t(x)&= x+a\int dv \int dy ~\Theta(x-y)\bfp(y,v,t). 
\end{split}
\label{trans:X<->x-hf-ap}
\end{align} 

\section{Proof of Eq.~\eqref{def:X_q-cont}}
\label{prf:def:X_q-cont}
We start by rewriting  Eq.~\eqref{def:X_q-dis}
\begin{align}
\mtX_q(\mtt) &= \hat \mtx(\mtX_q(0)) +v_q \mtt +a\sum_{r\ne q} \Theta\left[\hat \mtx(\mtX_q(0))+v_q\mtt-\hat \mtx(X_r(0))-v_r\mtt \right], \\
&=\hat \mtx_0(\mtX_q) +v_q \mtt +a \int d\mt Y\int dv~\hat \fr(\mt Z,v,0)~\Theta\left[\hat \mtx_0(\mtX_q+v_q\mtt-\hat \mtx_0(\mt Y)-v\mtt \right],
\label{def:X_q-dis-a}
\end{align}
where we have used the empirical density from Eq.~\eqref{eq:fr->scaled-fr} and 
\begin{align}
 \hat \mtx_{\mtt}(\mt X_q)&=\mtX_q - a\int d\mt Y \int dv ~\hat \fr(\mt Y,v,\mtt)\Theta\left[\mtX_q-\mt Y\right]. \label{trans:X->x-dis-a}   
\end{align}
Now, approximating the integrals over $\mt Y$ and $v$ in both Eqs.~\eqref{def:X_q-dis-a} and \eqref{trans:X->x-dis-a} by integrals over the coarse-grained density $\hat {\bar \fr}(\mt Y,v,0)$, we get 
\begin{align}
\mtX_q(\mtt) 
&=\hat {\bar \mtx}_0(\mtX_q(0)) +v_q \mtt +a \int d\mt Y\int dv~\hat {\bar\fr}(\mt Y,v,0)~\Theta\left[\hat {\bar \mtx}_0(\mtX_q(0))+v_q\mtt-\hat {\bar \mtx}_0(\mt Y)-v\mtt \right]
\label{def:X_q-dis-a-2}\\
\text{where,}~~ 
\hat {\bar \mtx}_{\mtt}(\mtX_q)&=\mtX_q - a\int d\mt Y \int dv ~\hat {\bar \fr}(\mt Y,v,\mtt)\Theta\left[\mtX_q-\mt Y\right]. \label{trans:X->x-dis-a-2}
\end{align}
Now using the scaling form in Eq.~\eqref{f_r-scaling}, we get 
\begin{align}
X_q(t) 
&=x_0(X_q(0)) +v_q t + a \int d Y\int dv~\fr(Y,v,0)~\Theta\left[x_0(X_q(0))+v_qt-x_0( Y)-vt \right]
\label{def:X_q-dis-a-3}\\
\text{where,}~~ 
x_t(X_q)&=\frac{\hat {\bar \mtx}_{\mtt}(\mtX_q)}{\ell}=X_q - a\int dY \int dv ~ \fr( Y,v,t)\Theta\left[X_q-Y\right], \label{trans:X->x-dis-a-2}
\end{align}
as in Eq.~\eqref{def:X_q-cont}.


\section{Derivation of the results in Eqs.~\eqref{dv_t-cond-av-2-dt} and \eqref{ex:mcj_f^gge}}
\label{derv:barv_t-and-D_t}
From Eq.~\eqref{def:X_q-cont} we write
\begin{align}
\begin{split}
X_{t+dt}(X_q,v_q) &= x_0(X_q) +v_q (t+dt) \\
&~~~+a\int dY \int dv~\fr(Y,v,0)\Theta\left(x_0(X_q)+v_q(t+dt)-x_0(Y)-v(t+dt)\right).
\end{split}
\label{def:X_q-cont-a}
\end{align}
We first note that $x_t(X_t(X_q,v_q))=x_0(X_q)+v_qt$ where the transformations $x_t(X)$ and the inverse transformation $X_t(x)$ are given in Eqs.~\eqref{trans:X->x-hf} and \eqref{trans:x->X-hf}, respectively. This suggests us to write
\begin{align}
\begin{split}
X_{t+dt}(X_q,v_q) &= x_t(X_t(X_q,v_q)) +v_q dt \\
&~~~+a\int dy \int dv~\fp(y,v,0)\Theta\left(x_t(X_t(X_q,v_q))+v_qdt-y-v(t+dt)\right),
\end{split}
\label{def:X_q-cont-a-1}
\end{align}
where in the last term on the right hand side we have used the transformation $y=x_0(Y)$. Noting $\fp(z+vt,v,0)=\fp(z,v,t)$, we rewrite the above equation as
\begin{align}
\begin{split}
X_{t+dt}(X_q,v_q) &= x_t(X_t(X_q,v_q)) +v_q dt \\
&~~~+a\int dz \int dv~\fp(z,v,t)\Theta\left(x_t(X_t(X_q,v_q))+v_qdt-y-vdt\right),
\end{split}
\label{def:X_q-cont-a-2}
\end{align}
Changing again to the hard rod coordinates $Z=X_t(z)$ at time $t$, we finally get 
\begin{align}
\begin{split}
X_{t+dt}(X_q,v_q) &= x_t(X_t(X_q,v_q)) +v_q dt \\
&~~~+a\int dZ \int dv~\fr(Z,v,t)\Theta\left(x_t(X_t(X_q,v_q))+v_qdt-x_t(Z)-vdt\right),\\
&=X_{dt}(X_t(X_q,v_q),v_q).
\end{split}
\label{def:X_q-cont-a-3}
\end{align}
Hence the displacement in duration $dt$ can be written as 
\begin{align}
dX_{t}(X_t(X_q,v_q),v_q)=X_{dt}(X_t(X_q,v_q),v_q)-X_t(X_q,v_q). \label{X_q(t+dt)}
\end{align}
Introducing the shorthand notation $X_t^q=X_t(X_q,v_q)$,
the displacement can be written as 
\begin{align}
 X_{dt}(X_t^q,v_q) &= x_t(X_t^q) +v_q dt +a\int dY \int dv~\fr(Y,v,t)\Theta\left(x_t(X_t^q)+v_q dt-x_t(Y)-v dt\right).
\label{def:dX_t-1}    
\end{align}
Note that Eq.~\eqref{def:dX_t-1} is exactly same as Eq.~\eqref{def:X_q-cont} except that we are now focusing on the displacement of the $q^{\rm th}$ quasiparticle in  duration $dt$ starting from position $X_t^q$ with velocity $v_q$ at time $t$. Hence to compute the conditional mean and variance of the displacement $dX_t(X_t^q,v_q)=X_{dt}(X_t^q,v_q)-X_t^q$ made in duration $dt$, we follow exactly the same procedure as done in sec.~\ref{sec:qp-dyna}.
We start by expanding  to quadratic orders in fluctuations, we get 
\begin{subequations}
\label{ex:dX_t-expanded}
\begin{align}
dX_t(Z,v) \approx dX_t^{\rm eu}(Z,v) +\delta X_t(Z,v) + \widehat{dX}_t(Z,v), \label{ex:dX_t-expanded-1}
\end{align}
where
\begin{align}
dX_{t}^{\rm eu}(Z,v) = \bx_t(Z) +v dt + a\int dY \int du~\bfr(Y,u,t)\Theta\left(\mc U^t_{dt}(Z,v;Y,u)\right) -X_t^q, \label{X_eu(t)-a}
 \end{align}
with 
\begin{align}
   \mc U^t_{dt}(Z,v;Y,u) = \bx_t(Z) +v {dt} - \bx_t(Y)-u {dt}, 
   \label{def:mcU^t_tau}
\end{align}
\begin{align}
\begin{split}
   \delta X_{t}(Z,v)&= \dx_t(Z) +a \int dY\int du~\Theta \left(\mc U^t_{dt}(Z,v;Y,u)\right) \dfr(Y,u,t),  \\
   &+a \int dY \int du~ \delta \left( \mc U^t_{dt}(Z,v;Y,u) \right) \big( \dx_t(Z) - \dx_t(Y)\big)  \bfr(Y,u,t), 
   \end{split}
   \label{def:delX_(X_q,v_q)(t)-hr-pic}
\end{align}
and
\begin{align}
\begin{split}
 \widehat{dX}_t(Z,v) =&~a  \int dY \int du~\delta\big(\mc U^t_{dt}(Z,v;Y,u)\big)~
 \big(\dx_t(Z) -\dx_t(Y)\big) ~\dfr(Y,u,t)  \\
 &~ + \frac{a}{2} \int dY \int du~\delta'\big(\mc U^t_{dt}(Z,v;Y,u)\big)~\bfr(Y,u,t)~
  \big(\dx_t(Z) -\dx_t(Y)\big )^2. 
  \end{split}
  \label{def:hat_dX_t}
\end{align}
\end{subequations}
As done in Eq.~\eqref{def:cond_<X-t>-1}, the conditional mean $ \lla dX_{t}|X_t^q,v_q\rra$ can be computed in a similar way and one finds 
\begin{subequations}
\label{dX_t-cond-av-0}
\begin{align}
 \lla dX_{t}|X_t^q,v_q\rra=dX_{t}^{\rm eu}(X_t^q,v_q) + \ovl{dX}_{t}(X_t^q,v_q), \label{dX_t-cond-av-1}
\end{align}
where $dX_t^{\rm eu}(X_t^q,v_q)$ is given in Eq.~\eqref{X_eu(t)-a} and 
 \begin{align}
 \ovl{dX}_{t}(X_t^q,v_q)=&~\frac{\lla \dfr(X_t^q,v_q,t)\delta X_t(X_t^q,v_q)\rra}{\bfr(X_t^q,v_q,t)} + \lla \widehat{dX}_t(X_t^q,v_q)\rra, 
 \label{def:<X_t(X_q,v_q)>-hr-pic}
 \end{align}
with
\begin{align}
 \lla \widehat{dX}_t(X_t^q,v_q)\rra=&a  \int dY \int du~\delta\big(\mc U^t_{dt}(X_t^q,v_q;Y,u)\big)~
 \big \langle \big(\dx_t(X_t^q) -\dx_t(Y)\big) ~\dfr(Y,u,t) \big \rangle   \label{def:<X_t(X_q,v_q)>-hr-pic-1}\\
 & + \frac{a}{2} \int dY \int du~\delta'\big(\mc U^t_{dt}(X_t^q,v_q;Y,u)\big)~\bfr(Y,u,t)~
 \left \langle \big(\dx_t(X_t^q) -\dx_t(Y)\big )^2\right \rangle. \notag 
\end{align}
\end{subequations}
and 
\begin{align}
    \dx_t(X) = -a \int dY \int du~\Theta(X-Y)~\dfr(Y,u,t). \label{def:dx_t(X)}
\end{align}
Note that $dX_{t}^{\rm eu}(X_t^q,v_q)$ in Eq.~\eqref{X_eu(t)-a} is the displacement of the quasiparticle in duration $dt$ starting from position $X_t^q$ at time $t$, if one considers contribution only from the mean fluid flow evolving according to the  Euler equation.
We will later see  that the correction $\ovl{dX}_{t}$ to the Euler displacement in Eq.~\eqref{dX_t-cond-av-1} appears due to the LR correlation of the density fields and hence will be of $O(1/\ell)$. Together, $dX_t^{\rm eu}$ and $\ovl{dX}_t$ provide the mean displacement of the $q^{\rm th}$ quasiparticle in duration $dt$ starting from position $X_t^q$ at time $t$. 
The above discussion allows us to write the infinitesimal displacement in Eq.~\eqref{ex:dX_t-expanded-1}  as
\begin{align}
 dX_t(X_t^q,v_q) \simeq    dX_{t}^{\rm eu}(X_t^q,v_q) + \ovl{dX}_{t}(X_t^q,v_q) + \delta X_t(X_t^q,v_q),
\end{align}
where the first two terms combined contribute to the mean and the third term $\delta X_{t}(X_t^q,v_q)$ is noise with zero mean, {\it i.e.,} $\lla \delta X_{t}(X_t^q,v_q) \rra =0$. 


Clearly, for $dt \to 0$ one has $dX_{t}^{\rm eu}(X_t^q,v_q) \to 0$. Expanding the right hand side of Eq.~\eqref{X_eu(t)-a} to $O(dt)$ one finds that  
\begin{align}
dX_{t}^{\rm eu}(X_t^q,v_q) \simeq v_{\rm eff}(X_t^q,v_q,t) dt + O(dt^2), 
\label{eq:dX_t^eu-fnl}
\end{align}
where $v_{\rm eff}(X_t^q,v_q,t)$ is given in Eq.~\eqref{def:v_eff}. 

Recall from sec.~\ref{sec:lr-C} and \ref{app:eq:<dfp-dfp>(0)-hrIC} that for both the initial conditions $\msc C(X,v;Y,u;t)$ [see Eq.~\eqref{C_r-st}] is sum of a singular part representing  GGE correlation and a non-singular part containing the LR correlation.
Since $\ovl{dX}_{t}$ depends linearly on the density correlations, one can write the contributions  from $\msc C^{\rm r}_{\rm gge}$ and $\msc C^{\rm r}_{\rm lr}$ separately 
\begin{align}
  \ovl{dX}_{t}(X_t^q,v_q) = \ovl{dX}_{t}^{\rm gge}(X_t^q,v_q) +  \ovl{dX}_{t}^{\rm lr}(X_t^q,v_q). \label{ovl(dX_t)=sum_2terms}
\end{align}
Similarly the variance has two contributions to the variance 
\begin{align}
\begin{split}
 \lla \dX_{t}^2(Z,v)\rra 
 &= \left[\lla \delta X_{t}^2(Z,v)\rra^{\rm gge} + \lla \delta X_{t}^2(Z,v) \rra^{\rm lr} \right]. 
 \end{split} 
 \label{var(dX_t)=sum_2terms}
\end{align}
It can be shown that for $dt \to 0$, both $\ovl{dX}_{t}(Z,v) \to 0$ and $\lla \delta X_{t}^2(Z,v) \rra \to 0$. In fact, to leading order in $dt$ one finds  (see \ref{app:def:mcj} and \ref{derv:variance-a})
\begin{subequations}
\label{def:mcj}
    \begin{align}
\ovl{dX}_{t}^{\rm gge}(Z,v) &= \frac{1}{\ell}~\ovl{\mc{j}}_{\rm d}^{\rm gge}(Z,v,t)~dt +O(dt^2), \label{mcj_c1^gge}\\    
\ovl{dX}_{t}^{\rm lr}(Z,v) &= \frac{1}{\ell}~\ovl{\mc{j}}_{\rm d}^{\rm lr}(Z,v,t)~dt +O(dt^2), \label{mcj_c1^lr}\\   
\lla \delta X_{t}^2(Z,v) \rra^{\rm gge} &= \frac{a^2}{\ell}~D_t(Z,v)~dt +O(dt^2), \label{mcj_f^gge}\\
\lla \delta X_{t}^2(Z,v) \rra^{\rm lr} &= 0 ~ + ~O(dt^2). \label{mcj_f^lr}
\end{align}
\end{subequations}
Adding the above contributions from ${\rm gge}$ and ${\rm lr}$ part of correlations, in Eqs.~\eqref{ovl(dX_t)=sum_2terms} and \eqref{var(dX_t)=sum_2terms} we get the results announced in Eqs.~\eqref{dv_t-cond-av-2-dt} and \eqref{ex:mcj_f^gge}.



\section{Derivation of the results in Eqs.~\eqref{mcj_c1^gge} and \eqref{mcj_c1^lr}}
For convenience, we start by rewriting $\ovl{dX}_{t}(X_t^q,v_q)$ from Eq.~\eqref{def:<X_t(X_q,v_q)>-hr-pic} here
\label{app:def:mcj}
 \begin{align}
\bfr(X_t^q,v_q,t) &\ovl{dX}_{t}(X_t^q,v_q)=~\underbrace{\lla \dfr(X_t^q,v_q,t)\delta X_t(X_t^q,v_q)\rra}_{T_1} , \label{def:<X_t(X_q,v_q)>-hr-pic-a} \\
 &+\underbrace{a \bfr(X_t^q,v_q,t) \int dY \int du~\delta\big(\mc U^t_{dt}(X_t^q,v_q;Y,u)\big)~
 \big \langle \big(\dx_t(X_t^q) -\dx_t(Y)\big) ~\dfr(Y,u,t) \big \rangle}_{T_2}  \cr
 & + \underbrace{\frac{a}{2}\bfr(X_t^q,v_q,t) \int dY \int du~\delta'\big(\mc U^t_{dt}(X_t^q,v_q;Y,u)\big)~\bfr(Y,u,t)~
 \left \langle \big(\dx_t(X_t^q) -\dx_t(Y)\big )^2\right \rangle
}_{T_3}. \notag
\end{align}
 To prove the results in Eq.~\eqref{def:mcj}, we observe from Eq.~\eqref{def:<X_t(X_q,v_q)>-hr-pic-a} that one would  need  $\lla \dx_t(X)\dfr(Y,u,t)\rra$ and $\lla \dx_t(X)\dx_t(Y)\rra$, which can essentially be obtained by performing appropriate integrals on the basic correlation function 
$\lla \dfr(X,v,t)\dfr(Y,u,t)\rra$. We recall from Eq.~\eqref{corr-scling+structure} that this correlation function is the sum of two terms -- a GGE part and a LR part:
\begin{align}
\lla \dfr(X,v,t)\dfr(Y,u,t)\rra = \frac{1}{\ell}\delta(X-Y)\msc{C}^{\rm r}_{\rm gge}(X,u,v) + \frac{1}{\ell}\msc{C}^{\rm r}_{\rm lr}(X,v;Y,u;t),
\label{C_r-st-a}
\end{align}
where $\msc C^{\rm gge}$ and $\msc C^{\rm lr}$ are explicitly given in Eqs.~\eqref{mscC_gge-<dfrdfr>-hpIC} and \eqref{mscC_lr-<dfrdfr>-hpIC}, respectively. Hence, as mentioned earlier, all terms $T_1,T_2$ and $T_3$ individually have two parts $T_i=T_i^{\rm gge}+T_i^{\rm lr}$, one coming from the GGE correlation and the other from the LR correlation. We present the computation of the two parts separately for each term.  
\subsection{Contribution from LR part of the correlation}
We first note that the discontinuous structure (see  Eq.~\eqref{ex:C^lr-discont}) of correlation 
\begin{align}
 \msc C^{\rm r}_{\rm lr}(X,v;Y,u;t) = \Theta(Y-X) \msc C^{\rm r,+}_{\rm lr}(X,v,Y,u,t) + \Theta(X-Y)\msc C^{\rm r,-}_{\rm lr}(X,v,Y,u,t),
 \label{ex:C^lr-discont-a}
\end{align}
will be particularly useful in computing the contribution from the LR part of the correlation. 

\paragraph{Evaluation of $T_1^{\rm lr}$:} Using the explicit expression of $\delta X_t(X_t^q,v_q)$ from Eq.~\eqref{def:delX_(X_q,v_q)(t)-hr-pic}, we write 
\begin{align}
   T_1=&  \lla \dfr(X_t^q,v_q,t)\dx_t(X_t^q) \rra 
   +a \int dY\int du~\Theta \left(\mc U^t_{dt}(X_t^q,v_q;Y,u)\right)\lla \dfr(X_t^q,v_q,t)\dfr(Y,u,t)\rra_{\rm lr}, \notag \\
   &+a \int dY\int du~\bfr(Y,u,t) \delta \left(\mc U^t_{dt}(X_t^q,v_q;Y,u)\right) \lla \dfr(X_t^q,v_q,t)\big(\dx_t(X_t^q) -\dx_t(Y)\big) \rra. 
   \label{def:T_1_lr}
\end{align}
Expanding in $dt$, one writes
\begin{align}
   T_1=  \left[ T_1\right]_{dt \to 0}
   &+a~dt \int du~\frac{v_q-u}{1-a\brr(X_t^q,t)}\Bigg( \Big\{ \lla \dfr(X_t^q,v_q,t)\dfr(Y,u,t)\rra \label{def:T_1_lr-expand-dt} \\
   &~+ ~\partial_Y\lf\frac{\bfr(Y,u,t)}{1-\brr(Y,t)}  \lla \dfr(X_t^q,v_q,t)\big(\dx_t(X_t^q) -\dx_t(Y)\big) \rra \rf\Big\}_{Y=Y_{dt}^q(u,t)} \Bigg)_{dt \to 0} + O(dt^2), 
   \notag
\end{align}
where 
\begin{align}
   Y_{dt}^q(u,t)=\bX_t\big(\bx_t(X_t^q)+(v_q-u)dt\big). \label{def:Y_dt^q} 
\end{align}
In this part we are interested to compute the contribution $T_1^{\rm lr}$ coming form LR correlation. Since the LR part of correlation $\msc C^{\rm r}_{\rm lr}(X,v;Y,u;t)$ is not singular, it is easy to see that $\left[T_{1}^{\rm lr}\right]_{dt \to 0}=0$.

Noting $\dx_t(Y)= -a \int dZ \int dw~ \Theta(Y-Z)\dfr(Z,w,t)$ and using the correlation in Eq.~\eqref{C_r-st-a}, it is easy to show that \cite{hubner2025hydrodynamics}
\begin{align}
  \ell \lla \dfr(X,v,t)\big(\dx_t(X) -\dx_t(Y)\big) \rra_{\rm lr}  =
  -a \int_Y^{X}dZ~\int dw~\msc C^{{\rm r,sgn}(Z-X)}_{\rm lr}(X,v;Z,w;t).
  \label{def:<dfX(dxX-dxY)>_lr}
\end{align}
We  insert the LR part of the correlation from Eq.~\eqref{C_r-st-a} and  the above expression, inside the integrands in the second and third terms of Eq.~\eqref{def:T_1_lr-expand-dt}, and after some simplifications we find, 
\begin{align}
\begin{split}
 T_1^{\rm lr} = \frac{dt}{\ell} a\int du\int du' \frac{v_q-u}{1-a\brr(X_t^q,t)} &\lf \delta(u-u') + \frac{a\bfr(X_t^q,u,t)}{1-a\brr(X_t^q,t)}\rf \\ 
 &~~~\times~\msc C^{{\rm r,sgn}(v_q-u)}_{\rm lr}(X_t^q,v_q;X_t^q,u';t)~~ +~~ O(dt^2).
 \end{split}
 \label{ex:ellT_1}
\end{align}

\paragraph{Evaluation of $T_{23}^{\rm lr}=T_2^{\rm lr}+T_3^{\rm lr}$:}
Performing the integral over $Y$ in Eq.~\eqref{def:<X_t(X_q,v_q)>-hr-pic-a}, we rewrite $T_{23}$ 
\begin{align}
 T_{23}=&~a \bfr(X_t^q,v_q,t) \int du~\Big[\frac{1}{1-\brr(Y,t)}
 \big \langle \big(\dx_t(X_t^q) -\dx_t(Y)\big) ~\dfr(Y,u,t) \big \rangle \Big]_{Y=Y_{dt}^q(u,t)}  \label{def:T_23} \\
 & + \frac{a}{2}\bfr(X_t^q,v_q,t) \int du~\Big[\frac{1}{1-\brr(Y,t)} \partial_Y\lf \frac{\bfr(Y,u,t)}{1-\brr(Y,t)}~
 \left \langle \big(\dx_t(X_t^q) -\dx_t(Y)\big )^2\right \rangle\rf \Big]_{Y=Y_{dt}^q(u,t)}, \notag 
\end{align}
where $Y_{dt}^q(u,t)$ is given in Eq.~\eqref{def:Y_dt^q}.
We would like to develop $T_{23}$ to $O(dt)$ and for that we expand 
$Y_{dt}^q(t) = X_t^q + \frac{v_q-u}{1-\brr(X_t^q,t)}~dt + O(dt^2)$. Using this expansion in Eq.~\eqref{def:T_23}, one can write 
\begin{align}
T_{23}=& \left[T_{23}\right]_{dt \to 0} \notag \\
   &~~+ a dt ~\bfr(X_t^q,v_q,t) \int du~\frac{v_q-u}{1-\brr(X_t^q,t)}\Bigg( \Big\{
    \partial_Y\Big[\frac{1}{1-\brr(Y,t)}
 \big \langle \big(\dx_t(X_t^q) -\dx_t(Y)\big) ~\dfr(Y,u,t) \big \rangle \Big] \notag \\
 &~~ + \frac{1}{2}  \partial_Y\Big[\frac{1}{1-\brr(Y,t)} \partial_Y\lf \frac{\bfr(Y,u,t)}{1-\brr(Y,t)}~
 \left \langle \big(\dx_t(X_t^q) -\dx_t(Y)\big )^2\right \rangle\rf \Big]
   \Big\}_{Y=Y_{dt}^q(u,t)} \Bigg)_{dt \to 0}
\label{def:T_23-O(dt)}
\end{align}
Each of the terms on the right hand side on the above equation have two parts -- one from GGE correlation and other from LR correlation. Once again using non-singular nature of $\msc C^{\rm r}_{\rm lr}(X,v;Y,u;t)$, it is easy to see that $\left[T_{23}^{\rm lr}\right]_{dt \to 0}=0$. However, as will see in the next section, there are non-zero contribution to the $O(dt^0)$ term coming from GGE correlation.  Now we evaluate the LR contribution to  $T_{23}^{\rm lr}$ at $O(dt)$. For that we need to insert the LR contribution of the correlations present in the integrands of Eq.~\eqref{def:T_23}. Once again using the non-singular feature of the LR correlation, one can simplify $T_{23}^{\rm lr}$
\begin{align}
\begin{split}
T_{23}^{\rm lr}=& a dt ~\bfr(X_t^q,v_q,t) 
    \int du~\int du'~\frac{v_q-u}{(1-\brr(X_t^q,t))^2}\Big[\delta(u-u')+\frac{a\bfr(X_t^q,u,t)}{1-\brr(X_t^q,t)}\Big]  \\
 &~~~\times~~\lf\Big\{\partial_Y\big \langle \big(\dx_t(X_t^q) -\dx_t(Y)\big) ~\dfr(Y,u',t) \big \rangle_{\rm lr} \Big] 
   \Big\}_{Y=Y_{dt}^q(u,t)}\rf_{dt \to 0} + O(dt^2).
 \end{split}  
 \label{def:T_23^lr-O(dt)}
\end{align}
Similar to Eq.~\eqref{def:<dfX(dxX-dxY)>_lr}, for $Y \to X$ one can write 
\begin{align}
   &\lla \dfr(Y,u',t)\big(\dx_t(X) -\dx_t(Y)\big) \rra_{\rm lr}  =
  -\frac{a}{\ell} \int_Y^{X}dZ~\int dw~\msc C^{{\rm r,sgn}(Y-Z)}_{\rm lr}(Z,w;Y,u';t)\notag \\
&~~~~~~~~~~~~~~~~
= -\frac{a}{\ell} \int_Y^{X}dZ~\int dw~\big[\Theta(Y-Z)\msc C^{{\rm r,+}}_{\rm lr}(Z,w;Y,u';t) + \Theta (Z-Y)\msc C^{{\rm r,-}}_{\rm lr}(Z,w;Y,u';t)\big],\displaybreak[3]\notag \\
&~~~~~~~~~~~~~~~~
= -\frac{a}{\ell} \int_Y^{X}dZ~\int dw~\big[\Theta(Y-X)\msc C^{{\rm r,+}}_{\rm lr}(Z,w;Y,u';t) + \Theta (X-Y)\msc C^{{\rm r,-}}_{\rm lr}(Z,w;Y,u';t)\big],
  \label{def:<dfY(dxX-dxY)>_lr}
\end{align}
Taking the derivative with respect to $Y$, we get  
\begin{align}
\begin{split}
  \partial_Y \lla \dfr(Y,u',t)\big(\dx_t(X) -\dx_t(Y)\big) \rra_{\rm lr}  =& \frac{a}{\ell} \Theta(X-Y)\int dw~\msc C^{{\rm r,-}}_{\rm lr}(Y,w;Y,u';t),  \\ 
  & +\frac{a}{\ell} \Theta(Y-X) \int dw~\msc C^{{\rm r,+}}_{\rm lr}(Y,w;Y,u';t),  \\
  & - \frac{a}{\ell}\int_Y^{X}dZ~\int dw~\partial_Y\msc C^{{\rm r,sgn}(Y-Z)}_{\rm lr}(Z,w;Y,u';t).
  \end{split}
  \label{def:<dfY(dxX-dxY)>_lr}
\end{align}
In the $Y \to X$ limit, the third term on the right hand side of the above equation goes to zero, whereas the contributions from the first two terms depend on whether $Y \to X^+$ or 
$Y \to X^-$. Since $Y_{dt}^q(u,t)=\bX_t\big(\bx_t(X_t^q)+(v_q-u)dt\big)$, for $v_q>u$, $Y \to X^+$ and for $v_q<u$, $Y \to X^-$ as $dt \to 0$. Using these facts, we find
\begin{align}
\lf\Big\{\partial_Y \lla \dfr(Y,u',t)\big(\dx_t(X) -\dx_t(Y)\big) \rra_{\rm lr} \Big\}_{Y=Y_{dt}^q(u,t)}\rf_{dt \to 0}
= \frac{a}{\ell}\int dw~\msc C^{{\rm r, sgn}(v_q-u)}_{\rm lr}(X_t^q,w;X_t^q,u';t).   
\end{align}
Inserting this result in Eq.~\eqref{def:T_23^lr-O(dt)} and simplifying one gets
\begin{align}
\begin{split}
T_{23}^{\rm lr}= \frac{dt}{\ell}~~\frac{a^2 ~\bfr(X_t^q,v_q,t)}{(1-\brr(X_t^q,t))^2} 
    \int du~\int du'~&(v_q-u)\Big[\delta(u-u')+\frac{a\bfr(X_t^q,u,t)}{1-\brr(X_t^q,t)}\Big], \\ 
   &~\times~~ \int dw~\msc C^{{\rm r, sgn}(v_q-u)}_{\rm lr}(X_t^q,w;X_t^q,u';t) + O(dt^2). 
\end{split}
\label{def:T_23^lr}
\end{align}
Finally, inserting $T_1^{\rm lr}$ and $T_{23}^{\rm lr}$ from Eqs.~\eqref{ex:ellT_1} and \eqref{def:T_23^lr}, respectively, in Eq.~\eqref{def:<X_t(X_q,v_q)>-hr-pic-a} and simplifying, one gets 
\begin{align}
 \ovl{dX}_t^{\rm lr}(X_t^q,v_q)=  \frac{dt}{\ell}~\ovl{\mc j}_{\rm d}^{\rm lr}(X_t^q,v_q) + O(dt^2),
 \label{def:ovl(dX)_t^lr-a}
\end{align}
as announced in Eq.~\eqref{mcj_c1^lr}, where
\begin{align}
\ovl{\mc j}_{\rm d}^{\rm lr}(X_t^q,&v_q,t) = ~\frac{a}{\bfr(X_t^q,v_q,t)(1-\brr(X_t^q,t))} ~\int du~\int du'\int dw~(v_q-u)    \label{def:mcj_d^lr-a} \\
    &~~\times~~\left[\delta(u-u')+\frac{a\bfr(X_t^q,u,t)}{1-\brr(X_t^q,t)}\right]~\left[\delta(v_q-w)+\frac{a\bfr(X_t^q,v_q,t)}{1-\brr(X_t^q,t)}\right]~ \msc C^{{\rm r, sgn}(v_q-u)}_{\rm lr}(X_t^q,w;X_t^q,u';t). \notag
\end{align}

\subsection{Contribution from GGE part of the correlation:}
To compute the GGE parts of $T_1$ and $T_{23}$, we will have the use the following GGE correlations. From Eqs.~\eqref{C_r-st-a} and \eqref{mscC_gge-<dfrdfr>-hpIC}, we have 
\begin{subequations}
\label{ex:gge-corrs-a}
\begin{align}
\lla \dfr(X,v,t)\dfr(Y,u,t)\rra_{\rm gge} = \frac{1}{\ell}\delta(X-Y) \lsq \delta(u-v)\bfr(X,v,t) -a(2-a\rho(X,t))\bfr(X,v,t)\bfr(Y,u,t)\rsq. 
\end{align}
By performing appropriate integrals of this correlation, it is easy to obtain the other necessary correlations \cite{hubner2025hydrodynamics}
\begin{align}
 \lla \dfr(X,v,t)\drr(Y,t)\rra_{\rm gge}=& \frac{1}{\ell}\delta(X-Y)\bfr(X,v,t)(1-a\brr(X,t))^2, \\  
  \lla \drr(X,t)\drr(Y,t)\rra_{\rm gge}=& \frac{1}{\ell}\delta(X-Y)\brr(X,t)(1-a\brr(X,t))^2, \\  
 \lla \dfr(X,v,t)\dx_t(Y)\rra_{\rm gge}=&-\frac{a}{\ell}\bfr(X,v,t)\Theta( Y-X)(1-a\brr(X,t))^2,\\
  \lla \dx_t(X)\dx_t(Y)\rra_{\rm gge}=& \frac{a^2}{\ell} \Gamma_t(\min(X,Y)).
\end{align}
where
\begin{align}
\Gamma_t(X)=\int dZ~\Theta(X-Z) \brr(Z,t)(1-a\brr(Z,t))^2.
\label{def:Gamma_t}
\end{align}
\end{subequations}
Using these correlations in Eqs.~\eqref{def:T_1_lr-expand-dt} and \eqref{def:T_23-O(dt)}, one can show that 
\begin{align}
  [T_1^{\rm gge}]_{dt \to 0}=&   -\frac{a^2}{2\ell}\bfr(X_t^q,v_q,t) \int du~{\rm sgn}(v_q-u)~\bfr(X_t^q,u,t), \\
  [T_{23}^{\rm gge}]_{dt \to 0}=&   \frac{a^2}{2\ell}\bfr(X_t^q,v_q,t) \int du~{\rm sgn}(v_q-u)~\bfr(X_t^q,u,t).
\end{align}
Hence $\ovl{dX}_t(X_t^q,v_q)$ in Eq.~\eqref{def:<X_t(X_q,v_q)>-hr-pic-a} is equal to $0$ for $dt \to 0$. This implies $\ovl{dX}_t(X_t^q,v_q)= dt \frac{\ovl{\mc j}_{\rm d}(X_t^q,v_q,t)}{\ell} + O(dt^2)$ with $\ovl{\mc j}_{\rm d}=\ovl{\mc j}_{\rm d}^{\rm gge}+\ovl{\mc j}_{\rm d}^{\rm lr}$ as was mentioned in Eqs.~\eqref{mcj_c1^gge} and \eqref{mcj_c1^lr}. In order to obtain $\ovl{\mc j}_{\rm d}$, one now needs to evaluate GGE contribution to the terms linear in $dt$ of $T_1$ and $T_{23}$ in Eqs.~\eqref{def:T_1_lr-expand-dt} and \eqref{def:T_23-O(dt)}, which can be done by inserting the correlations from Eq.~\eqref{ex:gge-corrs-a} and performing the manipulations. The calculations of these contributions are provided in \cite{hubner2025hydrodynamics} in detail. We here provide only the final results:
\begin{align}
 T_1^{\rm gge} =& [T_1^{\rm gge}]_{dt \to 0} + \frac{dt}{\ell}~\frac{a^2\bfr(X_t^q,v_q,t)}{2}\Big[(1-a\brr(Y,t))\partial_{Y}\frac{\int du~|v_q-u|\bfr(Y,u,t)}{1-\brr(Y,t)}\Big]_{Y\to X_t^q} 
 + O(dt^2) \\
 \begin{split}
 T_{23}^{\rm gge} =&[T_{23}^{\rm gge}]_{dt \to 0} + \frac{dt}{\ell}~\frac{a^2\bfr(X_t^q,v_q,t)}{2} 
 \Bigg[\frac{\partial_{Y}\int du~|v_q-u|\bfr(Y,u,t)}{1-a\brr(Y,t)} \\ 
 &~~~~~~~~~~~~~~~~~~~~~~~
 +a \brr(Y,t) \partial_Y\lf \frac{\int du~|v_q-u|\bfr(Y,u,t)}{1-a\brr(Y,t)}\rf\Bigg]_{Y\to X_t^q}
 + O(dt^2)
 \end{split}
\end{align}
Adding these two contributions and using $[T_1^{\rm gge}]_{dt \to 0}+[T_{23}^{\rm gge}]_{dt \to 0}=0$, one, after some manipulations gets
\begin{align}
 \ovl{dX}_t^{\rm gge}(X_t^q,v_q)=  \frac{dt}{\ell}~\ovl{\mc j}_{\rm d}^{\rm gge} (X_t^q,v_q,t)
 + O(dt^2),
 \label{def:ovl(dX)_t^gge-a}
\end{align}
where
\begin{align}
\ovl{\mc j}_{\rm d}^{\rm gge} (Y,v_q,t)=\frac{a^2}{2} \Bigg[\frac{\partial_{Y}\int du~|v_q-u|\bfr(Y,u,t)}{1-a\brr(Y,t)} 
 + \partial_Y\lf \frac{\int du~|v_q-u|\bfr(Y,u,t)}{1-a\brr(Y,t)}\rf\Bigg]. \label{def:ovl(mcj)_d-a}
\end{align}
Adding the two contributions from Eqs.~\eqref{def:ovl(dX)_t^lr-a} and \eqref{def:ovl(dX)_t^gge-a}, one gets the total $\ovl{dX}_t(X_t^q,v_q)$ to $O(dt)$.

\section{Derivation of the results in Eqs.~\eqref{mcj_f^gge} and \eqref{mcj_f^lr}}
\label{derv:variance-a}
We start with the following correlation
\begin{align}
   \mb C_{t,t'}(Z_q,v_q;Z_r,v_r) = \lla \dX_{t}(Z_q,v_q)\dX_{t'}(Z_r,v_r)\rra,
   \label{def:mbC_tt'-1}
\end{align}
where $\dX_{t}(Z_q,v_q)$ is defined in Eq.~\eqref{def:delX_(X_q,v_q)(t)-hr-pic}. Explicitly, this correlation reads
\begin{align}
 \mb C_{t,t'}(Z_q,v_q;Z_r,&v_r) = {\lla \dx_t(Z_q)\dx_{t'}(Z_r) \rra} \displaybreak[3] \cr
 &~~+ {a}\int dYdv~\Theta\lf\mc U_{{dt}}^t(Z_q,v_q;Y,v)\rf \lla \dfr(Y,v,t)\dx_{t'}(Z_r) \rra \displaybreak[3]\cr
 &~~+{a}\int dYdv~\Theta\lf\mc U_{{dt}}^{t'}(Z_r,v_r;Y,v)\rf \lla \dfr(Y,v,t')\dx_{t}(Z_q) \rra \displaybreak[3]\cr
 &~~+ {a}\int dYdv~\delta\lf\mc U_{{dt}}^t(Z_q,v_q;Y,v)\rf \lla \lf\dx_t(Z_q)-\dx_t(Y)\rf\dx_{t'}(Z_r) \rra \bfr(Y,v,t) \displaybreak[3] \cr
 &~~+ {a}\int dYdv~\delta\lf\mc U_{{dt}}^{t'}(Z_r,v_r;Y,v)\rf \lla \lf\dx_{t'}(Z_r)-\dx_{t'}(Y)\rf\dx_{t}(Z_q) \rra \bfr(Y,v,t') \displaybreak[3] \cr
 &~~+ {a^2}\int dYdv\int dZdu~\Theta\lf\mc U_{{dt}}^t(Z_q,v_q;Y,v)\rf\Theta\lf\mc U_{{dt}}^{t'}(Z_r,v_r;Z,u)\rf \displaybreak[3] \label{ex:C_tt'-a} \\ 
  &~~~~~~~~~~~~~~~~\times~~\lla \dfr(Y,v,t)\dfr(Z,u,t')\rra \displaybreak[3] \cr
  &~~+ {a^2}\int dYdv\int dZdu~\Theta\lf\mc U_{{dt}}^t(Z_q,v_q;Y,v)\rf\delta\lf\mc U_{{dt}}^{t'}(Z_r,v_r;Z,u)\rf \displaybreak[3] \cr
  &~~~~~~~~~~~~~~~~\times~~\lla \lf\dx_{t'}(Z_r)-\dx_{t'}(Z)\rf\dfr_{t}(Y,v) \rra \bfr(Z,u,t') \displaybreak[3] \cr
  &~~+ {a^2}\int dYdv\int dZdu~\delta\lf\mc U_{{dt}}^t(Z_q,v_q;Y,v)\rf\Theta\lf\mc U_{{dt}}^{t'}(Z_r,v_r;Z,u)\rf \displaybreak[3] \cr
  &~~~~~~~~~~~~~~~~\times~~\lla \lf\dx_{t}(Z_q)-\dx_{t}(Y)\rf\dfr_{t'}(Z,u) \rra \bfr(Y,v,t') \displaybreak[3] \cr
  &~~+ {a^2}\int dYdv\int dZdu~\delta\lf\mc U_{{dt}}^t(Z_q,v_q;Y,v)\rf\delta\lf\mc U_{{dt}}^{t'}(Z_r,v_r;Z,u)\rf \displaybreak[3] \cr
  &~~~~~~~~~~~~~~~~\times~~\lla \lf\dx_{t}(Z_q)-\dx_{t}(Y)\rf\lf\dx_{t'}(Z_r)-\dx_{t'}(Z)\rf \rra \bfr(Y,v,t)  \bfr(Z,u,t'), \displaybreak[3] \notag
\end{align}
where $\mc U^t_{dt}(Z,v;Y,u)$ is given in Eq.~\eqref{def:mcU^t_tau}. Using the fact that $\lla \dfr(Z,v,t)\dfr(Y,u,t')\rra$ does not have any singularity anywhere, for $t\ne t'$, in the ${dt} \to 0$ limit, one finds
\begin{align}
\lim_{{dt} \to 0}\mb C_{t,t'}(Z_q,v_q;Z_r,v_r)
=0. 
\end{align}
Next, we aim to find the $O(dt)$ part of $ \mb C_{t,t'}(Z_q,v_q;Z_r,v_r)$. For that we define the following `derivative' for a function $H_t(dt)$
\begin{align}
\partial_{dt} H_t(dt) = \frac{H_t(dt)- [H_t(dt)]_{dt \to 0}}{dt}.
\label{def:partial_dt}
\end{align}
It is easy to show that,
\begin{align}
\begin{split}
\partial_{dt} \mc U^t_{dt}(Z_q,v_q;Y,v) =& (v_q-v) + O(dt), \\ 
\delta \lf \mc U^t_{dt}(Z_q,v_q;Y,v)\rf =& \frac{\delta\lf Y -\bX_t(\bx_t(Z_q)+v_q{dt}-v{dt})\rf}{1-a\brr(Y,t)}, \\
\partial_{dt} \delta \lf \mc U^t_{dt}(Z_q,v_q;Y,v)\rf =& - \frac{(v_q-v)}{1-\brr(Y,t)} \partial_Y \lf \frac{\delta\lf Y -\bX_t(\bx_t(Z_q)+v_q{dt}-v{dt})\rf}{1-a\brr(Y,t)}\rf + O(dt), \\
\partial_Y\dx_t(Y) =& -a\drr(Y,t).
\end{split}
\label{useful-relations-a}
\end{align}
Using these results, one can show that for $t \ne t'$,
\begin{align}
\big[\partial_{dt} \mb C_{t,t'}(Z_q,v_q;Z_r,v_r)\big]_{{dt} \to 0} =0,~~\text{for}~~t \ne t'.    
\label{re:dC_tt=0-a}
\end{align}
The $t=t'$ case has to be treated separately.

\subsection{The equal time ($t=t'$) case:} 
We first note that similar to other correlations $\mb C_{t,t}$ is also a sum of two parts: $\mb C_{t,t}=\mb C_{t,t}^{\rm gge}+\mb C_{t,t}^{\rm lr}$. Since  the LR part of correlation $\msc C^{\rm r}_{\rm lr}(X,v;Y,u;t)$ is non-singular, one can follow the same steps of calculation as done for the $t\ne t'$ case and show that 
\begin{align}
 \mb C_{t,t}^{\rm lr}(Z_q,v_q;Z_r,v_r)\big{|}_{{dt} \to 0} =&0, ~~
 \partial_{dt} \mb C_{t,t}^{\rm lr}(Z_q,v_q;Z_r,v_r)\big{|}_{{dt} \to 0} =0,
 \label{ex:C_tt^lr-dt->0-a}
\end{align}
which  proves the result announced in Eq.~\eqref{mcj_f^lr}.

We now need to focus on the GGE contribution. Using the explicit forms of the correlations from  Eq.~\eqref{ex:gge-corrs-a} in Eq.~\eqref{ex:C_tt'-a}, we first rewrite 
$\mb C_{tt}^{\rm gge}$ as 
\begin{align}
\label{ex:C_tt^gge}
\ell \mb C^{\rm gge}_{t,t}(Z_q,&v_q;Z_r,v_r) = ~~{a^2} \Gamma_t(\min(Z_q,Z_r)) \displaybreak[3] \cr
& {-a^2} \int dY\int dv \Theta\lf \mc U_{dt}^t(Z_q,v_q;Y,v)\rf \bfr(Y,v,t)(1-a\brr(Y,t))^2 \Theta(Z_r-Y) \displaybreak[3] \notag \\
& {-a^2} \int dZ\int du \Theta\lf \mc U_{dt}^t(Z_r,v_r;Z,u)\rf \bfr(Z,u,t)(1-a\brr(Z,t))^2 \Theta(Z_q-Z) \displaybreak[3] \notag \\
& {+a^3} \int dY\int dv \delta\lf \mc U_{dt}^t(Z_q,v_q;Y,v)\rf \bfr(Y,v,t)\lsq \Gamma_t(\min(Z_q,Z_r))- \Gamma_t(\min(Z_r,Y))  \rsq,  \displaybreak[3]  \\
& {+a^3} \int dZ\int du \delta\lf \mc U_{dt}^t(Z_r,v_r;Z,u)\rf \bfr(Z,u,t)\lsq \Gamma_t(\min(Z_q,Z_r))- \Gamma_t(\min(Z_q,Z))  \rsq, \displaybreak[3]  \notag\\
& {+a^2} \int dY\int dv\int dZ\int du \Theta\lf \mc U_{dt}^t(Z_q,v_q;Y,v)\rf \Theta\lf \mc U_{dt}^t(Z_r,v_r;Z,u)\rf \delta(Y-Z)\msc C_{\rm gge}(Y,v,u) \displaybreak[3] \notag \\
& {-a^3} \int dY\int dv\int dZ\int du \Theta\lf \mc U_{dt}^t(Z_q,v_q;Y,v)\rf \delta\lf \mc U_{dt}^t(Z_r,v_r;Z,u)\rf \displaybreak[3] \notag \\
&~~~~~\times~~\bfr(Y,v,t)\bfr(Z,u,t)(1-a\brr(Y,t))^2 \lsq \Theta(Z_r-Y)- \Theta(Z-Y)  \rsq \displaybreak[3] \notag \\
& {-a^3} \int dY\int dv\int dZ\int du \delta\lf \mc U_{dt}^t(Z_q,v_q;Y,v)\rf \Theta\lf \mc U_{dt}^t(Z_r,v_r;Z,u)\rf \displaybreak[3] \notag \\
&~~~~~\times~~\bfr(Y,v,t)\bfr(Z,u,t)(1-a\brr(Y,t))^2 \lsq \Theta(Z_q-Z)- \Theta(Y-Z)  \rsq \displaybreak[3] \notag \\
& {+a^4} \int dY\int dv\int dZ\int du \delta\lf \mc U_{dt}^t(Z_q,v_q;Y,v)\rf \delta\lf \mc U_{dt}^t(Z_r,v_r;Z,u)\rf \displaybreak[3] \notag \\
&~~~~~\times~~\lsq \Gamma_t(\min(Z_q,Z_r))- \Gamma_t(\min(Z_r,Y)) -\Gamma_t(\min(Z_q,Z)) +\Gamma_t(\min(Y,Z))\rsq, \displaybreak[3] \notag 
\end{align}
where $\msc C_{\rm gge}$ is given in Eq.~\eqref{mscC_gge-<dfrdfr>-hpIC}.
Recalling $\mc U_{dt \to 0}^t(Z_q,v_r,;Y,v)=\bx_t(Z_q) - \bx_t(Y)$ and using the relations in Eq.~\eqref{useful-relations-a}, and performing some tedious manipulations, we get
\begin{align}
\label{ex:C_tt^gge-dt->0-a-2}
\lim_{dt \to 0}\ell \mb C^{\rm gge}_{t,t}(Z_q,v_q;&Z_r,v_r) =0
\end{align}
Combing the results from Eqs.~\eqref{ex:C_tt^lr-dt->0-a} and \eqref{ex:C_tt^gge-dt->0-a-2}, we have 
\begin{align}
\lim_{dt \to 0 } \mb C_{t,t}(Z_q,v_q;&Z_r,v_r)=0,~~\text{for}~~-\infty \leq (Z_q, Z_r) \leq \infty ~~\text{and}~~t>0.
\label{ex:C_tt-dt->0-a}
\end{align}
This suggests 
\begin{align}
 \mb C_{t,t}(&Z_q,v_q;Z_r,v_r) = dt \lf  \partial_{dt}\mb C^{\rm gge}_{t,t}(Z_q,v_q;Z_r,v_r)\rf_{dt \to 0} + O(dt^2). \label{ex:mbC_tt-O(dt)}  
\end{align}
Next, we evaluate $\lf\partial_{dt}\mb C^{\rm gge}_{t,t}(Z_q,v_q;Z_r,v_r)\rf_{dt \to 0}$ and for that we first compute the `derivative' -- 
\begin{align}
\label{ex:dmbC^gge_tt-a}
 \ell\partial_{dt}\mb C^{\rm gge}_{t,t}(&Z_q,v_q;Z_r,v_r) = {-a^2} \int dv~ (v_q-v)(1-a\brr(Z_q,t))\bfr(Z_q,v,t) \Theta\lf Z_r-Z_q\rf   \displaybreak[3] \notag \\
&~ {-a^2} \int du~ (v_r-u)(1-a\brr(Z_r,t))\bfr(Z_r,u,t) \Theta\lf Z_q-Z_r\rf   \displaybreak[3] \notag \\
&~{+a^3} \int dv~ \frac{v_q-v}{1-a\brr(Z_q,t)}\Big[ \partial_Y \lf  \frac{\bfr(Y,v,t)}{1-a\brr(Y,t)} \big\{\Gamma_t(\min(Z_q,Z_r)) - \Gamma_t(\min(Z_r,Y))\big\} \rf\Big]_{Y=Y_{dt}^q(v,t)}  \displaybreak[3] \notag \\
&~{+a^3} \int du~ \frac{v_r-u}{1-a\brr(Z_r,t)}\Big[ \partial_Z \lf  \frac{\bfr(Z,u,t)}{1-a\brr(Z,t)} \big\{\Gamma_t(\min(Z_q,Z_r)) - \Gamma_t(\min(Z_q,Z))\big\} \rf\Big]_{Z=Y_{dt}^r(u,t)}  \displaybreak[3] \notag \\
&~{+a^2} \int dv\int du~ \frac{v_q-v}{1-a\brr(Z_q,t)}\Theta\lf Y_{dt}^r(u,t)- Y_{dt}^q(v,t)\rf~\msc C_{\rm gge}(Z_q,v,u) \displaybreak[3] \notag \\
&~{+a^2} \int dv\int du~ \frac{v_r-u}{1-a\brr(Z_r,t)}\Theta\lf Y_{dt}^q(v,t)- Y_{dt}^r(u,t)\rf~\msc C_{\rm gge}(Z_r,v,u) \displaybreak[3] \notag \\
&~{-a^3} \int dv\int du~ \frac{(v_q-v)\bfr(Z_q,v,t)\bfr(Z_r,u,t)}{(1-a\brr(Z_r,t))}(1-a\brr(Z_q,t)) \displaybreak[3] \notag \\
&~~~~~~~~~~~~~~~~~~~~~\times~\big[\Theta\lf \mc U_{dt}^t(Z_r,v;Z_q,v_q)\rf-\Theta\lf \mc U_{dt}^t(Z_r,v_r;Z_q,v_q)+(v-u)dt\rf\big] \displaybreak[3] \notag \\
&~{-a^3} \int dv\int du~ \frac{(v_r-u)\bfr(Z_q,v,t)\bfr(Z_r,u,t)}{(1-a\brr(Z_q,t))}(1-a\brr(Z_r,t)) \displaybreak[3] \notag \\
&~~~~~~~~~~~~~~~~~~~~~\times~\big[\Theta\lf \mc U_{dt}^t(Z_q,u;Z_r,v_r)\rf-\Theta\lf \mc U_{dt}^t(Z_q,v_q;Z_r,v_r)+(u-v)dt\rf\big] \displaybreak[3] \notag \\
&~{-a^3}\int dY \int dv\int du~ \frac{(v_r-u)\bfr(Y,v,t)}{(1-a\brr(Z_r,t))}(1-a\brr(Y,t))^2 \Theta\lf \mc U_{dt}^t(Z_q,v_q;Y,v)\rf \displaybreak[3]  \\
&~~~~~~~~~~~~~~~~~~~~~\times~\Big[ \partial_Z \lf  \frac{\bfr(Z,u,t)}{1-a\brr(Z,t)} \big\{\Theta(Z_r-Y) - \Theta(Z-Y)\big\} \rf\Big]_{Z=Y_{dt}^r(u,t)} \displaybreak[3] \notag \\
&~{-a^3} \int dZ \int dv\int du~ \frac{(v_q-v)\bfr(Z,u,t)}{(1-a\brr(Z_q,t))}(1-a\brr(Z,t))^2 \Theta\lf \mc U_{dt}^t(Z_r,v_r;Z,u)\rf \displaybreak[3] \notag \\
&~~~~~~~~~~~~~~~~~~~~~\times~\Big[ \partial_Y \lf  \frac{\bfr(Y,v,t)}{1-a\brr(Y,t)} \big\{\Theta(Z_q-Z) - \Theta(Y-Z)\big\} \rf\Big]_{Y=Y_{dt}^q(v,t)} \displaybreak[3] \notag \\
&~{+a^4} \int dv\int du~ \frac{v_q-v}{1-a\brr(Z_q,t)}\frac{\bfr(Z_r,u,t)}{1-a\brr(Z_r,t)} \displaybreak[3] \notag \\
&~~~~~~~\times~\Bigg[ \partial_Y \Big(  \frac{\bfr(Y,v,t)}{1-a\brr(Y,t)} \big\{\Gamma_t(\min(Z_q,Z_r)) - \Gamma_t(\min(Z_r,Y)) \displaybreak[3] \notag \\ 
&~~~~~~~~-\Gamma_t\lf\min(\bX_t(\bx_t(Z_r)+(v_r-u)dt),Z_q)\rf \notag \\ 
&~~~~~~~~~~~~~~~~~~~~~~+\Gamma_t\lf\min(\bX_t(\bx_t(Z_r)+(v_r-v)dt),Y)\rf\big\} \Big)\Bigg]_{Y=Y_{dt}^q(v,t)}  \displaybreak[3] \notag \\
&~{+a^4} \int dv\int du~ \frac{v_r-u}{1-a\brr(Z_r,t)}\frac{\bfr(Z_q,v,t)}{1-a\brr(Z_q,t)} \displaybreak[3] \notag \\
&~~~~~~~\times~\Bigg[ \partial_Z \Big(  \frac{\bfr(Z,u,t)}{1-a\brr(Z,t)} \big\{\Gamma_t(\min(Z_q,Z_r)) - \Gamma_t(\min(Z_q,Z)) \displaybreak[3] \notag \\ 
&~~~-\Gamma_t\lf\min(\bX_t(\bx_t(Z_q)+(v_q-v)dt),Z_r)\rf \notag \\ 
&~~~~~~~~~~~~~~~~~~~~~~+\Gamma_t\lf\min(\bX_t(\bx_t(Z_q)+(v_q-v)dt),Z)\rf\big\} \Big)\Bigg]_{Z=Y_{dt}^r(u,t)}.  \displaybreak[3] \notag 
\end{align}
where recall from Eq.~\eqref{def:Y_dt^q} that $Y_{dt}^q(v,t)=\bX_t\big(\bx_t(Z_q)+(v_q-u)dt\big)$,  $\mc U_{dt}^t(Y,v;Z,u)$ is given in Eq.~\eqref{def:mcU^t_tau}, and the transformation $\bX_t(x)$ is given in Eq.~\eqref{trans:x->X-hf} with $\fp(y,v,t)$ inside the integrand replaced by $\bfp(y,v,t)$. One can prove that 
\begin{align}
  \lf \ell \partial_{dt}\mb C^{\rm gge}_{t,t}(Z_q,v_q;Z_r,v_r) \rf_{dt \to 0} =0,~~\text{for}~~Z_q \ne Z_r. 
  \label{ex:dmbC^gge_tt-dt->0-a-x_ne_y}
\end{align}
\paragraph{\textbf{\textit{Proof:}}} 
\begin{align}
\ell\partial_{dt}\mb C^{\rm gge}_{t,t}(&Z_q,v_q;Z_r,v_r) = {-a^2} \int dv~ (v_q-v)(1-a\brr(Z_q,t))\bfr(Z_q,v,t) \Theta\lf Z_r-Z_q\rf   \displaybreak[3] \notag \\
&~ {-a^2} \int du~ (v_r-u)(1-a\brr(Z_r,t))\bfr(Z_r,u,t) \Theta\lf Z_q-Z_r\rf   \displaybreak[3] \notag \\
&~{-a^3} \int dv~ (v_q-v) \bfr(Z_q,v,t) \brr(Z_q,t)\Theta(Z_r-Z_q) \displaybreak[3] \notag \\
&~{-a^3} \int du~ (v_r-u) \bfr(Z_r,u,t) \brr(Z_r,t)\Theta(Z_q-Z_r) 
\displaybreak[3] \notag \\
&~{+a^2} \int dv~ (v_q-v)\bfr(Z_q,v,t)(1-a\brr(Z_q,t))\Theta\lf Z_r-Z_q\rf \displaybreak[3] \notag \\
&~{+a^2} \int du~ (v_r-u)\bfr(Z_r,u,t)(1-a\brr(Z_r,t))\Theta\lf Z_q-Z_r\rf \displaybreak[3] \notag \\
&~{-a^3} \int dv~\frac{(v_q-v)\bfr(Z_q,v,t)\brr(Z_r,t)}{(1-a\brr(Z_r,t))}(1-a\brr(Z_q,t)) \big[\Theta\lf Z_r-Z_q\rf-\Theta\lf Z_r-Z_q)\rf\big] \displaybreak[3] \notag \\
&~{-a^3} \int du~ \frac{(v_r-u)\brr(Z_q,t)\bfr(Z_r,u,t)}{(1-a\brr(Z_q,t))}(1-a\brr(Z_r,t)) \big[\Theta\lf Z_q-Z_r\rf-\Theta\lf Z_q-Z_r\rf\big] \displaybreak[3] \notag \\
&~{+a^3}\int dY \int du~ \frac{(v_r-u)\brr(Y,t)\bfr(Z_r,u,t)}{(1-a\brr(Z_r,t))^2}(1-a\brr(Y,t))^2 \Theta\lf Z_q-Y\rf  \delta(Z_r-Y) \displaybreak[3] \notag \\
&~{+a^3} \int dZ \int dv~ \frac{(v_q-v)\bfr(Z_q,v,t)\brr(Z,t)}{(1-a\brr(Z_q,t))^2}(1-a\brr(Z,t))^2 \Theta\lf Z_r-Z\rf  \delta(Z_q-Z)  \displaybreak[3] \notag \\
&~{+a^4} \int dv\int du~ \frac{v_q-v}{1-a\brr(Z_q,t)}\frac{\brr(Z_r,t)}{1-a\brr(Z_r,t)}\frac{\bfr(Z_q,v,t)}{1-a\brr(Z_q,t)} \notag \\
&~~~~~~~~~~~~~~~~~~~~~~~~~~~~~~~\times~~\big[ \Theta(Z_r-Z_q)\Gamma_t'(Z_q)- \Theta(Z_r-Z_q)\Gamma_t'(Z_q)\big] \displaybreak[3] \notag \\
&~{+a^4} \int dv\int du~ \frac{v_r-u}{1-a\brr(Z_r,t)}\frac{\bfr(Z_q,v,t)}{1-a\brr(Z_q,t)} \frac{\bfr(Z_r,u,t)}{1-a\brr(Z_r,t)} \notag \\
&~~~~~~~~~~~~~~~~~~~~~~~~~~~~~~~\times~~
\Big[  \Gamma'_t(Z_r)\Theta\lf Z_q-Z_r\rf
-\Gamma'_t(Z_r)\Theta\lf Z_q-Z_r\rf \Big].  \displaybreak[3] \notag 
\end{align}
Simplifying, we get 
\begin{align}
&\ell\partial_{dt}\mb C^{\rm gge}_{t,t}(Z_q,v_q;Z_r,v_r)
=0, ~~\text{for}~~Z_q\ne Z_r.\notag
\end{align}

\paragraph{\textbf{\textit{The $\bm{Z_q=Z_r}$ case:}}} 
We therefore now focus on the case $Z_q=Z_r$.
Following a similar calculation we find 
\begin{align}
\label{ex:dmbC^gge_tt-dt->0-a-1}
\begin{split}
\ell \lf\partial_{dt}\mb C^{\rm gge}_{t,t}(Z_q,v_q;Z_q,v_r)\rf_{dt \to 0} 
&~= ~a^2 \int dv \frac{(v_q-v)\bfr(Z_q,v,t)}{1-a\brr(Z_q,t)}\big[\Theta(v_r-v_q) - \Theta(v-v_q) \big] \\
&~~~~~+a^2 \int du \frac{(v_r-u)\bfr(Z_q,u,t)}{1-a\brr(Z_q,t)}
\big[\Theta(v_q-v_r)- \Theta(u-v_r)\big]. 
\end{split}
\end{align}
We now analyze the two cases: (1) $v_q>v_r$ and (2) $v_q<v_r$, separately. For case (1), we can rewrite 
\begin{align}
\label{ex:dmbC^gge_tt-dt->0-a-2}
\ell &\lf\partial_{dt}\mb C^{\rm gge}_{t,t}(Z_q,v_q;Z_q,v_r)\rf_{dt \to 0}  \displaybreak[3] \notag \\ 
&~~~~~= \frac{a^2}{2} \Big[\underset{\st{1}}{\int du \frac{(v_r-u)\bfr(Z_q,u,t)}{1-a\brr(Z_q,t)}\Theta(v_r-u)} + \underset{\st{2}}{\int dv \frac{(v-v_q)\bfr(Z_q,v,t)}{1-a\brr(Z_q,t)} \Theta(v-v_q)}, \displaybreak[3] \notag \\
&~~~~~~~~~~~~
+\underset{\st{3}}{\int dv \frac{(v-v_r)\bfr(Z_q,v,t)}{1-a\brr(Z_q,t)} \Theta(v-v_r)}
-\underset{\st{4}}{\int du \frac{(u-v_q)\bfr(Z_q,u,t)}{1-a\brr(Z_q,t)}\Theta(u-v_q)}  \displaybreak[3]  \\
&~~~~~~~~~~~~
\underset{\st{1}}{\int du \frac{(v_r-u)\bfr(Z_q,u,t)}{1-a\brr(Z_q,t)}\Theta(v_r-u)} + \underset{\st{2}}{\int dv \frac{(v-v_q)\bfr(Z_q,v,t)}{1-a\brr(Z_q,t)} \Theta(v-v_q)}, \displaybreak[3] \notag \\
&~~~~~~~~~~~~
+\underset{\st{5}}{\int dv \frac{(v_q-v)\bfr(Z_q,v,t)}{1-a\brr(Z_q,t)} \Theta(v_q-v)}
-\underset{\st{6}}{\int du \frac{(v_r-u)\bfr(Z_q,u,t)}{1-a\brr(Z_q,t)}\Theta(v_r-u)} \displaybreak[3] \notag \\
&~~~~~~~~~~~~~~~~~~~~~~
-\lf~ \st{3}-\st{4}+\st{5}-\st{6}~\rf \Big] \displaybreak[3] \notag \\
&~~~~~= \frac{a^2}{2} \big[ \lf~\st{1}+\st{3}-\st{4}+\st{2} ~\rf 
+ \lf~\st{1}+\st{2}-\st{6}+\st{5} ~\rf - \lf~ \st{3}-\st{4}+\st{5}-\st{6}~\rf \big]. \displaybreak[3] \notag 
\end{align}
Finally, we have 
\begin{align}
\label{ex:dmbC^gge_tt-dt->0-a-vq>vr}
\ell \lf\partial_{dt}\mb C^{\rm gge}_{t,t}(Z_q,v_q;Z_q,v_r)\rf_{dt \to 0} 
&= ~\frac{a^2}{2} \Big[\int du \frac{|v_r-u|\bfr(Z_q,u,t)}{1-a\brr(Z_q,t)} +\int du \frac{|v_q-u|\bfr(Z_q,u,t)}{1-a\brr(Z_q,t)} \displaybreak[3] \notag \\ 
&~~~~ - \left \{ \int du \frac{(v_q-u)\bfr(Z_q,u,t)}{1-a\brr(Z_q,t)} -\int du \frac{(v_r-u)\bfr(Z_q,u,t)}{1-a\brr(Z_q,t)} \right\} \Big], \notag \\
&= \frac{a^2}{2} \Big[ D_t(Z_q,v_q)+D_t(Z_q,v_r)-\frac{(v_q-v_r)\brr(Z_q,t)}{1-a\brr(Z_q,t)}\Big],~~\text{for}~~v_q>v_r.
\end{align}
where
\begin{align}
 D_t(X,v) = \int du ~\frac{|v-u|\bfr(X,u,t)}{1-a\brr(X,t)}.
 \label{def:D_t(X,v)}
\end{align}
Similarly, one can show that
\begin{align}
\label{ex:dmbC^gge_tt-dt->0-a-vq<vr}
\ell &\lf\partial_{dt}\mb C^{\rm gge}_{t,t}(Z_q,v_q;Z_q,v_r)\rf_{dt \to 0}  = \frac{a^2}{2} \Big[ D_t(Z_q,v_q)+D_t(Z_q,v_r)-\frac{(v_r-v_q)\brr(Z_q,t)}{1-a\brr(Z_q,t)}\Big],~~\text{for}~~v_q<v_r.
\end{align}
Combining the results of Eqs.~\eqref{ex:dmbC^gge_tt-dt->0-a-vq>vr} and \eqref{ex:dmbC^gge_tt-dt->0-a-vq<vr}, we have 
\begin{align}
\label{ex:dmbC^gge_tt-dt->0-a-fn}
\ell &\lf\partial_{dt}\mb C^{\rm gge}_{t,t}(Z_q,v_q;Z_q,v_r)\rf_{dt \to 0}  = \frac{a^2}{2} \Big[ D_t(Z_q,v_q)+D_t(Z_q,v_r)-\frac{|v_r-v_q|\brr(Z_q,t)}{1-a\brr(Z_q,t)}\Big],
\end{align}
which, along with Eqs.~\eqref{ex:C_tt-dt->0-a} and \eqref{ex:dmbC^gge_tt-dt->0-a-x_ne_y}, proves the result in Eq.~\eqref{mcj_f^gge} with $D_t(Z,v)$ given in Eq.~\eqref{ex:mcj_f^gge}.
Using the result from Eqs.~\eqref{ex:dmbC^gge_tt-dt->0-a-fn} and ~\eqref{ex:dmbC^gge_tt-dt->0-a-x_ne_y} in Eq.~\eqref{ex:mbC_tt-O(dt)}, we get the following  relation 
\begin{subequations}
\label{ex:dmbC^gge_tt-dt->0-a-sm}
\begin{align}
\label{ex:dmbC^gge_tt-dt->0-a-sm-1}
\bigg[\frac{\mb C_{t,t'}(Z,v;Z',u)}{dt}\bigg]_{dt \to 0} = \frac{\delta(t-t')}{\ell}
\begin{cases}
    0 &~~\text{for}~~Z\ne Z' \\
    \Sigma_t(Z,v,u) &~~\text{for}~~Z=Z' 
\end{cases}, 
\end{align}
where
\begin{align}
  \Sigma_t(Z,v,u)=\frac{a^2}{2} \Big[ D_t(Z,v)+D_t(Z,u)-|v-u|\frac{\brr(Z,t)}{1-a\brr(Z,t)}\Big]. 
  \label{ex:Sigma_t}
\end{align}
\end{subequations}

\section{Proof of Eq.~\eqref{cancellation} for ${\rm IC}_{\rm fhp}$ initial state:}
\label{app:cancellation}
To prove this relation, we first decompose the correlation $\msc C^{\rm r}(X,v;Y,u;t)$ into asymmetric and symmetric parts, as 
\begin{subequations}
\label{decompose-C-sym-asymp-a}
  \begin{align}
\msc C^{\rm r}_{\rm lr}(X,v;Y,u;t) \overset{X \approx Y}{ =} 
\msc C^{\rm r,sym}_{\rm lr}(X,v,u,t) + \text{sgn}(X-Y)\msc C^{\rm r,asym}_{\rm lr}(X,v,u,t),
\end{align}
where  
\begin{align}
\begin{split}
\msc C^{\rm r,sym}_{\rm lr}(X,v,u,t)=& ~a^2 \lf \partial_X \bfr(X,v,t)\rf \lf \partial_X \bfr(X,u,t)\rf \bFr(X,t)  \\
&~~~~~~~~~~~~~~~- \frac{a}{2}(1-a\brr(X,t))\partial_X\big( \bfr(X,v,t) \bfr(X,u,t)\big), 
\end{split}
\label{ex:C_sym}\\
\msc C^{\rm r,asym}_{\rm lr}(X,v,u,t)=& -\frac{a}{2}(1-a\brr(X,t)) \Big[ 
\bfr(X,u,t)\partial_X \bfr(X,v,t) -\bfr(X,v,t)\partial_X \bfr(X,u,t)\Big], \label{ex:C_asym} 
\end{align}
with $\bFr(X,t)=\int dZ \int dv~\Theta(X-Z)\bfr(Z,v,t)$.
\end{subequations}
Inserting the asymmetric part into the expression in Eq.~\eqref{def:mcj_d^lr-a}, we get 
\begin{align}
\mt{j}_{\rm lr}^{\rm asym}(Z,v,t)=& \ovl{\mc j}_{\rm d}^{\rm lr,asym}(Z,v,t)\bfr(Z,v,t) \\
=& \frac{a^2}{2} ~\int du~\int du'\int dw~(v-u) \displaybreak[3] \notag \\ 
    &~~\times~~\left[\delta(u-u')+\frac{a\bfr(Z,u,t)}{1-\brr(Z,t)}\right]~\left[\delta(v-w)+\frac{a\bfr(Z,v,t)}{1-\brr(Z,t)}\right] \displaybreak[3] \notag \\
    &~~~~\times~~{\rm sgn}(v-u) \Big[ 
\bfr(Z,u',t)\partial_X \bfr(Z,w,t) -\bfr(Z,w,t)\partial_X \bfr(Z,u',t)\Big], \displaybreak[3] \notag \\
=& \frac{a^2}{2} \int du~\frac{|v-u|}{1-a\brr(Z,t)}\Big[ \bfr(Z,v,t)\partial_Z\bfr(Z,u,t) -\bfr(Z,u,t)\partial_Z\bfr(Z,v,t)\Big] \displaybreak[3] \notag \\
=& -\frac{1}{2}\int du~\msc D(v,u)\partial_Z \bfr(Z,u,t), \label{mtJ_lr_asym}
\end{align}
where 
\begin{align}
 \msc{D}(v,u) = \frac{1}{1-\brr(Z,t)}\lf \delta(u-v) \int dw~|v-w|~\bfr(Z,w,t) - |u-v|~ \bfr(Z,v,t)\rf.  
 \label{def:mscD-a}
\end{align} 
We now evaluate 
\begin{align}
  \mt j_{\rm gge}(Z,v,t)= \ovl{\mc j}_{\rm d}^{\rm gge}(Z,v,t)\bfr(Z,v,t) - \frac{a^2}{2}\partial_Z \big(D_t(Z,v)\bfr(Z,v,t) \big). 
\end{align}
Inserting the expression of $\ovl{\mc j}_{\rm d}^{\rm gge}(Z,v,t)$ from Eq.~\eqref{def:ovl(mcj)_d-a} and $D_t(Z,v)$ from Eq.~\eqref{def:D_t(X,v)}, we get
\begin{align}
 \mt j_{\rm gge}(Z,v,t)=&    \frac{a^2\bfr(Z,v,t)}{2} \lsq\frac{\partial_{Z}\int du~|v-u|\bfr(Z,u,t)}{1-a\brr(Z,t)} 
 + \partial_Z\lf \frac{\int du~|v-u|\bfr(Z,u,t)}{1-a\brr(Z,t)}\rf \rsq \displaybreak[3] \notag \\
 &~~~~~~~~~~~~~~~~~~~~~~-\frac{a^2}{2} \partial_Z \lf \bfr(Z,v,t) \int du ~\frac{|v-u|\bfr(Z,u,t)}{1-a\brr(Z,t)}\rf, \displaybreak[3] \notag \\
=&    \frac{a^2}{2}\int du~\frac{|v-u|}{1-a\brr(Z,t)}\Big[ \bfr(Z,u,t)\partial_Z\bfr(Z,v,t)-\bfr(Z,v,t)\partial_Z\bfr(Z,u,t)\Big], \displaybreak[3] \notag \\
=& \frac{1}{2}\int du~\msc D(v,u)\partial_Z \bfr(Z,u,t). \label{mtJ_gge}
\end{align}
Adding the results from Eqs.~\eqref{mtJ_lr_asym} and \eqref{mtJ_gge} yields 
$ \mt j_{\rm gge}(Z,v,t)+\mt{j}_{\rm lr}^{\rm asym}(Z,v,t)=0$ which proves the result in Eq.~\eqref{cancellation} for the initial state ${\rm IC}_{\rm fhp}$.


\section*{References}

\bibliographystyle{unsrt}
\bibliography{references-sp.bib}%
\end{document}